\documentclass[11pt,twoside,nohyper,preprint]{pallis}
\usepackage{rotating}
\usepackage{bm}
\usepackage{epsfig}
\usepackage{graphics}

\usepackage{fancyh}
\usepackage{epsfig}
\usepackage{amssymb}
\usepackage{latexsym}
\usepackage{times}
\usepackage{slashed}
\usepackage{upgreek}
\usepackage{amsmath}
\usepackage{amsfonts}
\usepackage{amsbsy}
\usepackage{amscd}
\usepackage{bbm}
\usepackage{cite}

\newcommand{\eec}{\end{center}}
\newcommand{\bec}{\begin{center}}

\newcommand{\eem}{\end{matrix}}
\newcommand{\bem}{\begin{matrix}}
\newcommand{\eeq}{\end{equation}}
\newcommand{\beq}{\begin{equation}}
\newcommand{\ba}{\begin{array}}
\newcommand{\ea}{\end{array}}
\newcommand{\bea}{\begin{eqnarray}}
\newcommand{\eea}{\end{eqnarray}}
\newcommand{\baq}{\begin{eqnarray}}
\newcommand{\eaq}{\end{eqnarray}}
\newcommand{\beqs}{\begin{subequations}}
\newcommand{\eeqs}{\end{subequations}}
\newcommand{\bel}{\begin{align}}
\newcommand{\eal}{\end{align}}

\newcommand\eqs[2]{Eqs.~(\ref{#1}) and (\ref{#2})}

\newcommand\eqss[3]{Eqs.~(\ref{#1}), (\ref{#2}) and (\ref{#3})}

\newcommand{\ftn}{\footnotesize}
\newcommand{\nsz}{\normalsize}
\newcommand{\ssz}{\scriptsize}
\newcommand{\TeV}{{\mbox{\rm TeV}}}

\newcommand{\GeV}{{\mbox{\rm GeV}}}

\newcommand{\EeV}{{\mbox{\rm EeV}}}
\newcommand{\PeV}{{\mbox{\rm PeV}}}

\newcommand{\YeV}{{\mbox{\rm YeV}}}

\newcommand{\hz}{{\mbox{\rm Hz}}}
\newcommand{\sFref}[2]{Fig.~\ref{#1}-{\small \sf ({#2})}}
\newcommand{\sEref}[2]{Eq.~(\ref{#1}{\small\sf {#2}})}

\newcommand{\etal}{{\it et al.\/}}

\def\to{\rightarrow}

\def\llgm{\left\lgroup}
\def\rrgm{\right\rgroup}
\def\lf{\left(}
\def\rg{\right)}
\newcommand\vev[1]{\langle {#1} \rangle}
\newcommand\vevl[1]{\left\langle {#1} \right\rangle}

\newcommand{\Gr}{\ensuremath{\widetilde{G}}}
\newcommand{\Yb}{\ensuremath{Y_{B}}}
\newcommand{\Yg}{\ensuremath{Y_{3/2}}}

\newcommand{\ks}{\ensuremath{k_\star}}
\newcommand{\Gsm}{\ensuremath{\mathbb{G}_{\rm SM}}}

\newcommand{\As}{\ensuremath{A_{\rm s}}}
\newcommand{\Ns}{\ensuremath{N_{\star}}}
\newcommand{\Ve}{\ensuremath{V}}
\newcommand{\Vhi}{\ensuremath{V_{\rm TI}}}
\newcommand{\dV}{\ensuremath{\Delta V}}

\newcommand{\Hhi}{\ensuremath{H_{\rm TI}}}

\def\ve{\varepsilon}

\newcommand{\Ms}{\ensuremath{M_{\rm S}}}
\newcommand{\ca}{\ensuremath{{\rm a}}}
\newcommand{\mP}{\ensuremath{m_{\rm P}}}

\newcommand{\Ggut}{\ensuremath{\mathbb{G}}}

\newcommand{\Glr}{\ensuremath{\mathbb{G}_{\rm LR}}}
\newcommand{\Glra}{\ensuremath{\mathbb{G}_{\rm L1R}}}

\newcommand{\grbl}{\ensuremath{SU(2)_{\rm R}\times U_{B-L}}}
\newcommand{\suc}{\ensuremath{SU(3)_{\rm C}}}
\newcommand{\sur}{\ensuremath{SU(2)_{\rm R}}}
\newcommand{\sul}{\ensuremath{SU(2)_{\rm L}}}
\newcommand{\ubl}{\ensuremath{U(1)_{B-L}}}
\newcommand{\ur}{\ensuremath{U(1)_{\rm R}}}

\newcommand{\dphi}{\ensuremath{{\rm I}_-}}

\newcommand{\lm}{\ensuremath{\lambda_\mu}}
\newcommand{\ml}{\ensuremath{m_\lambda}}
\newcommand{\mwr}{\ensuremath{{M_{W^\pm_{\rm R}}}}}
\newcommand{\tra}{\ensuremath{{\sf T}^{\rm a}_{\rm R}}}
\newcommand{\tbl}{\ensuremath{{\sf T}_{BL}}}
\newcommand{\wra}{\ensuremath{{W^{\rm a}_{\rm R\mu}}}}
\newcommand{\wrpm}{\ensuremath{{W^{\pm}_{\rm R}}}}

\newcommand{\aS}{\ensuremath{{\rm a}_S}}

\newcommand{\Gsn}{\ensuremath{\Gamma_{\rm I-}}}
\newcommand{\msn}{\ensuremath{m_{\rm I-}}}

\newcommand{\mgr}{\ensuremath{m_{3/2}}}
\newcommand{\am}{\ensuremath{{\rm a}_{3/2}}}

\newcommand{\hd}{{\ensuremath{H_d}}}
\newcommand{\hu}{{\ensuremath{H_u}}}
\newcommand{\hdm}{{\ensuremath{H_d^-}}}
\newcommand{\hup}{{\ensuremath{H_u^+}}}
\newcommand{\hdn}{{\ensuremath{H^0_d}}}
\newcommand{\hun}{{\ensuremath{H^0_u}}}
\newcommand{\nci}{{\ensuremath{\nu_i^c}}}

\newcommand{\lci}{{\ensuremath{l_i^c}}}

\newcommand{\ncit}{{\ensuremath{\tilde\nu_i^c}}}
\newcommand{\ecit}{{\ensuremath{\tilde e_i^c}}}

\newcommand{\ns}{\ensuremath{n_{\rm s}}}
\newcommand{\as}{\ensuremath{\alpha_{\rm s}}}
\newcommand{\sni}{\ensuremath{\nu^c_i}}
\newcommand{\ssni}{\ensuremath{\tilde \nu_i^c}}

\newcommand{\rhni}{\ensuremath{\nu^c_i}}
\newcommand{\wrhn[1]}{\ensuremath{\nu^c_{#1}}}

\newcommand{\ogw}{\ensuremath{\Omega_{\rm GW}h^2}}

\newcommand{\Dex}{\ensuremath{\Delta_{\rm c\star}}}
\newcommand{\diag}{\ensuremath{{\sf diag}}}

\newcommand{\fp}{\ensuremath{f_{\rm T}}}

\newcommand{\Trh}{\ensuremath{T_{\rm rh}}}

\newcommand{\sg}{\ensuremath{\sigma}}
\newcommand{\se}{\ensuremath{\widehat\sigma}}
\newcommand{\sex}{\ensuremath{\widehat{\sigma}_*}}
\newcommand{\sef}{\ensuremath{\widehat{\sigma}_{\rm f}}}

\newcommand{\sgx}{\ensuremath{\sigma_\star}}
\newcommand{\sgc}{\ensuremath{\sigma_{\rm c}}}

\newcommand{\sgf}{\ensuremath{\sigma_{\rm f}}}
\newcommand{\ld}{\ensuremath{\lambda}}

\newcommand{\Ld}{\ensuremath{\Lambda}}

\newcommand{\eph}{\ensuremath{\epsilon}}

\newcommand{\ldu}{\ensuremath{\uplambda}}

\def\ssb{\leavevmode\hbox{$\diagup$\kern-12pt\ftn\scshape
susy}}

\newcommand{\GNsn}{\ensuremath{{\Gamma}_{{\rm I}_-\to \nu_i^c}}}
\newcommand{\Ghsn}{\ensuremath{{\Gamma}_{{\rm I}_-\to \hu\hd}}}

\newcommand{\Eref}[1]{Eq.~(\ref{#1})}
\newcommand{\Sref}[1]{Sec.~\ref{#1}}
\newcommand{\Fref}[1]{Fig.~\ref{#1}}
\newcommand{\Tref}[1]{Table~\ref{#1}}
\newcommand{\cref}[1]{Ref.~\cite{#1}}
\newcommand{\crefs}[1]{Refs.~\cite{#1}}

\def\th{{\theta}}

\def\thn{{\vartheta_{\nu}}}
\def\thp{{\vartheta}}
\def\thpb{{\bar\vartheta}}

\def\Ka{K\"{a}hler potential}
\def\Kaa{K\"{a}hler}

\def\Kap{K\"{a}hler potential}

\newcommand{\plk}{{\it Planck}}

\newcommand{\bdhh}{{\ensuremath{\normalsize I{\kern-2.9pt H}}}}

\newcommand\lin[2]{\mbox{$\llgm\bem #1& #2\eem\rrgm$}}

\newcommand\stl[2]{\mbox{$\llgm\bem #1\cr #2\eem\rrgm$}}

\newcommand{\mrh[1]}{\ensuremath{M_{#1\nu^c}}}
\newcommand{\mrha}{\ensuremath{M_{1\nu^c}}}

\newcommand{\mntau}{\ensuremath{m_{\nu_\tau}}}

\newcommand{\phc}{\ensuremath{\Phi}}
\newcommand{\phcb}{\ensuremath{\bar\Phi}}
\newcommand{\what}{\ensuremath{\widehat}}
\newcommand{\wtilde}{\ensuremath{\widetilde}}

\newcommand{\nsu}{\ensuremath{N_0}}
\newcommand{\nk}{\ensuremath{N}}

\def\aal{{\bar\alpha}}
\def\bbet{{\bar\beta}}
\def\al{{\alpha}}
\def\bt{{\beta}}
\def\gm{{\gamma}}
\def\ggm{{\bar\gamma}}
\def\aal{{\bar\alpha}}
\def\bbet{{\bar\beta}}

\def\bi{{b_l}}
\def\di{{d_l}}
\def\ali{{\alpha_l}}

\def\alaz{{\alpha_{1Z}}}
\def\albz{{\alpha_{2Z}}}
\def\alcz{{\alpha_{3Z}}}
\def\alu{{\alpha_{\rm U}}}
\def\bali{{\bar\alpha_l}}
\def\balu{{\bar\alpha_{\rm U}}}
\def\bbi{{\bar b_l}}
\def\cx{{\cal X}}

\def\cy{{\cal Y}}

\def\bcy{{\cal\bar Y}}

\def\mun{{M_{\rm U}}}
\def\bmu{{\bar M_{\rm U}}}
\def\tu{{t_{\rm U}}}
\def\tZ{{t_{Z}}}
\def\btu{{\bar t_{\rm U}}}
\def\tint{{t_{\rm I}}}
\def\mint{{M_{\rm I}}}
\def\mtpm{{m_{T\pm}}}
\def\rd{{r_{d}}}
\def\brd{{\bar r_{db}}}

\newcommand{\mcs}{\ensuremath{{\mu_{\rm cs}}}}
\newcommand{\gmcs}{\ensuremath{{G\mu_{\rm cs}}}}

\newcommand{\rms}{\ensuremath{{r_{\rm ms}}}}
\newcommand{\srms}{\ensuremath{{r_{\rm ms}^{1/2}}}}

\def\nano{{\sf\small NG15}}

\newcommand{\nhc}{\ensuremath{\nu^c_\Phi}}
\newcommand{\nhcb}{\ensuremath{\bar\nu^c_\Phi}}

\newcommand{\ehc}{\ensuremath{e^c_\Phi}}
\newcommand{\ehcb}{\ensuremath{\bar e^c_\Phi}}

\newcommand{\btd}{\ensuremath{\bar T}}
\newcommand{\btdm}{\ensuremath{\bar T_-}}
\newcommand{\btdp}{\ensuremath{\bar T_+}}
\newcommand{\btdn}{\ensuremath{\bar T_0}}
\newcommand{\tdn}{\ensuremath{T_0}}
\newcommand{\tdm}{\ensuremath{T_-}}
\newcommand{\tdp}{\ensuremath{T_+}}

\newcommand{\dnp}{\ensuremath{\delta\nu_{\Phi+}}}
\newcommand{\dnm}{\ensuremath{\delta\nu_{\Phi-}}}
\newcommand{\dT}{\ensuremath{\delta T_0}}
\newcommand{\dnhc}{\ensuremath{\delta\nu_\Phi^c}}
\newcommand{\dnhcb}{\ensuremath{\delta\bar\nu^c_\Phi}}

\newcommand{\lp}{\ensuremath{\lambda_{\Phi}}}
\newcommand{\lt}{\ensuremath{\lambda_T}}
\newcommand{\mt}{\ensuremath{M_T}}
\newcommand{\vt}{\ensuremath{v_{T}}}
\newcommand{\vx}{\ensuremath{v_{\Phi}}}
\newcommand{\wh}{\ensuremath{W_{\rm H}}}

\newcommand{\we}{\ensuremath{W_{\rm E}}}
\newcommand{\qq}{\ensuremath{{\rm I}_{-}}}
\newcommand{\Lg}{\ensuremath{\mathcal{L}}}

\newcommand{\hh}{{\ensuremath{
I{\kern-2.6pt h}}}}
\newcommand{\bhh}{{\ensuremath{\bar{
I{\kern-2.6pt h}}}}}

\newcommand{\tad}{\mbox{$\llgm\bem T_0/\sqrt{2}&T_+\cr T_-&-T_0/\sqrt{2}\eem\rrgm$}}
\newcommand{\tadb}{\mbox{$\llgm\bem \bar T_0/\sqrt{2}&\bar T_+\cr\bar T_-&-\bar T_0/\sqrt{2}\eem\rrgm$}}

\def\veps{\boldsymbol{\varepsilon}}
\newcommand{\Tr}{\mbox{\sf Tr}}
\newcommand{\tr}{{\mbox{\sf\ssz T}}}

\newcommand{\rcc}{\ensuremath{\mathcal{R}}}
\newcommand{\rce}{\ensuremath{{\mathcal{R}}}}
\newcommand{\geu}{\ensuremath{ g}}

\renewenvironment{subequations}{%
\refstepcounter{equation}%
\setcounter{parentequation}{\value{equation}}%
  \setcounter{equation}{0}
  \def\theequation{\thesection.\theparentequation{\sf\ftn \alph{equation}}}%
  \ignorespaces
}{%
  \setcounter{equation}{\value{parentequation}}%
  \ignorespacesafterend
}

\title{\LARGE\boldmath \bfseries\scshape T-Model Higgs Inflation
and \\ Metastable Cosmic Strings}

\author{\Large \bfseries\scshape C. Pallis\\
School of Civil Engineering, \\ Faculty of Engineering,\\
Aristotle University of Thessaloniki,  \\ GR-541 24 Thessaloniki,
GREECE\\ \vspace{3pt} \email{kpallis@auth.gr}}

\abstract{We present the formation of metastable cosmic strings
(CSs) in the context of a supersymmetric (SUSY) left-right model.
The spontaneous $\sur$ symmetry breaking occurs during a stage of
T-model (Higgs) inflation (TI) driven by an $\sur$ triplet
superfield which inflates away the produced monopoles. The
subsequent breaking of the remaining $\ur\times\ubl$ symmetry,
triggered due to an instability arising in the system of a pair of
$\sur$ doublet superfields, leads to the production of CSs. TI is
based on a quartic potential, is consistent with data thanks to
the adopted hyperbolic \Kaa\ geometry and may be followed by
successful non-thermal leptogenesis. The decay of the produced CSs
interprets the recent observations from PTA experiments on the
stochastic background of gravitational waves with values of the
superpotential coupling constants close to $10^{-6}-10^{-8}$ and
symmetry-breaking scales a little lower than the SUSY grand
unified theory scale. A solution to the $\mu$ problem of the MSSM
is also accommodated provided that $\mu$ is two to three orders of
magnitude lower than the gravitino mass. The issue of the gauge
coupling unification is also discussed.

\\ \\ {\ftn\sffamily {\scshape Keywords}:  Cosmology, Inflation, Supersymmetric Models} \\
{\ftn\sffamily {\scshape PACS codes}:  98.80.Cq, 12.60.Jv,
95.30.Cq, 95.30.Sf}}

\begin{document}

\setcounter{page}{1} \pagestyle{fancyplain}

\addtolength{\headheight}{.5cm}

\rhead[\fancyplain{}{ \bf \thepage}]{\fancyplain{}{\sc TI and
Metastable CSs}} \lhead[\fancyplain{}{\sc
\leftmark}]{\fancyplain{}{\bf \thepage}} \cfoot{}

\section{Introduction}\label{intro}

The discovery of a \emph{gravitational wave} ({\sf\small GW})
background around the nanohertz frequencies announced from several
\emph{pulsar timing array} ({\sf\small PTA}) experiments
\cite{pta,pta1,pta2} -- most notably the \emph{NANOGrav 15-years
results} ({\sf\small \nano}) \cite{nano, nano2} -- provide a novel
tool in exploring the structure of the early universe
\cite{ellisnano, roshan}. Given that the interpretation of this
signal in terms of the merging of supermassive black hole binaries
is somewhat disfavored \cite{nano1}, its attribution to
gravitational radiation \cite{wells, servant, pillado} emitted by
topologically unstable superheavy \emph{cosmic strings}
({\sf\small CSs}) -- which may arise after the end of inflation --
attracts a fair amount of attention \cite{buch, buch1, quasi,
so102, so103, so104, nasri, blfhi, nasri1, su5, leont, park, infl,
infl1, infl2, msg}. It is worth mentioning that in the large
majority of \emph{Grand Unified Theories} ({\sf \small GUTs}) the
formation of CSs results as an unavoidable effect \cite{rachel}
during their \emph{spontaneous symmetry breaking} ({\sf \small
SSB}) chains down to the \emph{Standard Model} ({\sf \small SM})
gauge group,
\beq \mathbb{G}_{\rm SM}  :=  SU(3)_{\rm C} \times SU(2)_{\rm L}
\times U(1)_Y. \label{gsm} \eeq

In particular, the observations can be interpreted if the CSs are
metastable \cite{buch,nano1} -- for a variant with quasi-stable
CSs see \cref{quasi}. This type of CSs arise from a SSB
\cite{gastro} of the form
\beq
\mathbb{G}\underset{\rm
MMs}{\xrightarrow{\hspace{0.4cm}\hspace{0.4cm}}} \mathbb{G}_{\rm
int} \times U(1)\underset{\rm
CSs}{\xrightarrow{\hspace{0.4cm}\hspace{0.4cm}}} \mathbb{G}_{\rm
f} ~~\mbox{with}~~\pi_1(\mathbb{G}/\mathbb{G}_{\rm
int})=\pi_1(\mathbb{G}/\mathbb{G}_{\rm
f})=I~~\mbox{but}~~\pi_1(\mathbb{G}_{\rm int} \times
U(1)/\mathbb{G}_{\rm f})\neq I, \label{chain}\eeq
where {\sf\small MMs} stands for \emph{magnetic monopoles} and
$\pi_n$ is the $n^{\rm th}$ homotopy group. In the first stage of
the chain above MMs are produced -- since
$\pi_2(\mathbb{G}/\mathbb{G}_{\rm int} \times U(1))\neq I$ --
whereas CSs are formatted in the latter stage.  This is possible
if, e.g., $\mathbb{G}_{\rm int}=U(1)'$ and $\mathbb{G}_{\rm
f}=U(1)''$ where $U(1)$ is not orthogonal to $U(1)'$ \cite{buch,
buch1, leont, park, infl1, infl2, infl}. The generated CSs are
topologically unstable due to the penultimate prerequisite in
\Eref{chain} which does not permit stable CSs in the full theory.
Trying to keep the dimensionality of the gauge groups involved as
minimal as possible we here identify $\mathbb{G}$ with the
left-right gauge group \cite{tfhi}
\beq \Glr:=SU(3)_{\rm C}\times SU(2)_{\rm L}\times SU(2)_{\rm
R}\times U(1)_{B-L},\label{glr}\eeq
and therefore, $\mathbb{G}_{\rm int}\times U(1)$ is specified as
\beq \Glra:= SU(3)_{\rm C} \times SU(2)_{\rm L} \times U(1)_{\rm
R} \times U(1)_{B-L}\label{glra} \eeq
whereas $\mathbb{G}_{\rm f}$ coincides with $\Gsm$ in \Eref{gsm}.
We may assume that $\Glr$ is a remnant of another more fundamental
gauge group -- such as $SO(10)$ \cite{so102,so103,so104} or the
Pati-Salam \cite{nasri1} -- which is already broken at higher
energies.



From cosmological point of view, the production of MMs in
\Eref{chain} is catastrophic. For this reason, MMs are to be
inflated away. However, they can appear on CSs via quantum pair
creation \cite{vile, gastro} causing them to decay emitting GWs
\cite{csdecay, csdc,wells,servant,pillado}. The dilution of MMs
can be realized, if $\Ggut$ is broken before or during inflation.
To liberate our proposal from our ignorance about the
preinflationary period we here seek the MMs production during
inflation as, e.g., in \cref{infl, infl1}. Among the various
inflationary models -- for reviews see \crefs{nsreview, review,
review1} --, we here focus on Higgs inflation which is directly
relied to SSB. Indeed, according to this model the ``higgsflaton''
\cite{kaloper} may play, at the end of its inflationary evolution,
the role of a Higgs field \cite{nmh, gut2, gut, gut1, gut3, gut4,
ighi,unit, jhep, nmhk, univ, sor,sor1,sor2} leading to SSB. More
specifically, we here concentrate on \emph{T-model} (Higgs)
\emph{Inflation} ({\small\sf TI}) \cite{tmodel} based on the
quartic potential \cite{sor}. This kind of Higgs inflation is
obtained in the presence of a pole \cite{sor1} of order two in the
inflaton kinetic term. Since the natural framework of a GUT is
\emph{Supersymmetry} ({\sf\small SUSY}) -- and its topical
extension, \emph{Supergravity} ({\sf\small SUGRA}) -- where the
gauge hierarchy problem is set under control, we analyze TI within
SUGRA following \cref{sor,sor2}. The aforementioned kinetic mixing
is elegantly achieved by adopting a \Kap\ which parameterizes
hyperbolic \Kaa\ geometry \cite{tkref, sor}. The corresponding
constant moduli-space scalar curvature depends on the prefactor of
the logarithm in the inflationary part of $K$ and is related to
the tensor-to-scalar ratio $r$. On the other hand, the
construction of the superpotential $W$ of the model can be
systematically worked out by imposing a global $R$ symmetry
\cite{fhi} -- not to be confused with the gauge $\sur$ or $\ur$
symmetries included in $\Glr$ and $\Glra$ respectively. The
reasons for the selection of the specific inflationary model are
the following:

\begin{itemize}

\item[{\sf\small (a)}] The Higgs fields are non zero during TI and
so the MMs produced in \Eref{chain} are automatically inflated
away, i.e., without need for selecting alternative inflationary
tracks as in the case of F-term hybrid inflation
\cite{tfhi,jean,ax,infl1};

\item[{\sf\small (b)}] The resulting inflationary observables are
``spontaneously'' compatible with observations \cite{plin, sor,
sor2} i.e., without need to invoke higher order terms in the \Ka\
as in \cref{infl1, infl2};

\item[{\sf\small (c)}] An instability can be easily embedded in
the inflationary path (a little after its termination) thanks to a
$W$ term which may lead to the formation of CSs -- cf.~\cref{unit,
univ, ighi}.

\end{itemize}

The accomplishment of the first stage of SSB in \Eref{chain}
guides us to consider as inflaton a $B-L$ neutral Higgs field in
the adjoint representation of $\sur$ -- i.e., a $\sur$ triplet
superfield --, whereas the second step of SSB requires the
inclusion of a pair of $B-L$ charged Higgs fields in the
fundamental representation of $\sur$ -- i.e., a pair of $\sur$
doublets. Due to the $R$ symmetry we introduce a pair of $\sur$
triplets and only one $\Glr$ singlet superfield, in contrast to
the similar model discussed in \cref{buch, buch1}. Moreover,
thanks to a specific $W$ coupling, an instability occurs which
allows for the realization of the second stage of SSB after the
end of TI. The \emph{vacuum expectation values} ({\sf\small
v.e.vs}) of the various Higgs fields may be close to its other for
more or less natural values of the relevant parameters. This
effect implies a proximity between the scales associated with the
formation of the MMs and CSs which assures that the metastability
factor $\rms$ \cite{vile,csdecay} is low enough so that the
nucleation rate of the CSs is not negligibly suppressed although
exponentially decreasing as a tunneling process.

The model predicts the presence of three right-handed neutrinos
$\sni$ which acquire intermediate scale masses from appropriate
$W$ terms and in turn, generate the tiny neutrino masses via the
type I seesaw mechanism. Furthermore, the out-of-equilibrium decay
of the $\sni$'s, which are produced during reheating, provides us
with an explanation of the observed \emph{baryon asymmetry of the
universe} ({\sf\small BAU}) \cite{plcp} via \emph{non-thermal
leptogenesis} ({\sf\small nTL}) \cite{lept, dreeslept, zhang}.
Consistency of the BAU, gravitino ($\Gr$) constraint
\cite{brand,kohri} and the neutrino data \cite{valle} can be
achieved in regions of the parameter space with low $r$ values.
Also, taking advantage of the adopted $R$ symmetry, the parameter
$\mu$ appearing in the mixing term between the two electroweak
Higgs fields in the superpotential of the \emph{Minimal SUSY
Standard Model} ({\sf\small MSSM}) can be explained as in
\crefs{dls, unit,univ,ighi} via the v.e.v of an inflaton-coupled
field which appears linearly in $W$ -- for other approaches to
this problem see \cref{mubaer}. The relevant coupling constant is
to be appropriately suppressed and so a mild hierarchy between
$\mu$ and $\Gr$ mass, $\mgr$, is predicted. As a consequence, our
model fits well with natural SUSY as it is defined in
\cref{natsusy}.

The remaining manuscript is built up as follows: In \Sref{md} we
describe our model. Following, we analyze the inflationary and the
post-inflationary era in \Sref{hi} and \ref{phi} respectively.
Then, we exhibit the relevant imposed constraints and restrict the
parameters of our model in \Sref{res}. Our conclusions are
summarized in Sec.~\ref{con}. An accommodation of the gauge
coupling unification within our model is presented in
Appendix~\ref{app} following \cref{magic, magic0}. Lastly, we
investigate the nature of the waterfall phase \cite{pbha, pbhb,
cles, man, lindefr} which connects TI and reheating in
Appendix~\ref{appw}.

Throughout, the complex scalar components of the various Higgs
superfields are denoted by the same superfield symbol, whereas the
scalar components of the matter chiral superfields are denoted
with tilde -- e.g., $\tilde q_i^c$ are the scalar super patterns
of $q_i^c$. Also, charge conjugation is denoted by a star ($^*$)
-- e.g., $|Z|^2=ZZ^*$ -- and the symbol $,Z$ as subscript denotes
derivation \emph{with respect to} ({\small\sf w.r.t}) $Z$. Unless
otherwise stated, we use units where the reduced Planck scale $\mP
= 2.43\cdot 10^{18}~\GeV$ is equal to unity.

\section{Model Set-up}\label{md}

We present in Sec.~\ref{md0} below the building blocks of our
model and then, in \Sref{md1} and \ref{md2}, we specify the
various parts of its superpotential and its \Ka, respectively.

\subsection{Superfield Content and Symmetries} \label{md0}

As already mentioned, we adopt the left-right symmetric gauge
group $\Glr$ in \Eref{glr}. We focus on a simplified version of
the model presented in \cref{tfhi}. The components, the
representations under $\Glr$ and the transformations under
$\Glr/U(1)_{B-L}$ of the various matter and Higgs superfields of
the model are presented in Table~\ref{tab1}. The model also
possesses two global ${U(1)}$ symmetries, namely an $R$ symmetry
and an accidental baryon-number ($B$) symmetry which leads to a
stable proton \cite{dls}. The corresponding charges are shown in
Table~\ref{tab1} too. For simplicity we do not impose Peccei-Quinn
symmetry and do not introduce extra Higgs fields charged under
$\sul$ as in \cref{tfhi}.

\renewcommand{\arraystretch}{1.3}

\begin{table}[!t]
\begin{center}
\begin{tabular}{|ccccc|}
\hline {\sc Superfields}&{\sc Representations}&{\sc
Transformations}&\multicolumn{2}{c|}{\sc Global }\\
\multicolumn{1}{|c}{}&{\sc Under \Glr}&{\sc Under
}&\multicolumn{2}{c|}{\sc Charges}\\
\multicolumn{2}{|c}{}&{$\Glr/U(1)_{B-L}$}&$R$&$B$\\\hline \hline
\multicolumn{5}{|c|}{\sc Matter Superfields}\\ \hline\hline
{$l_i=\lin{\nu_{i}}{e_{i}}$} &{$({\bf 1, 2, 1}, -1)$}&$l_iU_{\rm L}^{\dagger}$&$1$&$0$\\
{$q_i=\lin{u_{i}}{d_{i}}$} &{$({\bf 3, 2, 1}, 1/3)$}&$q_iU_{\rm
L}^{\dagger} U_{\rm C}^\tr$&$1$&$1/3$\\\hline
{$l^c_i=\stl{-\nu^c_{i}}{e^c_{i}}$} & {$({\bf 1, 1, 2}, 1)$}&$U_{\rm R}^\ast l^c_i$&$1$&$0$\\
{$q^c_i=\stl{-u^c_{i}}{d^c_{i}}$} & {$({\bf \bar 3, 1, 2},-1/3)$}
&$U_{\rm C}^\ast U_{\rm R}^\ast q^c_i$&$1$&$-1/3$\\
\hline\hline
\multicolumn{5}{|c|}{\sc Higgs Superfields}
\\ \hline\hline
{$S$} & {$({\bf 1, 1, 1},0)$}&$S$&$2$&$0$\\ \hline
{$\bar \Phi=\lin{\bar\nu^c_{\Phi}}{\bar e^c_{\Phi}}$}&$({\bf 1, 1,
2},-1)$&
$\bar\Phi U^\tr_{\rm R} $&$0$&$0$\\
{$\Phi=\stl{\nu^c_{\Phi}}{e^c_{\Phi}}$} &{$({\bf 1, 1, 2}, 1)$}& $U_{\rm R}^\ast\Phi $&$0$&$0$\\
\hline
{$T=\tad$} &{$({\bf 1, 1, 3},0)$}&$U_{\rm R} T U_{\rm R}^\tr$ &$0$&$0$\\
{$\btd=\tadb$} &{$({\bf 1, 1, 3},0)$}&$U_{\rm R} \btd U_{\rm
R}^\tr$ &$2$&$0$\\ \hline
{$\hh=\llgm\bem H_{u}&H_{d}\eem\rrgm$} & {$({\bf1, 2, 2},0)$}&$U_{\rm L}\hh U^\tr_{\rm R}$&$0$&$0$\\
\hline
\end{tabular}
\end{center}
\caption[]{\sl \small The components, representations
under~~$\Glr$ and transformations under~~$\Glr/U(1)_{B-L}$ as well
as the global charges of the superfields of the model. Note that
$U_{\rm C}\in SU(3)_{\rm C},~U_{\rm L}\in SU(2)_{\rm L},~U_{\rm
R}\in SU(2)_{\rm R}$ and $\tr~,\dagger$, and $\ast$ stand for the
transpose, the hermitian conjugate, and the complex conjugate of a
matrix respectively. The color index is suppressed.}\label{tab1}
\end{table}
\renewcommand{\arraystretch}{1.}

The $i$th generation $(i=1,2,3)$ left-handed lepton and quark
superfields  -- with the color index suppressed -- are denoted as
$l_i$ and $q_i$ whereas the anti-lepton and anti-quark superfields
are represented as $l^c_i$ and $q^c_i$ respectively. In the
simplest version of the model -- cf. \cref{tfhi} --, the
electroweak Higgs doublets $H_u$ and $H_d$ which couple to the up-
and down-type quarks respectively belong to the bidoublet
superfield $\hh$. The SSB of $\Glr$ down to $\Gsm$ is implemented
in two stages, as required by the scenario of metastable CSs, with
the aid of of a pair of $\sur$ triplets, $T$ and $\btd$, and a
pair of $\sur$ doublets $\phcb$ and $\phc$. In both steps, a
$\Glr$ singlet $S$, plays an auxiliary role.  Namely, in the first
step, $\Glr$ is broken to $\Glra$ leading to the formation of MMs.
This SSB occurs due to the v.e.vs which the $\sur$ triplet $T$
acquires in the $B-L$ neutral direction. In the second step, the
SSB of $U(1)_{B-L}\times U(1)_{\rm R}$ contained in $\Glra$ occurs
via the v.e.vs of the Higgs superfields $\phc$ and $\phcb$ leading
to the production of CSs. $U(1)_{B-L}$ is not orthogonal to
$U(1)_{\rm R}$ and so the CSs may split into segments having MMs
and anti-MMs at the ends. In the end of this chain, we obtain the
gauge symmetry of MSSM in \Eref{gsm}. The final transition to the
present vacuum occurs via the well-known radiative electroweak
SSB. Schematically, the steps of SSB in our setting can be
demonstrated as follows
\beq\begin{aligned}\Glr\times U(1)_{R}\times U(1)_{B}\times {\rm
SUGRA}&-\hspace{-0.15cm}\lf\mbox{\rm\ftn
TI}\rg\hspace{-0.15cm}\underset{\rm
MMs}{\xrightarrow{\hspace{0.65cm} T_0=\sg\hspace{0.65cm}}}\,
\Glra\times U(1)_{R}\times U(1)_{B}\times {\rm SUGRA}\\
&-\hspace{-0.15cm}\lf\mbox{\rm\ftn
CSs}\rg\hspace{-0.15cm}\underset{\vev{\phcb}=\vev{\phc}=\vx}{{\xrightarrow{\hspace{0.4cm}\vev{T_0}=\vt
\hspace{0.4cm}}}}\; \Gsm\times\mathbb{Z}_2^{R}\times
U(1)_{B}\times {\rm
SUSY}\\
&~\underset{\rm RCs}{{\xrightarrow{\hspace{0.8cm}\vev{\hdn},
\vev{\hun} \hspace{0.8cm}}}}\; \suc \times U(1)_{\rm EM}\times
U(1)_{B}, \label{chain1}\end{aligned}\eeq
where {\sf\small RCs} stands for \emph{radiative corrections} and
{\sf\small  EM} for \emph{ElectroMagnetism}. In the scheme above
the issue of SUSY breaking remains unspecified -- for related
attempts see \cref{ssb1,ssb2,ssb3,ssb4, davis,asfhi}. We can
suppose that it is arranged in a hidden sector not interacting
with the inflationary one. Note that soft SUSY breaking effects
explicitly break $U(1)_R$ to the $\mathbb{Z}_2^{R}$ which remains
unbroken. If it is combined with the fermion parity it yields the
well-known $R$-parity of MSSM \cite{martin}.

Let us recall here that a \sur\ triplet $T$ can be expanded in
terms of the 3 generators of $\sur$ $\tra$ with $\ca=1,2,3$  as
follows
\beq T=\sqrt{2}T^\ca \tra~~\mbox{with normalization}~~\Tr(\tra
{\sf T}_{\rm R}^{\rm b})=\delta^{\rm ab}/2,\eeq
where $\Tr$ denotes trace of a matrix. In accordance with the
matrix representation of $T$ in \Tref{tab1} we find
\beq\Tr|T|^2=\mbox{$\sum_\ca$}|T_\ca|^2=|\tdn|^2+|\tdm|^2+|\tdp|^2
~~\mbox{with}~~T_\pm=(T^1\pm iT^2)/\sqrt{2}\eeq
and similarly for $\btd$. On the other hand, the $B-L$ charge
generator is defined as
\beq \tbl=\sqrt{3}\diag\lf 1,1\rg/2\sqrt{2} \eeq
inspired by the ${\sf T}^{15}_{\rm C}$ generator of $SU(4)_{\rm
C}$ in the Pati-Salam gauge group \cite{nmh}. As a result, the SM
hypercharge $Q_Y$ is identified as the linear combination
$Q_Y=Q_{{\sf T}^3_{\rm R}}+Q_{(B-L)}/2$ where $Q_{{\sf T}^3_{\rm
R}}$ is the $\sur$ charge generated by $T^3_{\rm
R}=\diag\lf1,-1\rg/2$ and $Q_{(B-L)}$ is the $B-L$ charge.

\subsection{Superpotential}\label{md1}

The superpotential of our model respects totally the symmetries in
\Tref{tab1}. Most notably, it carries $R$ charge 2 and is linear
w.r.t. $S$ and $\btd$. It naturally splits into four parts:
\beq \label{wtot} W=W_{\rm H} +W_{\rm Y}+W_{\mu}+W_{\rm M},\eeq
where the content of each term is specified as follows:

\subparagraph{\sf\small (a)} $W_{\rm H}$ includes the Higgs
superfields which are involved in the breaking of $\Glr$ to $\Gsm$
and is given by
\beqs\beq W_{\rm H}=\lt S\lf \Tr\ T^2-M^2\rg+\lp S\phcb\phc+
M_T\Tr (\bar{T}T)-\ld \phcb \veps \btd\veps
\phc\>\>\>\mbox{with}\>\>\>\veps=\llgm\bem0&1\cr-1
&0\eem\rrgm\cdot \label{wh} \eeq
Here  $M$ and $M_T$ are mass parameters whereas $\lt$, $\lp$ and
$\lambda$ are dimensionless parameters. Note that $M$, $M_T$,
$\lt$, and $\ld$ can be made real and positive by redefinitions of
$S$, $\bar T$, $T$ and $\phcb\phc$ respectively -- note that only
one phase remains free after imposing the D flatness in the
$\phcb-\phc$ sector. The third dimensionless parameter $\lp$,
however, remains in general complex. For definiteness, we choose
this parameter to be real and positive too. In terms of the
components of the various superfields -- see \Tref{tab1} -- $\wh$
takes the form
\bea \nonumber \wh&=&\lt S\lf \tdn^2+2\tdp\tdm-M^2\rg+\lp
S(\nhcb\nhc+\ehcb\ehc)+\mt\lf\btdm\tdp+\btdp\tdm+\btdn\tdn \rg
\\&&-\ld\lf\ehcb\btdp\nhc+\nhcb\btdm\ehc+\btdn\lf\nhcb\nhc-\ehcb\ehc\rg/\sqrt{2}\rg. \label{whc}
\eea\eeqs
From the expression above we can appreciate the role of $\btd$ in
providing with intermediate-scale masses $\tdp$ and $\tdm$
consistently with the $R$ charges in \Tref{tab1}. The charged
components of $\btd$ acquire masses through its coupling with
$\phc$ and $\phcb$.

\subparagraph{\sf\small (b)} $W_{\rm Y}$ contains the Yukawa
interactions between the Higgs and the matter superfields and is
given by
\beq  W_{\rm Y}=y_{ijQ} q_i\hh q_j^c+ y_{ijL} l_i\hh l_j^c,
\label{wy}\eeq
where $y_{ijQ}$ and $y_{ijL}$ are, respectively, the Yukawa
coupling constants of the quarks and lepton with the Higgs
superfield $\hh$. From Eq.~(\ref{wy})  we can readily derive the
superpotential terms of the MSSM endowed with some partial Yukawa
unification as follows
\bea W_{\rm Y} = -y_{ijQ}H_u^\tr \veps {Q}_i u^c_j+y_{ijQ}H_d^\tr
\veps Q_{i}d^c_{j}+y_{ijL}H_d^\tr\veps {L}_ie^c_j -y_{ijL}H_u^\tr
\veps L_i\nu^c_j, \label{wmssm}\eea
where $\sul$ doublet quark  lepton and Higgs superfields are
defined respectively as
\beq
Q_i=\lin{u_i}{d_i}^\tr,\>\>L_i=\lin{\nu_i}{e_i}^\tr,\>\>\hu=\lin{\hup}{\hun}^\tr\>\>\mbox{and}\>\>
\hu=\lin{\hdn}{\hdm}^\tr. \label{QL} \eeq
The partial Yukawa unification in \Eref{wmssm} can be moderately
violated with the inclusion of extra \sul\ non-singlet Higgs
fields as demonstrated in \cref{tfhi}.

\subparagraph{\sf\small (c)} $W_{\mu}$ is relevant for the
generation of the $\mu$ term of the MSSM and is given by
\beq\label{wmu} W_{\mu}=\frac12\lm S\Tr\left(\hh\veps
\hh^{\tr}\veps\right)=\lm SH_d^\tr\veps\hu\,.\eeq
Note that the selected $R$ assignments in \Tref{tab1} prohibit the
presence in $W$ of the bilinear Higgs term of MSSM which is
generated here via the v.e.v of $S$ as shown in \Sref{phi1} below.

\subparagraph{\sf\small (d)} $W_{\rm M}$ is responsible for
Majorana neutrino masses $\mrh[i]$ after the breaking of $G_{\rm
LR}$:
\bea \nonumber W_{\rm M}&=&\ld_{il^c}{(\phcb
l_i^{c})^2}/{\Ms}={\ld_{il^c}}(\ehcb e_i^c-\nhcb\nu_i^c)^2/{\Ms}.
\label{WM}\eea
Here, we work in the basis, where $\ld_{il^c}$ and so the
generated mass matrix $\mrh[i]$ -- see \Sref{phi4} below -- is
diagonal, real and positive. These masses, together with the Dirac
neutrino masses in Eq.~(\ref{wmssm}), lead to the light neutrino
masses via the well-known (type I)  seesaw mechanism. As in
\cref{tfhi} we use the string scale $\Ms=5\cdot10^{17}/\mP$ as an
effective scale to facilitate the implementation of nTL. For
effective scale equal to \mP, the resulting BAU turns out to be a
little less than the expectations -- see \Sref{res1} below. Note
that in the present setting it is not possible to avoid the
non-renormalizable coupling as in \cref{univ, ighi, unit, sor2},
where the $\sni$ are not included in $\sur$ doublets. As a
consequence, $\phcb$ is to be $\sur$ doublet in order to obtain
singlet coupling with $\lci$ and then the inclusion of $\phc$ is
imperative to cancel the gauge $B-L$ anomalies.

\subsection{K\"{a}hler Potential}\label{md2}

The \Ka\ respects the symmetries of the models and has the
following contributions
\beq \label{ktot} K=K_{\rm I}+K_{0},\eeq
from the inflationary and the non-inflationary sectors of the
model. Namely

\subparagraph{\sf\small (a)} $K_{\rm I}$ depends on the fields
involved in TI and has the form
\beq K_{\rm I}=-N\ln\lf1-\Tr|T|^2\rg+{N}\ln{(1-\Tr\
T^2)}/{2}+{N}\ln(1-\Tr\ T^{*2})/{2}.\label{ki}\eeq
As explained in \cref{sor, tkref} this type of \Ka s is tailor
made for T-model inflation in SUGRA, since the first term endows
the inflaton kinetic term with a pole of order two whereas the two
last terms do not influence the \Kaa\ metric -- see \Sref{hi1}
below -- but assure that the exponential prefactor of the SUGRA
potential -- see \Sref{hi1} -- becomes unity along the
inflationary track. As a consequence, the resulting inflationary
potential gets considerably simplified.

\subparagraph{\sf\small (b)} $K_{0}$ includes the fields different
than the inflaton. We adopt the form
\beq K_{0}=\nsu\ln\lf1+\frac{|S|^2+\Tr|\btd|^2+
|\phcb|^2+|\phc|^2+|\tilde l_i|^2+|\tilde q_i|^2+|\tilde
l^c_i|^2+|\tilde q^c_i|^2+\Tr
(\hh^\dagger\hh)}{\nsu}\rg\cdot\label{ko}\eeq
According to \cref{su11}, $K_{0}$ assures for $0<\nsu<6$ a
stabilization of the inflationary direction \cite{rube} w.r.t $S$
fluctuations during TI without invoking higher order terms. Note
that the last term of numerator above can be translated in terms
of the $\sul$ Higgs doublets as
\beq \Tr (\hh^\dagger\hh) = |\hu|^2+|\hd|^2. \eeq
Hereafter we express the electroweak Higgs in terms of $\hu$ and
$\hd$.

Extending the discussion in \cref{su11,sor} we can conclude that
$K$ in \Eref{ktot} enjoys the following global symmetry in the
moduli space
\beq SU(3,1)/(SU(3)\times U(1)) \times SU(21)/U(1),
\label{ksym}\eeq
where the first and second factor corresponds to the respective
contributions of $K$ in \Eref{ktot}. The scalar curvature of the
moduli space is constant and equal to ${\cal R}_{K}=-12/N+42/N_0$.

\section{Inflationary Period}\label{hi}

In Sec.~\ref{hi1} below we describe the SUGRA framework which is
used for the analysis of our model, and then, in \Sref{hi2}, we
determine the inflationary potential. We finally derive the
inflationary observables in \Sref{hi3}.

\subsection{Supergravity Framework}\label{hi1}

The part of the Einstein-frame action within SUGRA related to the
complex scalars has the form
\beq\label{action}  {\sf S}=\int d^4x \sqrt{-{
\mathfrak{g}}}\lf-\frac{1}{2}\rcc +K_{\al\bbet} \geu^{\mu\nu}D_\mu
Z^\al D_\nu Z^{*\bbet}-\Ve\rg\,, \eeq
where the fields involved in our inflationary and
post-inflationary scenario are
\bea Z^\al=S, \btd, T, \phcb, \phc, \hu, \hd, \tilde
l_i^c.\label{zs1}\eea
Here $\rce$ is the space-time Ricci scalar curvature,
$\mathfrak{g}$ is the determinant of the background
Friedmann-Robertson-Walker metric, $g^{\mu\nu}$ with signature
$(+,-,-,-)$. Also, $K_{\al\bbet}={K_{,Z^\al Z^{*\bbet}}}$ is the
(already mentioned above) \Kaa\ metric and $D_\mu$ is the gauge
covariant derivative. Focusing on the $\sur\times\ubl$ sector of
the model, $D_\mu$ operates non-trivially on the following Higgs
fields
\beqs\bea && D_\mu T=\partial_\mu T+ig[\tra, T]\wra,\label{Dmu1}\\
&&D_\mu \bar T =\partial_\mu \bar T +ig[\tra,\bar T]\wra,\label{Dmu2}\\
&& D_\mu \phc=\partial_\mu \phc+ig\tbl A_{BL\mu}\phc-ig\tra\wra\phc,\label{Dmu3}\\
&&D_\mu \phcb=\partial_\mu \phcb-ig\tbl A_{BL\mu}
\phcb+ig\tra\wra\phcb,\label{Dmu4}\eea\eeqs
where $\wra$ and $A_{BL\mu}$ are the $\sur\times\ubl$ gauge fields
associated with a (unified) gauge coupling constant $g$ whereas
$[{\sf A},{\sf B}]$ denotes commutator of the matrices ${\sf A}$
and ${\sf B}$. More details about the achievement of the gauge
coupling unification within our model are given in
Appendix~\ref{app}.

Finally, $\Ve$ in \Eref{action} is the SUGRA potential which
includes the contributions $V_{\rm F}$ and $V_{\rm D}$ from F and
D terms respectively. These can be calculated via the formula
\beq \Ve=V_{\rm F}+V_{\rm D}\>\>\>\mbox{with}\>\>\> V_{\rm
F}=e^{K}\left(K^{\al\bbet}D_\al W D^*_{\bbet} W^*-3{\vert
W\vert^2}\right)\>\>\>\mbox{and}\>\>\> V_{\rm D}=\frac{g^2}2\lf
\sum_\ca{\rm D}^{\rm a}_{\rm R}{\rm D}^{\rm a}_{\rm R}+{\rm
D}^2_{B-L}\rg,\label{Vsugra} \eeq
where we introduce the inverse of $K_{\al\bbet}$ from the relation
$K^{\al\bbet}K_{\al\ggm}=\delta^{\bbet}_{\ggm}$. Also $D_\al
W=W_{,Z^\al} +K_{,Z^\al}W$ is the K\"{a}hler-covariant derivative
of $W$ and ${\rm D}^{\rm a}_{\rm R}$ and ${\rm D}_{B-L}$ are the D
terms corresponding to $\grbl$ part of the $\Glr$. The fields in
\Eref{zs1} yield the following contributions
\beqs\begin{align} & {\rm D}^{\rm a}_{\rm R}=-g\lf 2 \Tr \lf
K_{,T}\tra T\rg+2  \Tr \lf K_{,\bar T}\tra \bar T\rg-
K_{,\phc}\tra \phc + \phcb \tra K_{,\phcb} + K_{,\tilde l_i^c}
\tra \tilde l_i^c\rg; \label{dra}\\ & {\rm D}_{B-L}=-g\lf
K_{,\phc} \tbl \phc - \phcb \tbl K_{,\phcb}+K_{,\tilde l_i^c} \tbl
\tilde l_i^c\rg. \label{dbl}
\end{align}\eeqs
For the expressions including the contributions of $T$ into ${\rm
D}^{\rm a}_{\rm R}$ we should take into account the identities
$$\Tr\lf T^\dagger\tra T\rg=-\Tr\lf T\tra
T^\dagger\rg~~\mbox{and}~~\Tr \lf T\tra T\rg =0.$$

\subsection{Inflationary Potential}\label{hi2}

We proceed to the derivation of the tree and the one-loop
corrected inflationary potential in \Sref{hi2a} and \ref{hi2b}
respectively after determining the inflationary path in
\Sref{hi2aa}.

\subsubsection{Inflationary Trajectory.}\label{hi2aa}

The inflationary potential can be derived from \Eref{Vsugra}
taking the following contributions from \eqs{wtot}{ktot}
\beq W=W_{\rm H}+W_{\rm M}+W_\mu~~\mbox{and}~~K=K_{\rm
I}+K_0(\tilde q_i=\tilde l_i=\tilde q_i^c=0).\label{wktot}\eeq
In addition, we determine a D-flat direction and express $V_{\rm
F}$ in term of the inflaton which is identified by $\tdn$
according to our strategy in \Eref{chain1}. More explicitly, we
parameterize the fields of our models as follows
\beq\label{hpar} \tdn={\sg} e^{i\th}\>\>\>\mbox{and}\>\>\>X^\al=
(X_1^\al+iX_2^\al)/{\sqrt{2}},\eeq
where $X^\al=S, \nhc, \nhcb, \ehc, \ehcb, \btdn, \btdm, \btdp,
\tdm, \tdp, \ncit, \ecit, \hup, \hdm, \hun, \hdn$. We below
investigate the implementation of TI driven by the real field
$\sg$ along the field configuration
%
%
\beq\begin{aligned} \label{inftr}
&S=\nhc=\nhcb=\btdn=\ncit=\th=0\\
\mbox{and}~~&\ehc=\ehcb=\btdm=\btdp=\tdm=\tdp=\ecit=\hup=\hdm=\hun=\hdn=0.
\end{aligned}\eeq
This selection assures the D flatness, since ${\rm D}_{\rm
R}^{\ca}={\rm D}_{BL}=0$ along the configuration above. In
particular, ${\rm D}_{\rm R}^{\ca}=0$ is justified by a general
argument proven in \cref{gut}. It may be also confirmed in our
case, if we calculate explicitly the contribution to ${\rm D}_{\rm
R}^{\ca}$ with $\ca=1$ and $\ca=2$ from $\tdn$ -- note that ${\rm
D}_{\rm R}^{3}$ turns out to be $\tdn$ independent. Namely, we
obtain
\beqs\bea {\rm D}_{\rm R}^1&=&-\frac{g}2\lf\sqrt{2}N\lf
(\tdm^*-\tdp^*)\tdn+{\rm c.c}\rg/(1-\Tr|T|^2)+\cdots\rg,\label{d2}\\
{\rm D}_{\rm R}^2&=&-\frac{g}2\lf\sqrt{2}N\lf
i(\tdp^*+\tdm^*)\tdn+{\rm c.c}\rg/(1-\Tr|T|^2)+\cdots\rg,\eea
\label{d3}\eeqs
where the ellipsis represents terms including components of
$\phcb, \phc$ and $\bar T$ which are fixed at the origin along the
path of \Eref{inftr}. The same trajectory assures the avoidance of
a possible runaway problem, since the term $-3{\vert W\vert^2}$ in
\Eref{Vsugra} vanishes thanks to the constraint $S=\bar T=0$
\cite{rube}.

\subsubsection{Tree-level Result.}\label{hi2a}

Along the trough in \Eref{inftr} the only surviving terms in
\Eref{Vsugra} are
\beq \label{Vhi0}\Vhi =e^{K}\lf K^{SS^*}\, |W_{,S}|^2+
K^{\btdn\btdn^*}\,
|W_{,\btdn}|^2\rg={\lt^2}\lf\sg^2-M^2\rg^2+\mt^2\sg^2, \eeq
where we take into account $K^{SS^*}=K^{\btdn\btdn^*}=1$. The
naive expectation that this model is observationally ruled out by
now \cite{plin} since $\Vhi=\Vhi(\sg)$ coincides with the one of
the quartic or quadratic power-low model is not correct, as
explained extensively in \cref{tmodel, sor}. Indeed, $\sg$ is not
canonically normalized and therefore, no safe conclusion can be
achieved without to take this effect into account. To accomplish
it we find that, along the configuration of \Eref{inftr},
$K_{\al\bbet}$ defined below \Eref{ki} takes the form
\beq \lf K_{\al\bbet}\rg=\diag\lf
N/\fp^2,N/\fp,N/\fp,\underbrace{1,...,1}_{18~\mbox{elements}}\rg
~~\mbox{with}~~\fp=1-\sg^2. \label{VJe3}\eeq
The kinetic terms of the various scalars in \Eref{action} can be
brought into the following form
\beqs\beq \label{K3} K_{\al\bbet}\dot Z^\al \dot
Z^{*\bbet}=\frac12\lf\dot{\se}^{2}+\dot{\what
\th}^{2}\rg+\frac12\lf\dot{\what X}_1^\al\dot{\what X}_1^\al
+\dot{\what{X}}_2^\al\dot{\what{X}}_2^\al\rg.\eeq
Here dot means derivative w.r.t the cosmic time and the
canonically normalized (hatted) fields are defined as follows
\beq  \label{cann}
\frac{d\se}{d\sg}=J=\frac{\sqrt{2N}}{\fp},\>\>\> \what{\th}=
J\sg\th,\>\>\> \what T_\gm=\sqrt{\frac{N}{\fp}}T_\gm,\>\>\>\what
X_1^\bt= X_1^\bt\>\>\>\mbox{and}\>\>\>\what{X}^\bt_2=
X_2^\bt,\eeq\eeqs
where $\gm=+,-$ and $X^\bt$ are $X^\al$ shown below \Eref{hpar}
without $\tdm$ and $\tdp$. From the first expression above we can
verify that in our model there is a kinetic pole of order two for
$\sg=1$ which generates the well-known relation between $\sg$ and
$\se$ including $\tanh$ \cite{tmodel, sor}. This is actually the
reason for the name of our inflationary model and not its
realization by the $T$ field as erroneously may be considered.
Thanks to the $\sg-\se$ relation, $\Vhi$ plotted in terms of $\se$
experiences a stretching for $\se>1$ which results to a plateau
facilitating, thereby, the establishment of TI for $\se\gg1$.
However, $\se\gg1$ can coexist with $\sg<1$ and so, we have the
opportunity to work in a $\sg$ regime consistent with the
foundation of SUGRA as an effective theory below $\mP=1$.

\subsubsection{Stability and 1-Loop RCs.}\label{hi2b}

To consolidate our inflationary models, we have to verify that the
inflationary direction in \Eref{inftr} is stable w.r.t the
fluctuations of the non-inflaton fields. To this end, we construct
the mass-squared spectrum of the various scalars included in
\Eref{inftr}. We find the expressions of the masses squared $\what
m^2_{\al}$ (where the hat is used only in cases with
$K_{\al\aal}\neq1$) and arrange them in \Tref{tab2}. These masses
are compared with the Hubble-parameter during TI,
\beq \Hhi=\lf\Vhi/3\rg^{1/2}~~\mbox{with}~~\Vhi\simeq\lt^2\sg^4
\label{hti}\eeq
the inflationary potential in \Eref{Vhi0} which is dominated by
the quartic power of $\sg$ since, as we see in \Sref{res2} below,
the scenario of metastable CSs works for $\mt\ll10^{-6}$.
Moreover, in the formulae of \Tref{tab2} we take into account
$M\ll\sg$, and the fact that $\mt\ld$ and $\mt^2$ are much less
than $\ld\lt$ and $\lt^2$ in the overall allowed parameter space
of our model. The presented formulas are rather accurate compared
to the exact results.

\renewcommand{\arraystretch}{1.3}
\begin{table}[!t]
\begin{center}
\begin{tabular}{|c||c|c|c|}\hline {\sc
Fields}&{\sc Eigenstates}& \multicolumn{2}{|c|}{\sc Masses Squared}\\
\hline\hline
\multicolumn{4}{|c|}{\sc SM-Singlet Components} \\ \hline\hline
1 real scalar&$\widehat\theta$&$\widehat m_{\theta}^2$&
\multicolumn{1}{|c|}{$6\Hhi^2$}\\
$2$ real scalars &$\bar T_{01}, \bar T_{02}$ & $m^2_{
\bar T_0}$&$3\Hhi^2(1+1/N_0)$\\
$2$ real scalars&$S_1, S_2$ &$ m_{S}^2$&\multicolumn{1}{|c|}{$6\Hhi^2(1/N_0-2\fp^2/N\sg^2)$}\\
$4$ real &$\frac1{\sqrt{2}}(\nu_{\Phi1}^c\pm\bar\nu_{\Phi1}^c)$ &
$ m^2_{\nu_{\phc\pm}}$&$3\Hhi^2(1+1/N_0\pm\lp/\lt\sg^2)$\\
scalars &$\frac1{\sqrt{2}}(\nu_{\Phi2}^c\pm\bar\nu_{\Phi2}^c)$&&\\
$6$ real scalars &$\ncit$ & $m^2_{i \nu^c}$&$3\Hhi^2(1+1/N_0)$
\\ \hline
$2$ Weyl  spinors& $\what \psi_{T0}, \what \psi_{S}$ & $\what
m^2_{ \psi0}$&{$2\fp^2\lt^2\sg^2/N$}\\ \hline\hline
\multicolumn{4}{|c|}{\sc SM-non-Singlet Components} \\
\hline\hline
2 real &$\frac1{\sqrt{2}}(T_{+1}-T_{-1})$ &$\widehat m_{
T\pm}^2$&\multicolumn{1}{|c|}{$M^2_{W^\pm_{\rm R}}+6\Hhi^2(1+1/N-1/N\sg^2)$}\\
scalars&$\frac1{\sqrt{2}}(T_{+2}+T_{-2})$ &&\\
$4$ real &$\frac1{\sqrt{2}}(e_{\Phi1}^c\pm\bar e_{\Phi1}^c)$ & $
m^2_{e_{\phc\pm}}$&$3\Hhi^2(1+1/N_0\pm\lp/\lt\sg^2)$\\
scalars &$\frac1{\sqrt{2}}(e_{\Phi2}^c\pm\bar e_{\Phi2}^c)$&&\\
$4$ real scalars &$\bar T_{\pm1}, \bar T_{\pm2}$ & $m^2_{
\bar T\pm}$&$3\Hhi^2(1+1/N_0)$\\
$6$ real scalars &$\tilde e^c_i$ & $m^2_{i
e^c}$&$3\Hhi^2(1+1/N_0)$
\\ \hline
2 gauge bosons &{$W_{\rm R}^\pm$}&{$M^2_{W^\pm_{\rm R}}$}&
\multicolumn{1}{|c|}{$2Ng^2\sg^2/\fp$}\\\hline
$6$ Weyl  & $\psi_{\bar T\pm}$ & $\what m^2_{\psi\bar T\pm}$&
${\mt^2\fp/N}$\\
spinors&$\ldu^\pm_{\rm R}, \widehat\psi_{T\pm}$&
$M_{\ldu\rm R}^2$&\multicolumn{1}{|c|}{$2Ng^2\sg^2/\fp$}\\
\hline
$8$ real &$\frac1{\sqrt{2}}(H^0_{\rm u1,2}\pm H^0_{\rm d1,2})$ &
$m^2_{H_{ud\pm}}$&$3\Hhi^2(1+1/N_0\pm\lm/\lt\sg^2)$\\
scalars &$\frac1{\sqrt{2}}(H^+_{u1,2}\pm\bar H^-_{d1,2})$&&\\
\hline
\end{tabular}\end{center}
\caption{\sl\small Mass-squared spectrum along the inflationary
trajectory of \Eref{inftr}. The indices $1$ and $2$ are referred
to the real and imaginary parts of the scalar fields according to
\Eref{cann}. To avoid very lengthy formulas we neglect terms
proportional to $M\ll\sg$ and $\mt^2$ or $\mt\ld$ compared to
$\lt^2\sg^2$ and $\lt\sg$.}\label{tab2}
\end{table}\renewcommand{\arraystretch}{1.}

As shown in \Eref{inftr} the stability of the inflationary path
has to be checked along 20 complex and 1 real directions, i.e.
along 41 real directions. In \Tref{tab2} we arrange these
directions in two groups taking as criterion whether the
components of the various fields are SM singlets or non-singlets.
We find 15 \emph{degrees of freedom} ({\sf\small d.o.f}) in the
first group and 24 d.o.f in the second group which are summed to
39 d.o.f. The residual two d.o.f are associated with the $2$
Goldstone bosons
\beq(T_{+1}+T_{-1})/\sqrt{2}
\>\>\mbox{and}\>\>(T_{+2}-T_{-2})/\sqrt{2},\label{gold}\eeq
which are absorbed by the two gauge bosons $W_{\rm R}^\pm$ which
become massive. In particular, the non-vanishing $\tdn$ values
trigger the SSB $SU(2)_{\rm R} \to U(1)_{\rm R}$. Therefore, 2 of
the 3 generators of $SU(2)_{\rm R}$ are broken, leading to the two
Goldstone bosons in \Eref{gold} which are ``eaten'' by the 2 gauge
bosons which become massive. As a consequence, 12 d.o.f of the
spectrum before the SSB -- 6 d.o.f corresponding to 3 complex
components of $T$ and 6 d.o.f corresponding to 3 massless gauge
bosons, $W_{\rm R}^\ca$ of $\sur$ -- are redistributed as follows:
4 d.o.f are associated with the real propagating scalars ($\sg,
\th$ and the orthogonal combinations of the states in \Eref{gold})
whereas the residual 2 d.o.f combine together with the $6$ d.o.f
of the initially massless gauge bosons to make massive the two
combinations of them $W_{\rm R}^{\pm}=(W^1\mp i W^2)/\sqrt{2}$.
Let us note here that, as in \cref{nmh}, the Goldstone bosons in
\Eref{gold} are not exactly massless since $V_{\rm TI,\sg}\neq0$.
These masses turn out to be $\what m_{T0}^2=3\fp\Hhi^2/2N\sg$ and
are ignored for the computation of the one-loop RCs below. This
subtlety is extensively discussed in a similar regime in
\cref{postma}.

From \Tref{tab2} it is evident that $0<\nsu\leq6$ assists us to
achieve $m^2_{S}>\Hhi^2$ -- in accordance with the results of
\cref{su11}. Note that these $\nsu$ values enhance, also, the
positive contributions to other ratios $m^2_{X^{\al}}/\Hhi^2$ too.
Indeed, we obtain $m^2_{X^{\al}}/\Hhi^2\gg1$ during the last
$50-60$ e-foldings of TI and so any inflationary perturbations of
the fields other than the inflaton are safely eliminated. On the
other hand, $m^2_{\nu_\phc-}$, $m^2_{e_\phc-}$ and $m^2_{H_{ud}-}$
include a negative contribution which may drive them into negative
values for
\beq \label{lmp} \lp>(1+1/\nsu)\lt\sg^2 ~~\mbox{and}~~
\lm>(1+1/\nsu)\lt\sg^2.\eeq
To assure the success of our scenario we demand:
\beq \label{lmpcon} \mbox{\sf\small (a)}~~\lm<\lp~~~\mbox{and}~~~
\mbox{\sf\small (b)}~~\lp<(1+1/\nsu)\lt\sgf^2.\eeq
\sEref{lmpcon}{a} is imposed so that the tachyonic instability of
the $\phc-\phcb$ system occurs first with the $\hu-\hd$ being
confined to zero. This way, $\nhc$ and $\nhcb$ start evolving
towards their v.e.vs -- see \Sref{phi1} below -- and trigger the
SSB of $\Glra$ down to $\Gsm$ whereas the remaining SSB of $\Gsm$
occurs afterwards during the radiative electroweak phase
transition -- see \Eref{chain1}. On the other hand,
\sEref{lmpcon}{b} implies that the tachyonic instability of the
$\phc-\phcb$ system  appears after the end of TI. Given that
$\sg=\sgx\simeq1\gg\sgf$, no tachyonic instability appears along
the inflationary path of \Eref{inftr} during TI. In other words,
if the aforementioned conditions on $\lp$ and $\lm$ are satisfied,
$m^2_{e_{\phc-}}$ and $m^2_{\nu_{\phc-}}$ develop a negative value
as $\sg$ crosses below a critical value,
\beq\sgc<\sgf~~\mbox{where}~~\sgc=\lf\lp\nsu/\lt(1+\nsu)\rg^{1/2},\label{sgc}\eeq
found from the condition $m^2_{e_{\phc-}}(\sgc)\simeq
m^2_{\nu_{\phc-}}(\sgc)=0$, whereas $m^2_{H_{du-}}(\sgc)>0$ and so
the $\hu-\hd$ system remains well stabilized for $\sg=\sgc$.

In \Tref{tab2} we also present the masses squared of the gauge
bosons, chiral fermions and gauginos of the model along the
direction of \Eref{inftr}. We must remark that the numbers of
bosonic and fermionic d.o.f should be equal in each group of
components presented in \Tref{tab2}. To establish this equality we
we have to take into account the massless fermions. In particular,
for the group of SM-singlet components we obtain the massless
fermions $\psi_{\tdn}, \psi_{\nhc}, \psi_{\nhcb}$ and $\sni$ which
yield 12 d.o.f. If we add them to 4 d.o.f we obtain 16 d.o.f which
are equal to the bosonic d.o.f if we add one d.o.f for the
inflaton $\tdn$ which is not listed in \Tref{tab2}. As regards the
second group of components, we obtain $24+2\cdot3+2=32$ bosonic
d.o.f which are equal to the fermionic ones if we notice that we
obtain 20 d.o.f from the massless states $\psi_{e^c_\phc},
\psi_{e^c_\phc}, \psi_{e_i^c}, \psi_{\hun}, \psi_{\hdn},
\psi_{\hup}, \psi_{\hdm}$ and $\uplambda^3$ together with $12$
d.o.f from the massive ones.

The mass spectrum is necessary in order to calculate the one-loop
RCs employing the well-known Coleman-Weinberg formula. This
formula can be self-consistently applied, if we consider SUGRA as
an effective theory with cutoff scale equal to $\mP$. To this end
we insert in aforementioned formula the masses which lie well
below $\mP$, i.e., all the masses arranged in \Tref{tab2} besides
$M_{\wrpm}$, $M_{\uplambda\rm R}$ and $\what m_{T\pm}$. Note that
although these masses satisfy the relation $2\what
m_{T\pm}+6M^2_{\wrpm}-8M_{\uplambda\rm R}=0$  yield a non-zero
contribution to RCs, as in similar models \cite{ighi, univ, jhep}.
Therefore, the one-loop RCs to $\Vhi$ read
\beq \dV_{\rm RC}= \dV_{\rm SM0}+\dV_{\rm SM\pm},\eeq
where the individual contributions, coming from the corresponding
sectors of Table~\ref{tab2}, are given by
\beqs\bea V_{\rm SM0}&=& {1\over64\pi^2}\lf \widehat m_{
\th}^4\ln{\widehat m_{\th}^2\over\Lambda^2} + 2m_{ \bar
T0}^4\ln{m_{\bar T0}^2\over\Lambda^2} +2
m_{S}^4\ln{m_{S}^2\over\Lambda^2}+2
m_{\nu_{\phc+}}^4\ln{m_{\nu_{\phc+}}^2\over\Lambda^2}\nonumber \right.\\
&+&\left.2 m_{\nu_{\phc-}}^4\ln{m_{\nu_{\phc-}}^2\over\Lambda^2}+
6m_{i\nu^c}^4\ln{m_{i\nu^c}^2\over\Lambda^2}-4m_{\widehat
m_{\psi_0}}^4 \ln{\widehat
m_{\psi0}^2\over\Lambda^2}\rg,\label{Vrc0}\eea
\bea V_{\rm SM\pm}&=&{1\over64\pi^2}\lf2m_{e_{\phc+}}^4
\ln{m_{e_{\phc+}}^2\over\Lambda^2}+2m_{e_{\phc-}}^4
\ln{m_{e_{\phc-}}^2\over\Lambda^2}+2m_{\bar T\pm}^4 \ln{m_{\bar
T\pm}^2\over\Lambda^2}+4m_{H_{ud+}}^4
\ln{m_{H_{ud+}}^2\over\Lambda^2}\right.\nonumber \\
&+&\left.4m_{H_{ud-}}^4 \ln{m_{H_{ud-}}^2\over\Lambda^2}
+6m_{ie^c}^4 \ln{m_{ie^c}^2\over\Lambda^2} -4m_{\psi\bar T\pm}^4
\ln{m_{\psi\bar T\pm}^2\over\Lambda^2} \rg~. \label{Vrcc}
\eea\eeqs
Here $\Lambda$ is a renormalization mass scale. The resulting
$\dV_{\rm RC}$ lets intact our inflationary outputs, provided that
$\Lambda$ is determined by requiring $\dV_{\rm RC}(\sgx)=0$ or
$\dV_{\rm RC}(\sgf)=0$. These conditions yield
$\Ld\simeq4.2\cdot10^{-5}-2.4\cdot10^{-4}$ and render our results
practically independent of $\Lambda$ since these can be derived
exclusively by using $\Vhi$ in \Eref{Vhi0} with the various
quantities evaluated at $\Ld$ -- cf. \cref{jhep}.

\subsection{Inflation Analysis}\label{hi3}

We here recall the results on the inflationary observables derived
in \cref{sor}. Namely, a period of slow-roll TI is determined by
the condition
\beq{\small\sf
max}\{\widehat\epsilon(\sg),|\widehat\eta(\sg)|\}\leq1,\label{src}\eeq
where the slow-roll parameters can be derived by applying the
standard formulae -- see e.g. \cref{review}. Restricting ourselves
to purely quartic TI, which requires $\mt^2\sg^2\ll\lt^2\sg^4$, we
find
\beqs\beq\label{sr}\epsilon= {1\over2}\left(\frac{V_{{\rm
TI},\se}}{\Vhi}\right)^2\simeq 4\frac{1-2\sg^2}{N\sg^2}
\>\>\>\mbox{and}\>\>\>\eta = \frac{V_{{\rm TI},\se\se}}{\Vhi}
\simeq 2\frac{3-8\sg^2}{N\sg^2}\,\cdot\eeq
From the condition in \Eref{src} we infer that TI terminates for
$\sg=\sgf$ such that
\beq \sgf\simeq\mbox{\small\sf
max}\left\{\frac{2}{\sqrt{8+N}},\sqrt{\frac{6}{16+N}}
\right\}.\label{sgap}\eeq\eeqs

The number of e-foldings $\Ns$ that the scale $\ks=0.05/{\rm Mpc}$
experiences during TI and the amplitude $\As$ of the power
spectrum of the curvature perturbations generated by $\sg$ can be
computed respectively using the standard formulae  \cite{review}
as follows
\begin{equation}
\label{Nhi}  {\sf (a)}~~\Ns=\int_{\sef}^{\sex}
d\se\frac{\Vhi}{\Ve_{\rm TI,\se}}\>\>\>\mbox{and}\>\>\>{\sf
(b)}~~\As^{1/2}= \frac{1}{2\sqrt{3}\, \pi} \;
\frac{\Vhi^{3/2}(\sex)}{\left|\Ve_{\rm TI,\se}(\sex)\right|},\eeq
where $\sgx~[\sex]$ being the value of $\sg~[\se]$ when $\ks$
crosses the inflationary horizon. Taking into account that
$1\simeq\sgx\gg\sgf$, we can express approximately $\Ns$ as a
function of $\sgx$ and then solve w.r.t $\sgx$ as follows
\begin{equation} \label{sgx}
\Ns\simeq\frac{N}4\frac{\sgx^2}{1-\sgx^2}\>\>\Rightarrow\>\>\sgx\simeq
2\sqrt{\frac{\Ns}{4\Ns+N}}\,.
\end{equation}
Inserting the last result into \sEref{Nhi}{b} we can obtain a
rather accurate estimation of $\lt$. In fact
\beq
\label{lda}\sqrt{\As}\simeq\:\frac{\lt}{\pi}\sqrt{\frac{2\Ns^3}{3N(4\Ns+N)}}
\>\>\Rightarrow\>\> \lt\simeq
\pi\sqrt{\frac{3N\As(4\Ns+N)}{2\Ns^3}}\cdot\eeq

Making use of the expression for $\sgx$ in \Eref{sgx} we may also
calculate the remaining inflationary observables -- i.e., the
spectral index $\ns$, its running $\as$ and the tensor-to-scalar
ratio $r$ -- via the relations
\beqs\baq \label{ns} && \ns=\: 1-6\epsilon_\star\ +\
2\eta_\star\simeq
1-\frac{3}{\Ns}+\frac{4}{4\Ns+N},\\
&& \label{as} \as =\:{2\over3}\left(4\eta_\star^2-(n_{\rm
s}-1)^2\right)-2\xi_\star\simeq-\frac{3}{\Ns^2}+\frac{16}{(4\Ns+N)^2},\>\>\>\>\>\>\>\>\\
&& \label{rs} r=16\epsilon_\star\simeq{4N}/{\Ns^2},\eaq\eeqs
where $\xi={\Ve_{\rm TI,\widehat\sg} \Ve_{\rm
TI,\widehat\sg\widehat\sg\widehat\sg}/\Ve_{\rm TI}^2}$ and the
variables with subscript $\star$ are evaluated at $\sg=\sgx$. We
remark that the dependence of $\ns$ and $\as$ on $N$ is very weak
whereas $r$ increases sharply with it consistently with the
predictions of the original version of T-model inflation
\cite{tmodel, rh}.

\section{Post-Inflationary Period}\label{phi}

We analyze here the post-inflationary completion of our model. We
first, in \Sref{phi0}, determine the SUSY vacuum and then, in
\Sref{phi00}, we provide an interpretation of the $\mu$ term of
MSSM. We below, in \Sref{phi1}, we derive the mass spectrum in the
SUSY vacuum, and compute the decay rate of the produced CSs in
\Sref{phi2}. Finally, we show how reheating -- see \Sref{phi3} --
and nTL -- see \Sref{phi4} -- can be processed consistently with
the $\Gr$ constraint and the low energy neutrino data. Hereafter
we restore units, i.e., we take $\mP=2.433\cdot10^{18}~\GeV$.

\subsection{SUSY Vacuum}\label{phi0}

Since the SUSY vacuum of the theory is expected to lie well below
$\mP$, it is legitimate to obtain the SUSY limit, $V_{\rm SUSY}$,
of $V$ in \Eref{Vsugra} taking into account $W$ and $K$ in
\Eref{wktot}. In particular, $V_{\rm SUSY}$ turns out to be
\cite{martin}
\beq \label{Vsusy} V_{\rm SUSY}= \widetilde K^{\al\bbet}
W_{,Z^\al} W^*_{,Z^{*\bbet}}+\frac{g^2}2 \lf \mbox{$\sum_{\rm
a}$}{\rm D}^\ca_{\rm R} {\rm D}^\ca_{\rm R}+{\rm
D}_{B-L}^2\rg,\eeq
where $\widetilde K$ is the limit of $K$ in \Eref{wktot} for
$\mP\to\infty$ which reads
\beq \label{Kquad}\widetilde K=N\Tr |T|^2-\frac{N}2(\Tr T^2+\Tr
T^{*2})+ \Tr |\bar T|^2+|\phc|^2+|\phcb|^2 +|S|^2+|\hu|^2+|\hd|^2
+|\tilde l_i^c|^2\,.\eeq
Given that the fields in the second line of \Eref{inftr} are
necessarily confined at zero during TI and in the SUSY vacuum we
here focus on the SM-singlet directions of $Z^\al$ in \Eref{zs1},
i.e., we take
\beq Z^\al=S, \tdn, \btdn, \nhc, \nhcb, \hun, \hdn, \ssni\,.\eeq
Upon substituting in \Eref{Vsusy} $W$ and $\widetilde K$ from
\eqs{wktot}{Kquad} we obtain the non-zero F-term contribution to
$V_{\rm SUSY}$ which reads
\bea \nonumber V_{\rm 0F}&
=&\left|\lt(\tdn^2-{M^2})+\lp\nhc\nhcb+{\lm}\hun\hdn\right|^2
+\left|2\lt S\tdn+\mt\btdn\right|^2/N
+\left|\mt\tdn-\ld\nhc\nhcb/\sqrt{2}\right|^2
\\&+& \left|\lp S\nhcb-\ld\btdn\nhcb/\sqrt{2}\right|^2+\left|\lp
S\nhc-\ld\btdn\nhc/\sqrt{2}+2\ld_{i\nu^c}\nhcb\tilde\nu^{c2}_i/\Ms\right|^2
+4\ld_{i\nu^c}^2|\bar\nu_\phc^{c2}\ssni|^2/\Ms^2\nonumber\\&+&\lm^2
|S|^2\lf\left|\hun\right|^2+\left|\hdn\right|^2\rg. \label{VF}\eea
Note that the D terms are kept zero along the trajectory of
\Eref{inftr} which includes the SUSY vacuum. From the last
equation, we find that the latter lies along the direction
\beqs\beq
\vev{S}=\vev{\btdn}=\vev{\hun}=\vev{\hdn}=\vev{\ssni}=0,\>\vev{\tdn}=\vt
\>\>\>\mbox{and}\>\>\>
|\vev{\nhc}|=|\vev{\nhcb}|=\vx\,,\label{vevs} \eeq where the
non-zero v.e.vs are given by
\beq
\label{vxt}\vx=\lf\frac{\sqrt{2}\mt\vt}{\ld}\rg^{1/2}\>\>\mbox{and}\>\>
\vt=\frac{\ml}{2}\lf\sqrt{1+\frac{4M^2}{\ml^2}}-1\rg\>\>\mbox{with}
\>\>\ml=\frac{\sqrt{2}\lp}{\ld\lt}\mt\,.\eeq\eeqs
The non-vanishing \vx\ value triggers the SSB pattern $U(1)_{\rm
R} \times U(1)_{B-L}\to U(1)_{Y}$ which allows for the formation
of CSs. Their metastability depends on the relative magnitudes of
$\vt$ and $\vx$, as shown in \Sref{phi2} below.

\subsection{Generation of the $\mu$ Term of MSSM}\label{phi00}

The contributions from the soft SUSY-breaking terms, although
negligible during TI, may shift slightly $\vev{S}$ from zero in
\Eref{vevs}. In fact, the relevant potential terms are
\beq V_{\rm soft}= \lf\lt A_T S \tdn^2 +\lp A_\phc S \nhcb\nhc-
{\rm a}_{S}S\lt M^2 + {\rm h. c.}\rg+
m_{\al}^2\left|X^\al\right|^2, \label{Vsoft} \eeq
where $X^\al=S, \tdn, \nhcb, \nhc, \hu$ and $\hd$, $m_{\al}\ll M,
A_T, A_\phc$ and $\aS$ are soft SUSY-breaking mass parameters and
we assume $\hu=\hd\simeq0$ during and after TI. Rotating $S$ in
the real axis by an appropriate $R$-transformation, choosing
conveniently the phases of $A_T, A_\phc$ and $\aS$ -- so as the
total low energy potential $V_{\rm tot}=V_{\rm SUSY}+V_{\rm soft}$
to be minimized -- and substituting in $V_{\rm soft}$ the $\tdn$,
$\nhc$ and $\nhcb$ values from \Eref{vevs}, we get
\beq \vev{V_{\rm tot}(S)}= \frac2{N}\lt^2 v^2_{T\phc}S^2-2\lt
M^2 \am \mgr S~~\mbox{with}~~\begin{cases}\am=1+\lp\vx^2/\lt M^2+\vt^2/M^2,\\
v^2_{T\phc}=2\vt^2+N\lp^2\vx^2/\lt^2.
\end{cases}\label{Vol} \eeq
Here the first term comes from the F-term SUSY potential with
$\nhc$, $\nhcb$ and $\tdn$ obtained their v.e.vs in \Eref{vevs} --
cf. \cref{univ, dls,unit,ighi} --, $\mgr$ is the $\Gr$ mass and we
assume
\beq \mgr\simeq|A_T|\simeq |A_\phc| \simeq|{\rm a}_{S}|.\eeq
The extermination condition of $\vev{V_{\rm tot}(S)}$ w.r.t $S$
leads to a non vanishing $\vev{S}$ as follows,
\beqs\beq \label{vevS}{d}\vev{V_{\rm tot}(S)}/{d S}
=0~~\Rightarrow~~\vev{S}= NM^2\am\mgr/2\lt v^2_{T\phc}.\eeq
The emerged $\mu$ parameter from $W_\mu$ in \Eref{wmu} is
\beq \label{mumgr}\mu =\lm
\vev{S}\simeq\lm\am\mgr\frac{\sqrt{N}\Ns}{2\sqrt{6\As}\pi}
\frac{M^2}{v^2_{T\phc}}, \eeq\eeqs
where we employed the $\lt-\sqrt{\As}$ condition in \Eref{lda}.
Given that $\sqrt{\As}\sim10^{-5}$, as we see in \Sref{res1}
below, $\mu$ turns out to be of the order of $\mgr$ for
$\lm\sim10^{-6}$, in accordance with similar findings in
\cref{univ, unit, ighi, sor2}.

\subsection{Mass Spectrum}\label{phi1}

We focus on the $\sur\times\ubl$ Higgs sector of our model which
consists of the superfields
\beq Z^\al=S, \btd, T, \phcb, \phc \label{zshiggs}\eeq
and we concentrate on $W=W_{\rm H}$ in \Eref{wh} and $K=\wtilde
K(\hu=\hd=l_i^c=0)$ in \Eref{Kquad}. Taking into account the
dimensionality of each superfield we infer that this sector of the
model includes 22+8=30 bosonic d.o.f and equal number of fermionic
d.o.f. In other words, we have 44 d.o.f associated with the chiral
superfields and 16 with the vector superfields. Since the vacuum
configuration in \Eref{vevs} is more structured than that during
TI in \Eref{inftr}, it is not doable to display mass eigenvalues
as done in \Tref{tab2}. Instead, we below expose the mass matrices
of the various sectors of the theory.

\subsubsection{Mass-Squared Matrices for Scalars.} To extract
the relevant mass spectrum in the vacuum of \Eref{vevs}, we first
separate the fields in \Eref{zshiggs} into SM singlet and
non-singlet as follows
\beq Z^\bt=S, \tdn, \btdn, \nhc, \nhcb~~\mbox{and}~~ Z^\gm= \tdp,
\tdm, \ehc, \ehcb, \btdp, \btdm. \label{bgdef}\eeq
The relevant matrices for both groups of fields in \Eref{bgdef}
have the following structure
\beq {\cal M}^2_{0}(Z^\al,Z^\aal)=\llgm\bem {\cal M}^2_{XY^*} &
{\cal M}^2_{XY}\cr {\cal M}^2_{X^*Y^*}&{\cal
M}^2_{X^*Y}\eem\rrgm,\label{m0}\eeq
where the $X$ and $Y$ denote subsets for the fields $Z^\bt$ and
$Z^\gm$ and the submatrices, in all cases considered, obey the
relations
\beq {\cal M}^2_{XY^*}={\cal M}^2_{X^*Y}=\vevl{V_{{\rm
SUSY},XY^*}}~~\mbox{and}~~{\cal M}^2_{XY}={\cal
M}^2_{X^*Y^*}=\vevl{V_{{\rm SUSY},XY}}\eeq
with $V_{\rm SUSY}$ given by \Eref{Vsusy}. After taking into
account the conditions in \Eref{vevs}, we find the total
mass-squared matrix of the scalars which can be divided into the
following disconnected parts:

\subparagraph{(a)} The matrix of the $S-\btdn$ sector which has
the form
\beq {\cal M}^2_{XY^*}(S,\btdn)=\llgm\bem
{4\lt^2\vt^2/N+2\lp^2\vx^2}&{\sqrt{2}\ld(\lt-\lp N)\vx^2/N}\cr
{\sqrt{2}\ld(\lt-\lp
N)\vx^2/N}&{\mt^2/N+\ld^2\vx^2}\eem\rrgm~~\mbox{and}~~ {\cal
M}^2_{XY}(S,\btdn)=0. \label{m01}\eeq From the diagonalization of
the relevant $4\times4$ matrix we find 4 real non-zero mass
eigenstates.

\subparagraph{(b)}\label{parminf} The matrices of the
$\nhcb-\nhc-\tdn$ sector which read
\beqs\beq {\cal M}^2_{XY^*}(\nhcb, \nhc, \tdn )=\llgm\bem
{m^2_{\nu\rm F}}&{m^2_{\nu\rm F}}&{m^2_{T\nu}}\cr
{m^2_{\nu\rm F}}&{m^2_{\nu\rm F}}& {m^2_{T\nu}}\cr
{m^2_{T\nu}}&{m^2_{T\nu}}&{m^2_{TT}}\eem\rrgm+{\cal
M}^2_{XY}(\nhcb, \nhc, \tdn),\label{m02}\eeq
where the various elements of the matrix above are
\beq m^2_{\nu\rm F}=\frac18(8\lp^2+4\ld^2)\vx^2,
~~m^2_{T\nu}=\frac{2\lp\lt\vt\vx}{\sqrt{N}}
-\frac{\ld\mt\vx}{\sqrt{2N}}~~\mbox{and}~~m^2_{TT}=\frac{\mt^2+4\lt^2\vt^2}{N},\label{mms}\eeq
whereas the matrix with the D-term contributions is
\beq{\cal M}^2_{XY}(\nhcb, \nhc, \tdn)=m^2_{\nu\rm D}\llgm\bem
{1}&{-1}&{0}\cr
{-1}&{1}& {0}\cr
{0}&{0}&{0}\eem\rrgm~~\mbox{with}~~m^2_{\nu\rm
D}=\frac58g^2\vx^2.\label{m03}\eeq\eeqs
From the matrices in \eqs{m02}{m03} we note that
\beq \det {\cal M}^2_{XY^*} = \det  {\cal M}^2_{XY}=0,\eeq
which implies that at least one eigenvalue of these matrices is
zero. Indeed, this corresponds to the Goldstone boson absorbed by
the neutral gauge boson $A^{\perp}$ of the model -- see below. In
addition, we obtain 1 real scalar from the D terms and 2 complex
scalars (i.e., 4 d.o.f) from the F terms. The latter contributes
to the reheating of the universe as shown in \Sref{phi3}.

\subparagraph{(c)} The matrices of the $\tdp-\tdm-\ehc-\ehcb$
sector which are found to be
\beqs\beq {\cal M}^2_{XY^*}(\tdp, \tdm, \ehc, \ehcb)=\llgm\bem
{{\mt^2\over N}+Ng^2\vt^2}&{0}&{m^2_{T^*e}+m^2_{Te}}&{0}\cr
{0}&{{\mt^2\over N}+Ng^2\vt^2}&{0}&{m^2_{T^*e}+m^2_{Te}}\cr
{m^2_{T^*e}+m^2_{Te}}&{0}&{\frac12(g^2+2\ld^2)\vx^2}&{0}\cr
{0}&{m^2_{T^*e}+m^2_{Te}}&{0}&{\frac12(g^2+2\ld^2)\vx^2}\eem\rrgm
\label{m04}\eeq and \beq {\cal M}^2_{XY}(\tdp, \tdm, \ehc,
\ehcb)=\llgm\bem
{0}&{-g^2N\vt^2}&{0}&{-m^2_{Te}}\cr
{-g^2N\vt^2}&{0}&{-m^2_{Te}}&{0}\cr
{0}&{-m^2_{Te}}&{0}&{-\frac12g^2\vx^2}\cr
{-m^2_{Te}}&{0}&{-\frac12g^2\vx^2}&{0}\eem\rrgm, \label{m05}\eeq
where the elements of the matrices above are
\beq m^2_{T^*e}=-\frac{\ld\mt\vx}{\sqrt{N}}~~\mbox{and}~~
m^2_{Te}=\sqrt{\frac{N}{2}}g^2\vt\vx\,. \eeq\eeqs
Upon diagonalization we obtain 2 zero eigenvalues -- which are
absorbed by the two charged gauge bosons $W_{\rm R}^\pm$ which
become massive after the SSB --, 2 real scalars due to D terms and
4 due to F terms.

\subparagraph{(d)} The matrix of the $\btdm -\btdp$ sector which
has the form
\beq {\cal M}^2_{XY^*}(\tdp,
\tdm)=\lf{\mt^2}/{N}+\ld^2\vx^2\rg\diag\lf1,1\rg~~\mbox{and}~~{\cal
M}^2_{XY}(\tdp, \tdm)=0.\label{m06}\eeq
Obviously, in this sector we obtain 4 massive real scalars with
mass eigenvalues
\beq m_{T\pm}=\lf{\mt^2}/{N}+\ld^2\vx^2\rg^{1/2}.\label{m06a}\eeq

In total, we obtain 19 d.o.f from the scalar sector of the model.

%

\subsubsection{Mass-Squared Matrix For Gauge Bosons.} This matrix can be found
from the kinetic terms which include the covariant derivatives
given in \Eref{Dmu1} -- (\ref{Dmu4}) as follows
\beqs\beq \lf{\cal M}^2_{1}\rg^{\rm AB}=g^2\vevl{\phc^\dagger {\sf
T}^{\rm A\tr} {\sf T}^{\rm B*}\phc+\phcb {\sf T}^{\rm A\tr} {\sf
T}^{\rm B*}\phcb^\dagger+N\Tr\lf [T^\dagger, {\sf T}^{\rm A}]
[{\sf T}^{\rm B\tr}, T]\rg}~~\mbox{with}~~{\rm A,
B}=1,...,4.\label{mgb1}\eeq
Here we employ the following ``unified'' description of the
$\sur\times\ubl$ generators
\bea{\sf T}^{\rm A}=\begin{cases}{\sf
T}_{\rm R}^\ca & \mbox{for}~ {\rm A}=\ca,\\
{\sf T}_{BL} & \mbox{for}~  {\rm A}=4,\end{cases} \nonumber\eea
which assists us to compactly represent ${\cal M}^2_{1}$. Its
final form is
\beq {\cal M}^2_{1}=\frac{g^2}{2} \llgm\bem
{2N\vt^2+\vx^2}&{-i\vx^2}&{0}&{0}\cr {i\vx^2}&{2N\vt^2+\vx^2}&
{0}&{0}\cr {0}&{0}&{\vx^2}&{-\sqrt{3}\vx^2/\sqrt{2}}\cr
{0}&{0}&{-\sqrt{3}\vx^2/\sqrt{2}}&{-3\vx^2/2}\eem\rrgm.
\label{mgb2}\eeq\eeqs
The diagonalization of the matrix above yields three eigenvalues
\beqs\beq M^2_{W^\pm_{\rm
R}}=g^2(2N\vt^2+\vx^2)\>\>\mbox{and}\>\>M^2_{A^\perp}=5g^2\vx^2/2,
\label{mgb3}\eeq
which correspond to the following eigenstates
\beq W^\mp_{\rm R}=\frac1{\sqrt{2}}\lf W^1_{\rm R}\pm iW^2_{\rm
R}\rg \>\>\mbox{and}\>\>A^{\perp}=-\sqrt{2\over5}W^3_{\rm
R}+\sqrt{3\over5}A_{BL}. \label{mgb4}\eeq\eeqs
Here $A^{\perp}$ is perpendicular to $A^{||}$ which remains
massless, and can be interpreted as the $B$ boson associated with
the $U(1)_Y$ factor of $\Gsm$. Therefore, we obtain $3\cdot3+2=11$
d.o.f from the sector of the gauge bosons. If we add these to the
19 d.o.f from the scalar sector we obtain 30 d.o.f which equal to
the initial number of bosonic d.o.f of the model.

\subsubsection{Mass Matrix for Fermions.} Although not directly
related to the aim of the paper, we  also derive for completeness
the mass matrix for fermions which has the following form
\beq {\cal M}_{1/2}=\llgm\bem 0_{4\times4} & {\cal M}_{1/2\rm
D}\cr {\cal M}^\tr_{1/2\rm D}&{\cal
M}_{1/2W}\eem\rrgm~~\mbox{where}~~\begin{cases}{\cal M}_{1/2\rm
D}=\vevl{{\rm D}_{,\what Z^\al}^{\rm A}},\\ {\cal
M}_{1/2W}=\vevl{W_{,\what Z^\al \what Z^\bt}}.\end{cases}
\label{m12}\eeq
and ${\what Z^\al}$ denotes canonically normalized $Z^\al$
determined from \Eref{VJe3} taking in to account that
$\vev{\sg}=\vt\ll\mP$. The contributions from D-terms are found to
be
\setcounter{MaxMatrixCols}{11}
\beqs\beq {\cal M}_{1/2\rm D}=g\llgm\bem
{0}&{0}&{-\vx\over\sqrt{2}}&{\vx\over\sqrt{2}}&{0}&{-\sqrt{N}\vt}&{\sqrt{N}\vt}&{0}&{0}&{0}&{0}\cr
{0}&{0}&{-i\vx\over\sqrt{2}}&{-i\vx\over\sqrt{2}}&{0}&{-i\sqrt{N}\vt}&{-i\sqrt{N}\vt}&{0}&{0}&{0}&{0}\cr
{0}&{\vx\over\sqrt{2}}&{0}&{0}&{-\vx\over\sqrt{2}}&{0}&{0}&{0}&{0}&{0}&{0}\cr
{0}&{-\sqrt{3}\vx\over2}&{0}&{0}&{\sqrt{3}\vx\over2}&{0}&{0}&{0}&{0}&{0}&{0}\eem\rrgm.\eeq
Taking the derivatives of $W$ w.r.t $Z^\al$ in the following order
$Z^\al=S,\nhcb, \ehc, \ehcb, \nhc, \tdp, \tdm, \tdn, \btdp, \btdm,
\btdn$ we find also the contributions from the $W$ terms which
take the form
\beq {\cal M}_{1/2W}=\llgm\bem
{0}&{\lp\vx}&{0}&{0}&{\lp\vx}&{0}&{0}&{2\lt\vt\over\sqrt{N}}&{0}&{0}&{0}\cr
{\lp\vx}&{0}&{0}&{0}&{0}&{0}&{0}&{0}&{0}&{0}&{-\ld\vx\over\sqrt{2}}\cr
{0}&{0}&{0}&{0}&{0}&{0}&{0}&{0}&{0}&{-\ld\vx}&{0}\cr
{0}&{0}&{0}&{0}&{0}&{0}&{0}&{0}&{-\ld\vx}&{0}&{0}\cr
{\lp\vx}&{0}&{0}&{0}&{0}&{0}&{0}&{0}&{0}&{0}&{-\ld\vx\over\sqrt{2}}\cr
{0}&{0}&{0}&{0}&{0}&{0}&{0}&{0}&{0}&{\mt\over\sqrt{N}}&{0}\cr
{0}&{0}&{0}&{0}&{0}&{0}&{0}&{0}&{\mt\over\sqrt{N}}&{0}&{0}\cr
{2\lt\vt\over\sqrt{N}}&{0}&{0}&{0}&{0}&{0}&{0}&{0}&{0}&{0}&{\mt\over\sqrt{N}}\cr
{0}&{0}&{0}&{-\ld\vx}&{0}&{0}&{\mt\over\sqrt{N}}&{0}&{0}&{0}&{0}\cr
{0}&{0}&{-\ld\vx}&{0}&{0}&{\mt\over\sqrt{N}}&{0}&{0}&{0}&{0}&{0}\cr
{0}&{-\ld\vx\over\sqrt{2}}&{0}&{0}&{-\ld\vx\over\sqrt{2}}&{0}&{0}&{\mt\over\sqrt{N}}&{0}&{0}&{0}
\eem\rrgm.\eeq\eeqs
Needless to say, the fermionic and bosonic d.o.f are equal to
22+8=30, where 22 d.o.f are associated with the chiral fermions
and 8 d.o.f correspond to gauginos.

\paragraph{} To obtain an independent verification for the correctness of
our computation,  we check the validity of the supertrace formula
which take, for our model, the form
\beq{\sf STr}{\cal M}^2=3(2M^2_{W^\pm_{\rm R}}+M^2_{A^\perp})-2\Tr
{\cal M}_{1/2}^\dagger{\cal M}_{1/2}+\Tr {\cal
M}^2_{0}=0.\label{strace}\eeq
We can convince ourselves that it is verified, if we do the
following replacements
\beqs\bea \Tr {\cal M}_{0}^2(Z^\bt)&=&4{\mt^2\over
N}+16{\lt^2\vt^2\over N}+\lf\frac52g^2+8\lp^2
+4\ld^2\rg\vx^2, \label{str1}\\
\Tr {\cal M}_{0}^2(Z^\gm)&=& 8{\mt^2\over
N}+8\ld^2\vx^2+2g^2(2N\vt^2+\vx^2), \label{str2}\\
\Tr \lf{\cal M}_{1/2}^\dagger{\cal M}_{1/2}\rg &=& 6{\mt^2\over
N}+ 8{\lt^2\vt^2\over
N}+2(2\lp^2+3\ld^2)\vx^2+g^2\lf8N\vt^2+9\vx^2\rg,\label{str3}\eea\eeqs
where we make use of the identification of $Z^\bt$ and $Z^\gm$ in
\Eref{bgdef}.

\subsection{Metastable CSs}\label{phi2}

The $U(1)_{\rm R}\times U(1)_{B-L}$ breaking which occurs for
$\sg\simeq\sgc$ causes the production of a network of
topologically unstable CSs which may be metastable. This network
has the potential to undergo decay via the Schwinger production of
MM--anti-MM pairs leading thereby to the generation of a
stochastic GW background. The tension $\mcs$ and the decay rate
per unit length of the CSs can be estimated by \cite{csdecay} --
for recent refinements see \cref{csdc} --
\begin{equation} \label{mucs} \mcs \simeq
4\pi\vx^2~~\mbox{and}~~\Gamma_{\rm dc}=\mcs
e^{-\pi\rms}/2\pi,\end{equation}
where the metastability factor $\rms$ is calculated via the
relation \cite{vile}
\begin{equation} \label{rms} \rms \simeq
m_{\rm M}^2/\mcs~~\mbox{with}~~m_{\rm
M}=4\pi\mwr/g^2,\end{equation}
the mass of the MMs generated by the SSB $SU(2)_{\rm R}\to\ur$.
Taking into account the expression of $\mwr$ in \Eref{mgb3} we
observe that the v.e.vs of both $T$ and $\phc-\phcb$ contribute to
$m_{\rm M}$ -- cf.~\cref{buch1}.

\subsection{Reheating}\label{phi3}

As shown in \Sref{phi4} -- see  Appendix~\ref{appw} too --, soon
after the end of TI, $\sg$ together with $\nhc$ and $\nhcb$ enter
into a phase of damped oscillations abound the minimum in
\Eref{vevs} reheating the universe. After diagonalizing the
matrices in \Eref{m02} and taking into account that the scalar
with mass due to D term is not energetically producible, we obtain
two mass-squared eigenvalues
\beq m_{\rm I\pm}^2=\frac12\lf2m^2_{\nu\rm F}+m_{TT}^2\pm
m^2_{TT\nu}\rg~~\mbox{with}~~m^2_{TT\nu}=\lf8m_{T\nu}^4+
(m_{TT}^2-2m^2_{\nu\rm F})^2\rg^{1/2},\eeq
where the various quantities involved are given in \Eref{mms}. The
eigenvalues above correspond to the complex fields ${\rm I}_\pm$,
given by
\beq \label{Itn} {\rm I}_\pm=\gm_{\nu\pm}\dnp+\gm_{T\pm}\what\dT
~~~\mbox{with}~~~\begin{cases}\gm_{\nu\pm}=\pm\lf{m^2_{T\nu\pm}}/{4m^2_{TT\nu}}\rg^{1/2}
\\\gm_{T\pm}=2\sqrt{2}m^2_{T\nu}/(8m^4_{T\nu}+m^2_{T\nu\pm})^{1/2}
\end{cases}\eeq
and $m^2_{T\nu\pm}= \pm2m^2_{\nu\rm F}\mp m_{TT}^2+m^2_{TT\nu}$.
Here the (complex) deviations of the fields $T$, $\nhc$ and
$\nhcb$ from their v.e.vs in \Eref{vevs} are denoted as $\dT$,
$\dnhc$ and $\dnhcb$ respectively and we have defined the complex
scalar fields
\beq\delta\nu_{\Phi\pm}=\lf\dnhc\pm\dnhcb\rg/\sqrt{2}~~\mbox{and}~~\what\dT=\sqrt{N}\dT.\eeq
Note that $\dnm$ does not acquire mass as it is the Goldstone
boson absorbed by $A^{\perp}$ in \Eref{mgb4} -- see paragraph (b)
of \Sref{parminf}.

The system of ${\rm I}_\pm$ fields settles into a phase of damped
oscillations abound the minimum in \Eref{vevs} reheating the
universe at a temperature which is exclusively determined by the
decay of $\qq$, as shown in the similar case analyzed in the
Appendix of \cref{tfhi}. Consequently, we have
\beq\Trh=
\left({72/5\pi^2g_{*}}\right)^{1/4}\lf\Gsn\mP\rg^{1/2}\>\>\>\mbox{with}\>\>\>\Gsn=\GNsn+\Ghsn\,.\label{Trh}\eeq
Here $g_{*}=228.75$ counts the MSSM effective number of
relativistic degrees of freedom and we take into account the
following decay widths
\beq \GNsn=\frac{\tilde\ld_{il^c}^2}{32\pi}\gamma_{{\rm
I}\nu-}^2\msn\lf1-\frac{4\mrh[
i]^2}{\msn^2}\rg^{3/2}\>\>\>\mbox{and}\>\>\>
\Ghsn=\frac{2\ld_{T\mu}^2}{16\pi}\gamma_{{\rm
I}T-}^2\msn,\label{Gs}\eeq
where $\gamma_{{\rm I}\nu-}$ and $\gamma_{{\rm I}T-}$ account for
the transition from $\delta\nu_{\Phi+}$ and $\what\dT$ to $\qq$
and may be brought into the form
\beqs\beq \gamma_{{\rm
I}\nu-}=\frac{\gm_{T+}}{\gm_{\nu-}\gm_{T+}-\gm_{\nu+}\gm_{T-}}\>\>\>\mbox{and}\>\>\>
\gamma_{{\rm
I}T-}=-\frac{\gm_{\nu+}}{\gm_{\nu-}\gm_{T+}-\gm_{\nu+}\gm_{T-}}.\eeq
Also the coupling constants which read
\beq\tilde\ld_{il^c}=\sqrt{2}\ld_{il^c}\frac{\vx}{\Ms}
\>\>\>\mbox{and}\>\>\>\ld_{T\mu}=\frac{2}{\sqrt{N}}\frac{\vt}{\msn}\lt\lm\eeq\eeqs
originate from the corresponding interaction Lagrangian terms
\beqs\bea\Lg_{\dphi\to \sni\sni}&=&-\frac12W_{{\rm
M},\nu_i^c\nu^c_i}\sni\sni\ +{\rm
h.c.}=-\sqrt{2}\ld_{il^c}\frac{\vx}{\Ms}\dnp\sni\sni\ +{\rm
h.c.},\\ \Lg_{\dphi\to \hu\hd}&=&-\left|\lf W_{\rm
H}+W_\mu\rg_{,S}\right|^2=-\frac{2}{\sqrt{N}}\frac{\vt}{\msn}\lt\lm\msn\what\dT
H_u^{*\tr}\veps H_d^* +{\rm h.c.}, \eea\eeqs
where $W_{\rm H}$, $W_\mu$ and $W_{\rm M}$ are given by
\eqss{whc}{wmu}{WM} respectively. Employing \Eref{Itn} we can
express $\dnp$ and $\what\dT$ in terms of $\qq$.

\subsection{Non-thermal Leptogenesis and Gravitino Constraint}\label{phi4}

Our post-inflationary scenario can be completed assuming that
$\qq$ decays  at least into one pair of $\rhni$ which is heavier
enough than $\Trh$. In particular, we seek the validity of the
following conditions
\beq\label{kin}
\msn\geq2\mrh[1]\>\>\>\mbox{and}\>\>\>\mrh[1]\gtrsim
10\Trh\>\>\>\mbox{with}\>\>\>\mrh[i]=2\ld_{il^c}\vev{\nhcb}^2/\Ms.\eeq
If the inequalities above are fulfilled, the out-of-equilibrium
decay of $\rhni$ generates a lepton-number-asymmetry yield which
is partially converted through the sphaleron effects into a yield
of the observed BAU \cite{lept, zhang}
\beq Y_B=-0.35\cdot{5\over2}{\Trh\over\msn}
{\GNsn\over\Gsn}\ve_L\,.\label{yb}\eeq
Assuming that the Majorana masses of $\nu^c_i$ are normally
hierarchical -- i.e., $\mrh[1]\ll \mrh[2],\mrh[3]$ -- and $\qq$
decays via \GNsn\ in \Eref{Gs} exclusively into $\wrhn[1]$, we can
obtain a maximal value for the lepton-number-asymmetry $\ve_L$
which is \cite{sasa, dreeslept}
\beq\label{el} \ve_L = -\frac {3}{8\pi}\frac{\mntau
\mrh[1]}{\vev{\hu}^2} \>\>\mbox{where}\>\> \mntau=\sqrt{\Delta
m^2_\oplus}=0.05\>{\rm eV} \eeq
is the mass of heaviest light neutrino $\nu_\tau$ which equals to
the square root of atmospheric neutrino mass squared difference
$\Delta m^2_\oplus$. Also, we set $\vev{\hu}=174~\GeV$ adopting
the large $\tan\beta$ regime of MSSM.

The required $\Trh$ in \Eref{yb} must be compatible with
constraints on the  $\Gr$ abundance, $Y_{3/2}$, at the onset of
\emph{nucleosynthesis} ({\sf\small BBN}), which is estimated to be
\beq\label{Ygr} Y_{3/2}\simeq 1.9\cdot10^{-22}\ \Trh/\GeV,\eeq
where we take into account only thermal production of $\Gr$, and
assume that $\Gr$ is much heavier than the MSSM gauginos.

\section{Constraining the Model Parameters}\label{res}

We below -- see \Sref{res2} -- present the allowed parameter space
of our model after imposing a number of constraints outlined in
\Sref{res1}.

\subsection{Imposed Constraints}\label{res1}

Our set-up must satisfy a number of observational requirements
specified below.

\subparagraph{\sf (a) Constraints on {\nsz $\Ns$} and {\nsz
$\As$}.} The quantities in \Eref{Nhi} has to be confronted with
the observational requirements \cite{plcp}
\beq\mbox{\small\sf
(a)}\>\>\Ns\simeq61.5+\frac14\ln\frac{\Vhi(\sex)^2}{g_{\rm
rh*}^{1/3}\Vhi(\sef)}\>\>\>\mbox{and}\>\>\>\mbox{\small\sf
(b)}\>\>\sqrt{\As}\simeq4.588\cdot10^{-5}.\label{ntot}\eeq
In deriving \sEref{ntot}{a} we assume, as in point (c) below, that
TI is followed in turn by a oscillatory phase with mean
equation-of-state parameter $w_{\rm rh}\simeq1/3$ -- which
corresponds to a quartic potential \cite{plin} --, radiation and
matter domination. We observe that $\Ns$ turns out to be
independent of $\Trh$.

\subparagraph{\sf (b) Constraints on {\nsz $\ns, \as$} and {\nsz
$r$}.} We take into account the latest data from \emph{Planck}
(release 4) \cite{plin}, baryon acoustic oscillations,
\emph{Cosmic Microwave Background} {\small \sf (CMB)} lensing and
BICEP/{\it Keck} \cite{gws}. Adopting the most updated fitting in
\cref{gws} we obtain approximately the following allowed margins
\begin{equation}  \label{nspl}
\mbox{\small\sf
(a)}\>\>\ns=0.965\pm0.0074\>\>\>\mbox{and}\>\>\>\mbox{\small\sf
(b)}\>\>r\leq0.032,
\end{equation}
at 95$\%$ \emph{confidence level} ({\sf\small c.l.}) with
$|\as|\ll0.01$.

\subparagraph{\sf (c) Constraints on {\nsz \gmcs} and {\nsz
\rms}.} The interpretation \cite{nano1} of the recent observations
\cite{nano,pta} dictates
\beq  4.3\cdot 10^{-8}\lesssim  \gmcs\lesssim 2\cdot
10^{-7}~~\mbox{for}~~8.21\gtrsim\sqrt{\rms}\gtrsim7.69\label{csnano}
\eeq
at $2\sigma$ where the upper bound originates from \cref{ligo} and
is valid for a standard cosmological evolution -- as described in
point (a) above -- and for CSs produced after TI. This scenario is
further supported by our analysis in Appendix~\ref{appw} which
reveals that the $\nhc-\nhcb$ post-inflationary phase transition
is fast enough and so, it does not generate any complementary
stage of inflation. Since the model predicts confined as well as
unconfined magnetic flux for the monopole, we adopt the results of
Table~4 in \cref{nano1} related to {\sc meta-l} model -- cf.
\cref{buch1}.

\subparagraph{\sf (d)  Constraints on {\nsz $\Yb$} and {\nsz
\Yg}.} $\Yb$ and \Yg\ in \eqs{yb}{Ygr} have to be compatible with
\plk\ \cite{plcp} and BBN data \cite{brand,kohri}, i.e.
\beqs\bea && ~~~~~~~~~~~~~~~~~~~~~~~~~~~~~~~~Y_B=\lf8.697\pm0.054\rg\cdot10^{-11}~~\mbox{at 95\% c.l.}\label{ybdat}\\
\mbox{and}\>\>&&\Yg\lesssim\left\{\bem
%
10^{-14}\hfill \cr
10^{-13}\hfill \cr \eem
\right.\>\>\>\>\mbox{for}\>\>\>\>\mgr\simeq\left\{\bem
0.69~\TeV\hfill \cr
10.6~\TeV\hfill \cr \eem
\right.\>\>\>\>\mbox{implying}\>\>\>\>\Trh\lesssim5.3\cdot\left\{\bem
%
10^{-2}~\EeV\,,\hfill \cr
10^{-1}~\EeV\,,\hfill \cr\eem
\right.\label{ygrdat}\eea\eeqs
where we assume that $\Gr$ decays with a tiny hadronic branching
ratio. The bounds above can be somehow relaxed in the case of a
stable $\Gr$.

\subsection{Results}\label{res2}

As deduced from \Sref{hi1} -- \ref{hi3}, our model depends on the
parameters
\bea
\nk,~\nsu,~M,~\mt,~\lt,~\lp,~\ld,~\lm~~\mbox{and}~~g.\label{para}
\eea
Despite that, from phenomenological point of view, any $\nk$ and
$\nsu$ value is possible we should keep in mind that integer
values are better motivated from theoretical point of view. In
particular, we use (and we do not mention it again) $\nsu=1$
throughout to maximize the relevant contribution to the effective
masses -- see \Tref{tab2}. Also, inspired by the gauge coupling
unification within MSSM we set $g=0.7$ -- cf. \cref{buch1}.
Although the precise unification is violated within our model due
to the presence of the superfields $\btd, T, \phcb$ and $\phcb$
with masses in the range from \EeV\ to \YeV -- see below --, we
expect that $g$ does not deviate a lot from its value above.
Nonetheless, a more elaborated approach to this point is presented
in Appendix~\ref{app}. The inflationary part of the model is
largely independent of $M$ and $\mt$ provided that $M\ll\mP$ and
$\mt\ll10^{-5}\mP$ as in our case. Enforcing \sEref{ntot}{b} fixes
$\lt$ at a value accurately calculated by \Eref{lda}. Employing it
as input in our code together with $\mt$, $\lp$ and $\ld$ we can
find $\ml$ and eliminate $M$ in favor of $\rms$. Indeed, taking
advantage from \eqs{mgb3}{rms} we end up with the relations for
$\vt$ and $M$
\beq \vt=\mt\frac{g^2\rms-4\pi}{4\sqrt{2}\pi\ld
N}~~\mbox{and}~~M=(\vt^2+\ml\vt)^{1/2}.\eeq
With selected $\rms$, $\gmcs$ is controlled by the variation of
$\ld$ and $\mt$ as shown by \eqs{mgb3}{rms}. On the other hand,
$\lp$ and $\lm$ are mainly constrained by \eqs{lmp}{lmpcon} and
have no sizable impact on the observables.

\begin{figure}[!t]\vspace*{-.12in}
\hspace*{-.19in}
\begin{minipage}{8in}
\epsfig{file=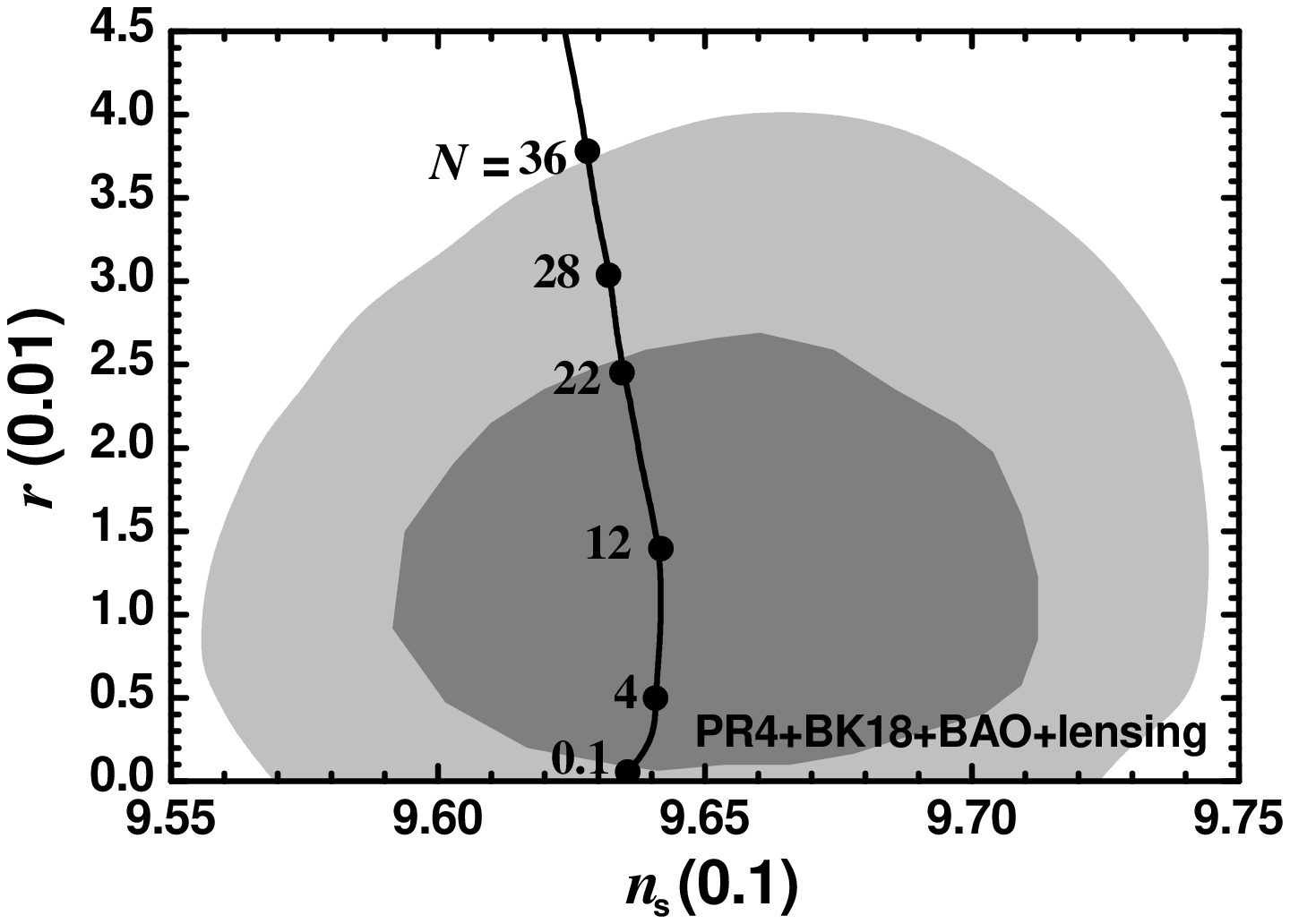,height=3.6in,angle=-90}
\hspace*{-1.2cm}
\epsfig{file=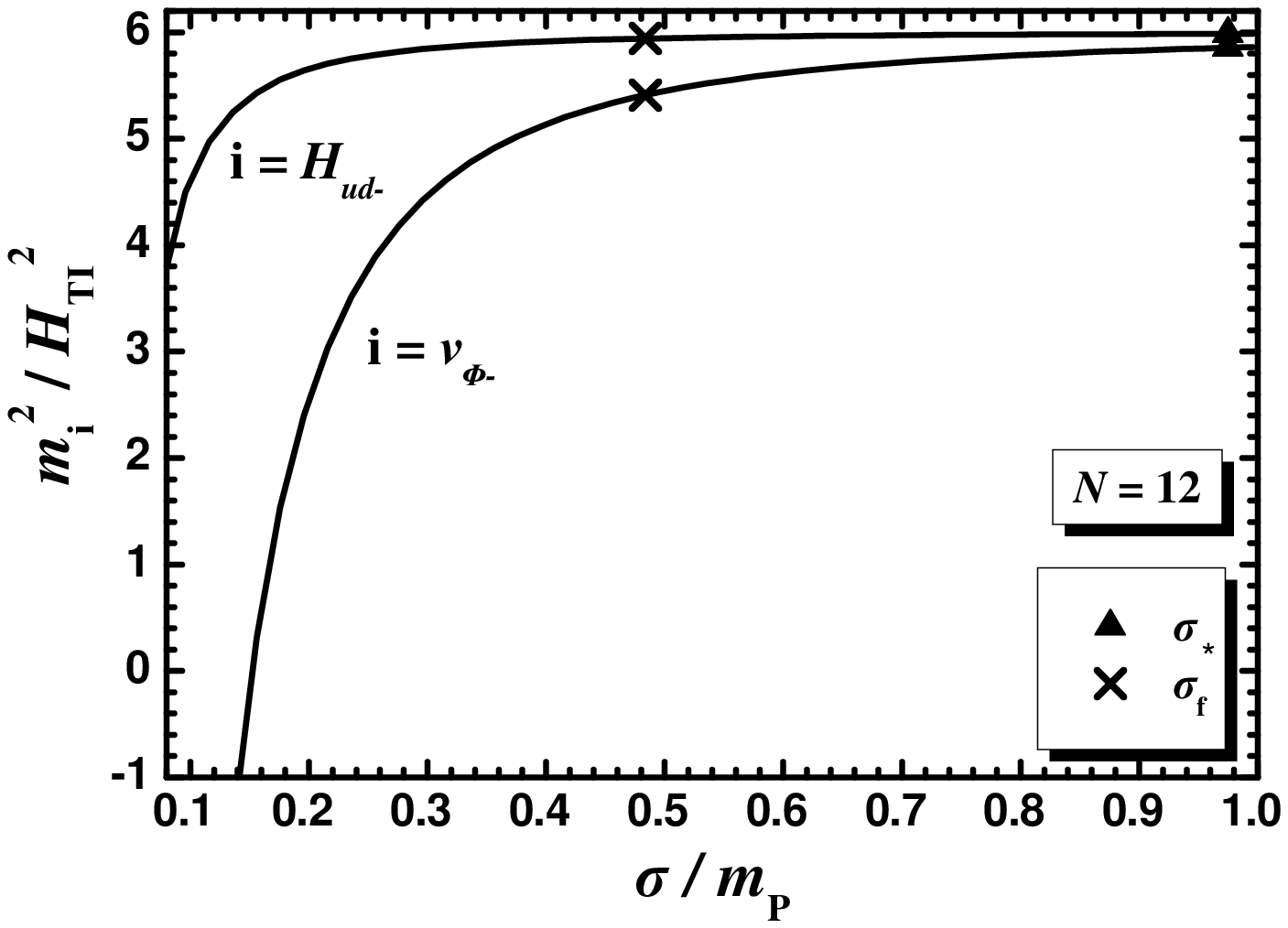,height=3.6in,angle=-90} \hfill
\end{minipage}\vspace*{-.3in}
\begin{flushleft}
\begin{tabular}[!h]{ll}
\hspace*{-.13in}
\begin{minipage}[t]{7.8cm}\caption[]{\sl\small Curve allowed by
\Eref{ntot} in the $\ns-r$ plane for various $N$'s indicated along
it for $\mt=10^{-9}\mP$, $\lp=10^{-6}$, $\ld=7.8\cdot10^{-7}$,
$\srms=8$ and $\lm=10^{-7}$. The marginalized joint $68\%$
[$95\%$] c.l. regions \cite{plin, gws} from PR4, {\sffamily\ftn
BK18}, BAO and lensing data-sets are depicted by the dark [light]
shaded contours.}\label{fig1a}\end{minipage}
&\begin{minipage}[t]{7.5cm}\caption[]{\sl\small The ratios
$m^2_{\rm i}/\Hhi^2$ as a function of $\sg$ with ${\rm
i}=\nu_{\phc-}$ and $H_{ud-}$ for $N=12$ and the remaining inputs
of \Fref{fig1a}. The values corresponding to $\sgx$ and $\sgf$ are
also depicted.}\label{fig1b}
\end{minipage}
\end{tabular}
\end{flushleft}\vspace*{-.11in}
\end{figure}

The available parameter space of the purely inflationary part of
our model is delineated in \Fref{fig1a}. We take $\mt=10^{-9}\mP$,
$\lp=10^{-6}$, $\lm=10^{-7}$, $\ld=7.8\cdot10^{-7}$ and $\srms=8$
(resulting to $\gmcs=10^{-7}$). Enforcing \Eref{ntot} with
$\Ns\simeq56$ we can estimate $\sgx$ and $\lt$ and obtain the
allowed curve in the $\ns-r$ plane by varying $N$. The result
displayed in \Fref{fig1a} is compared with the observational data
\cite{plin,gws}. We observe that $\ns$ remains essentially
independent from $N$ whereas $r$ increases with it, in accordance
with the analytic estimates in \eqs{ns}{rs}. More specifically, we
obtain
\beq \label{resi} 0.963\lesssim\ns\lesssim0.964,\>\>\>0.1\lesssim
N\lesssim 36\>\>\>\mbox{and}\>\>\> 0.0005\lesssim
{r}\lesssim0.039.\eeq
Regarding $\as$, it varies in the range $-(6.3-7.1)\cdot10^{-4}$
and so, TI is also consistent with the fitting of data with the
$\Lambda$CDM+$r$ model \cite{plin}. The proximity of $\sgx$ to $1$
signals a gentle tuning in the initial conditions since we obtain
$\Dex=1-\sgx\simeq(0.2-7)\%$ increasing with $N$.

One of the most impressive outputs of our proposal is the
stabilization of the $\phcb-\phc$ and $\hu-\hd$ sectors during TI
and their hierarchical destabilization after it so as the phase
transition, which causes the formation of CSs, to be automatically
triggered. To highlight further this achievement, we present in
\Fref{fig1b} the variations of $m^2_{\rm i}/\Hhi^2$ for
i$=\nu_{\phc-}$ and ${H_{ud-}}$ -- see \Tref{tab2} --  as
functions of $\sg$ fixing $N=12$ and employing the remaining
inputs of \Fref{fig1a}. We remark that $m^2_{\rm i}/\Hhi^2$ for
both i's are increasing functions of $\sg$ and remain larger than
unity for $\sgf\leq\sg\leq\sgx$, where $\sgx=0.975\mP$ and
$\sgf=0.48\mP$ are also depicted. After the end of TI, for
$\sg\leq\sgc=0.15\mP$ -- see \Eref{sgc} --, we obtain
$m^2_{\nu_{\phc-}}\leq0$ and therefore the $\phcb-\phc$ system
gets destabilized from the origin and starts rolling towards its
v.e.v in \Eref{vevs}. At the same time
$m^2_{\nu_{\phc-}}/\Hhi^2\ll m^2_{H_{ud-}}/\Hhi^2$ and so the
$\hu-\hd$ system remains well stabilized for all relevant $\sg$
values. Both prerequisites for the successful realization of this
picture in \Eref{lmpcon} are readily satisfied. From \Eref{mumgr}
we obtain $\mu/\mgr=0.04$ which hints towards natural SUSY
\cite{natsusy}.

\begin{figure}[t]\vspace*{-.12in}
 \hspace*{-.19in}\begin{minipage}{75mm}
\epsfig{file=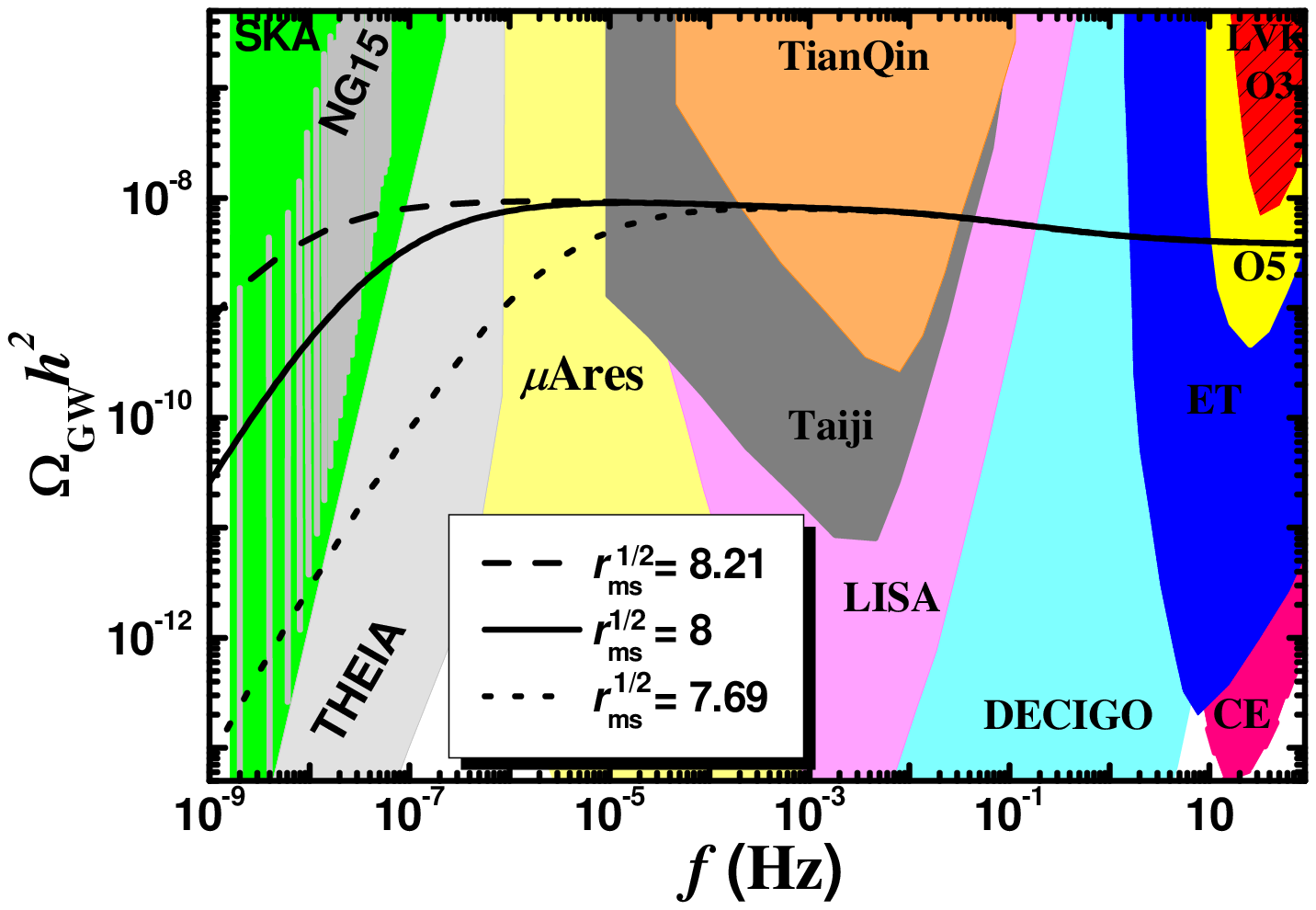,height=3.6in,angle=-90}
\end{minipage}
\hfill
\begin{minipage}{99mm}\renewcommand{\arraystretch}{1.2}
\vspace*{-.3in}\begin{center} {\small
\begin{tabular}{|c||c|c|c|}\hline
{\sc Model}&\multicolumn{3}{|c|}{$\srms$}\\\cline{2-4}
{\sc Parameters}&$7.69$&{$8$}&{$8.21$}\\ \hline\hline
$\ld/10^{-7}$&$7.35$&$7.8$&{$8.25$}\\\hline
$M/\YeV$&{$0.35$}&$0.363$&{$0.367$}\\
$\vt/\YeV$&{$0.255$}&$0.275$&{$0.284$}\\
$\vx/\YeV$&{$1.094$}&$1.1$&{$1.088$}\\ \hline
$M_{W^\pm_{\rm R}}/\YeV$&$1.16$&$1.21$&$1.23$ \\ \hline
\end{tabular}}
\end{center}\renewcommand{\arraystretch}{1.0}
\end{minipage}\vspace*{-.2in}
\hfill \caption{\sl\small GW spectra from the decay of CSs for
$N=12$, $\mt/\mP=10^{-9}$, $\lt=2.17\cdot10^{-5}$, $\lp=10^{-6}$
and various $\ld$ and $\srms$ values indicated in the Table of the
plot with fixed $\gmcs\simeq10^{-7}$. The shaded areas in the
background indicate the sensitivities of the current -- i.e.
NANOGrav \cite{nano} and LVK \cite{ligo} -- and future -- SKA
\cite{ska}, THEIA \cite{thia}, $\mu$Ares \cite{mares}, LISA
\cite{lisa}, Taiji \cite{tj}, TianQin \cite{tq}, BBO \cite{bbo},
DECIGO \cite{decig}, ET \cite{et} and CE \cite{ce} -- experiments.
The relevant values of the model parameters are listed in the
Table -- recall that $1~\YeV=10^{15}~\GeV$.}
\label{figgw}\end{figure}

Armed with the formulae presented in \cref{blfhi} -- which
originates from \cref{pillado} -- we compute the GWs spectrum,
$\ogw$, produced from the decay of the CSs and show it as a
function of the frequency $f$ in \Fref{figgw}. We use the same $N,
\mt, \lt$ and $\lp$ as in \Fref{fig1b} but we vary $\ld$ as shown
in the Table of \Fref{figgw}, where the resulting values of the
various scales appearing in our formulas are given too. The $\ld$
variation allows us to fix $\gmcs=10^{-7}$ for any selected $\rms$
within its $95\%$ c.l. interval of \Eref{csnano}. Namely, we show
$\ogw$ for $\srms$ to $7.69$ (dotted line) $8$ (solid line) and
$8.21$ (dashed line). We remark that as $\srms$ increases, the
increase of $\ogw$ becomes sharper and provide better fit to the
observations. Note that we use variable values for the effective
number of the relativistic d.o.f $g_*$ as mentioned in
\cref{blfhi} considering the MSSM sparticle spectrum for
temperatures above $10~\TeV$. As a consequence, we confirm the
slight reduction of $\ogw$ at high $f$ values mentioned in
\cref{geffant}. From the plot we may appreciate the necessity of
the upper bound on $\gmcs$ in \Eref{csnano} in order to fulfil the
upper bound from \cref{nano1} at a frequency $f_{\rm
LVK}\sim25~\hz$.  Shown are also in the plot examples of
sensitivities of possible future observatories \cite{ska, thia,
mares, lisa, tj, tq, bbo, decig, et, ce} which can test the
signals at various $f$ values.

To explore further the ranges of the parameters $\ld, \mt$ and
$\lp$ which render our proposal compatible with \Eref{csnano},
besides \eqs{ntot}{nspl}, we delineate in \Fref{fig2} the regions
allowed by the aforementioned three constraints for
$\srms\simeq8$, $\lm=5\cdot10^{-7}$ and $\lp=10^{-6}$. We depict
in \sFref{fig2}{a} the allowed area in the $N-\ld$ plane for
$\mt=10^{-9}\mP$ whereas in \sFref{fig2}{b}, {\small\sf (c)} and
{\small\sf (d)} the ones in the $N-\mt$, $N-M$ and $N-\vx$ planes
respectively for $\ld=10^{-6}$.  Needless to say, $\lt$ is found
consistently with \sEref{ntot}{b} as a function of $N$. The
boundaries of the allowed areas in \Fref{fig2} are determined by
the dashed [dot-dashed] lines corresponding to the upper [lower]
bound on $\gmcs$ in Eq.~(\ref{csnano}). The suitable $\gmcs$ is
achieved adjusting $\ld$ in \sFref{fig2}{a} or $\mt$ in the
remaining plots. We observe that $\gmcs$ increases by increasing
$\mt$ and decreasing $\ld$. Moreover, we remark that $\mt$ is much
lower than the value $10^{13}~\GeV$ which is required by
\sEref{ntot}{b} for quadratic inflation \cite{quad}. We also
notice that all the other scales besides $\mt$ lie at the $\YeV$
regions and increase with $\gmcs$. Taking into account the data
from \Fref{fig2} we find
\beq
\label{res1a}0.05\leq\mu/\mgr\leq0.23~~\mbox{and}~~0.39\leq\msn/\EeV\leq2.95.\eeq
Therefore, natural SUSY \cite{natsusy} remains a favorable output
within the whole parameter space of the model. On the other hand,
$\msn$ turns out to be of the order of \EeV -- recall that
$1~\EeV=10^9~\GeV$.

\begin{figure}[!t]\vspace*{-.12in}
\hspace*{-.19in}
\begin{minipage}{8in}
\epsfig{file=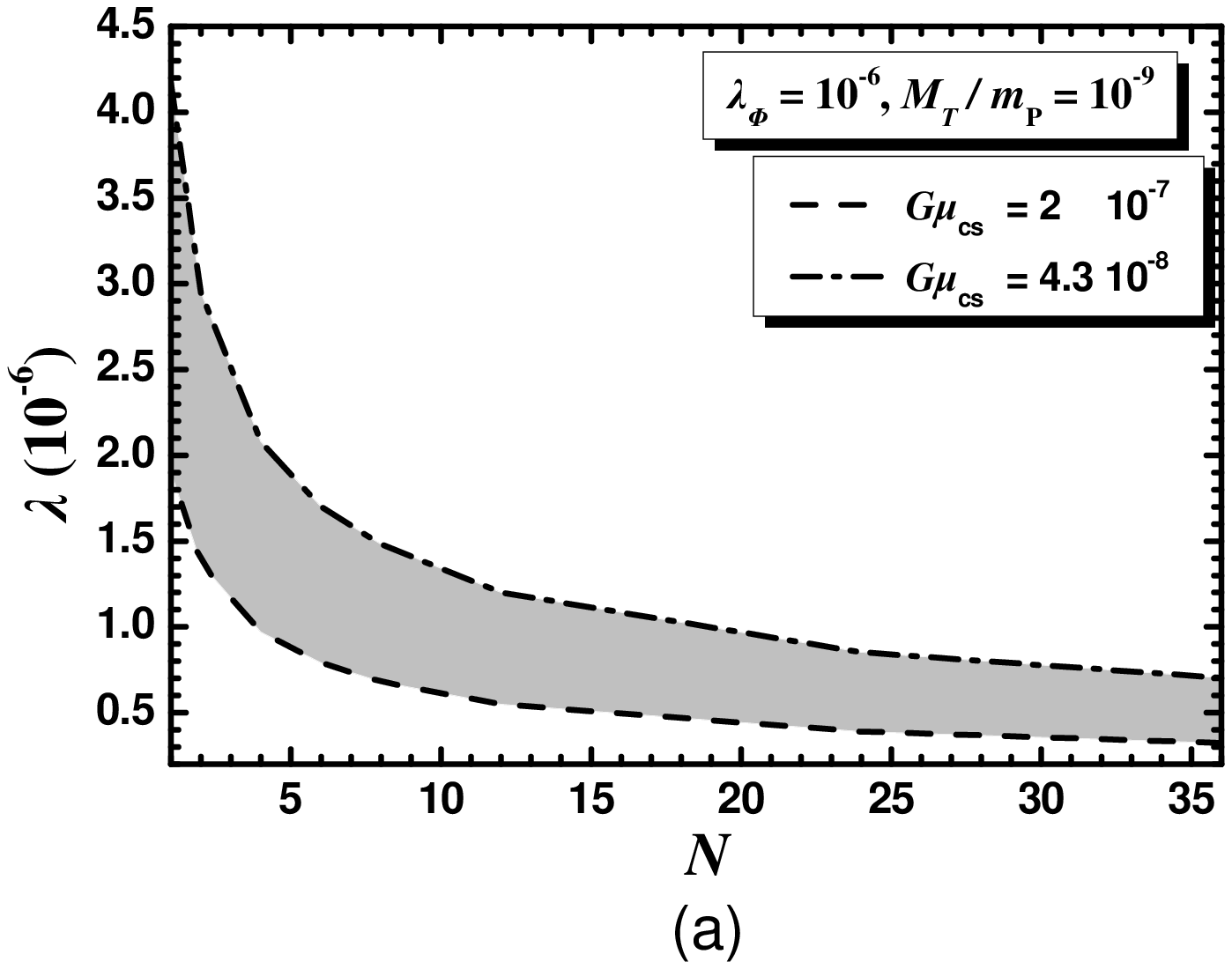,height=3.6in,angle=-90} \hspace*{-1.2cm}
\epsfig{file=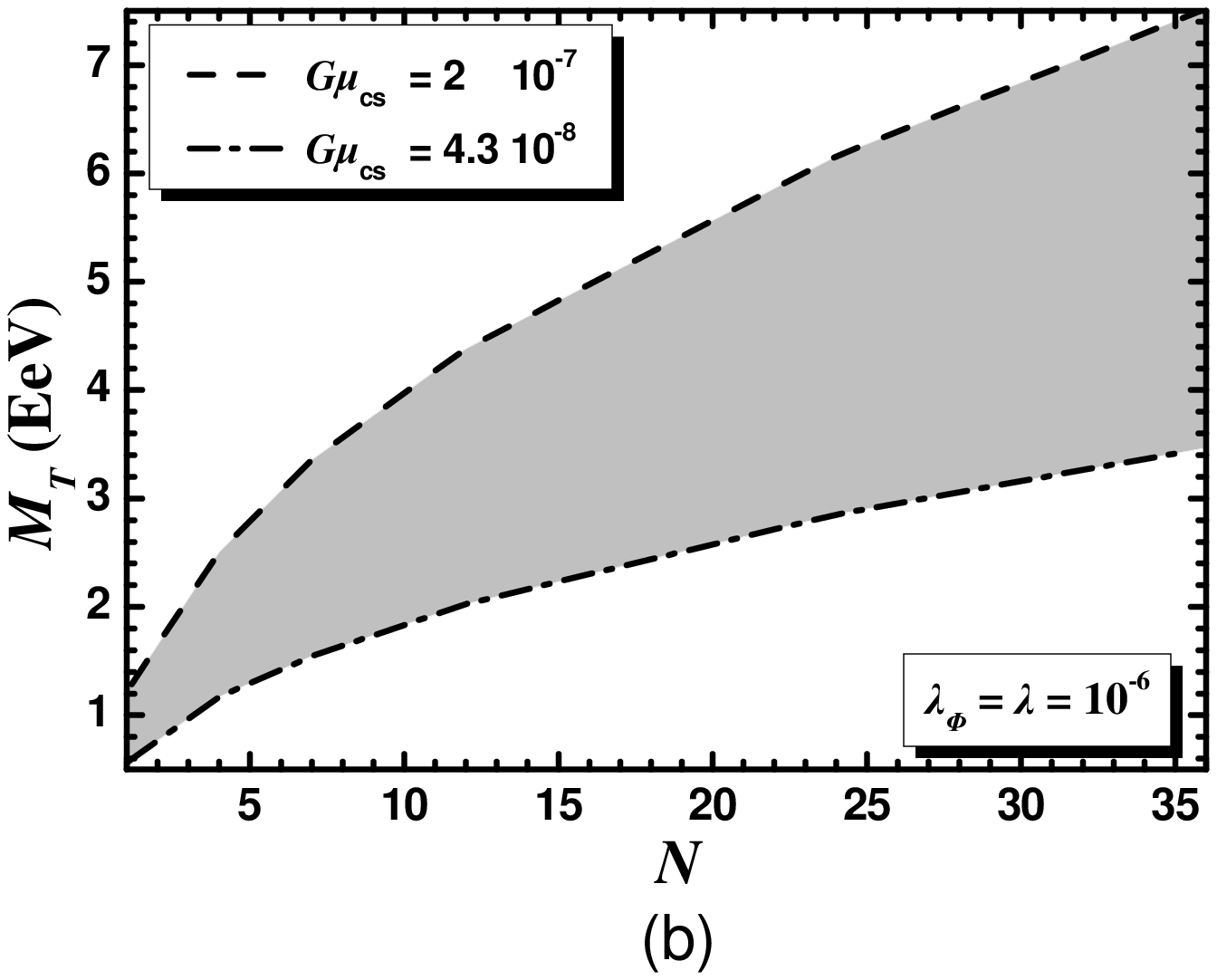,height=3.6in,angle=-90} \hfill\\
\epsfig{file=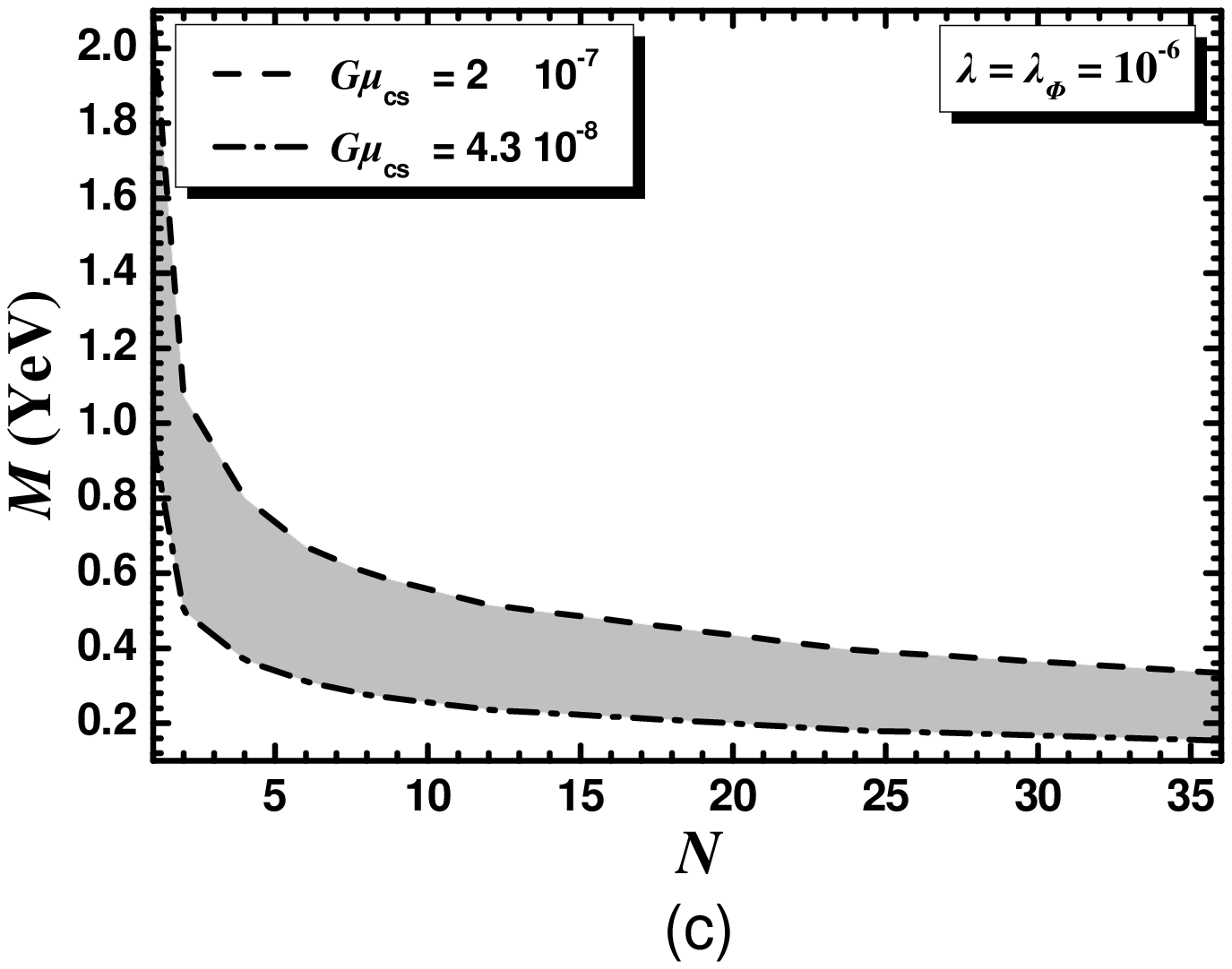,height=3.6in,angle=-90}
\hspace*{-1.2cm}
\epsfig{file=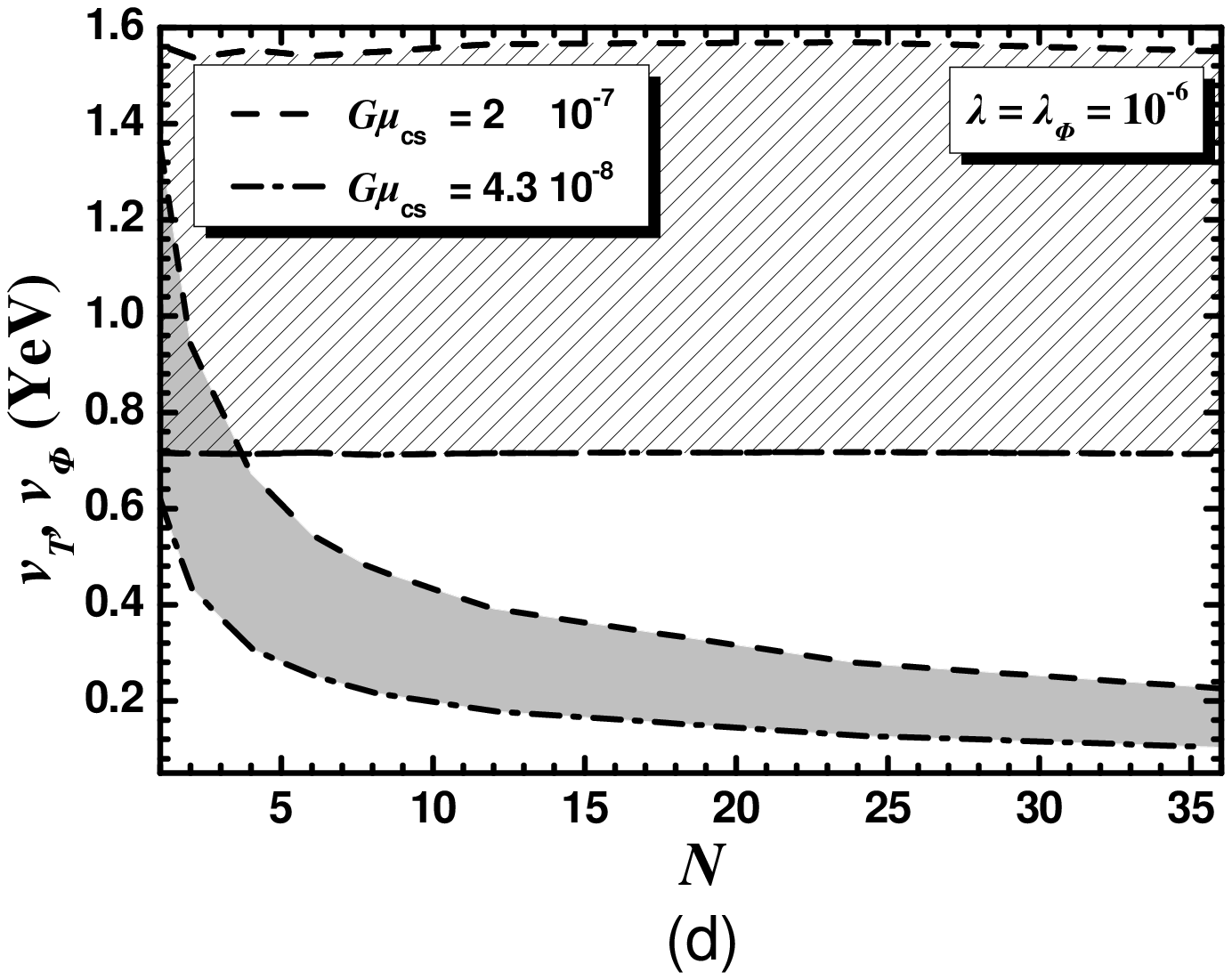,height=3.6in,angle=-90} \hfill
\end{minipage}
\hfill \caption{\sl\small Allowed (shaded) regions as determined
by Eqs.~(\ref{ntot}), (\ref{nspl}) and (\ref{csnano}) in the
$N-\ld$ {\sffamily\ftn (a)}, $N-\mt$ {\sffamily\ftn (b)}, $N-M$
{\sffamily\ftn (c)} and $N-\vx$ and {\sffamily\ftn (d)} plane for
$\srms\simeq8$, $\lm=5\cdot10^{-7}$ and $\lp=10^{-6}$. We also fix
$\mt=10^{-9}\mP$ {\sffamily\ftn (a)} or $\ld=10^{-6}$
{\sffamily\ftn (b), (c)} and {\sffamily\ftn (d)}. Hatched is the
allowed $N-\vt$ region. The conventions adopted for the boundary
lines are also shown. }\label{fig2}
\end{figure}

Our final task is to find out if there are portions of the
parameter space compatible with \eqs{ybdat}{ygrdat} too. Since the
decrease of $N$ and the increase of $M$ and $\gmcs$ increase
$\msn$, the fulfillment of \Eref{ybdat} can be facilitated by
relatively low $N$ and large $M$ and $\gmcs$ values -- see
\Eref{yb}. From our running we remark that the constraint above is
satisfied for $\msn\geq3~\EeV$ and $\Trh\geq1~\PeV$. Moreover,
$\lm$ has to be low enough so that the branching ratio in
\Eref{yb} does not suppress the final result. Taking these
fundamental observations into account we construct the curves in
\Fref{fig4} which are allowed by all the constraints of
\Sref{res2}. We fix for both panels in this figure $\lm=10^{-8}$.
Along the dashed and the dot-dashed curves we fix $\mt=1~\EeV$,
$\gmcs=10^{-7}$ and $\lp=10^{-6}$ and $5\cdot10^{-6}$
respectively. We remark that the allowed $N$ values are lower than
$6.5$. To allow for larger $N$ values we employ larger $\gmcs$ and
$\mt$ values but lower $\lp$, i.e., $\gmcs=1.25\cdot10^{-7}$,
$\mt=10~\EeV$ and $\lp=10^{-7}$ to obtain the solid line which is
extended up to $N=13.5$. From \sFref{fig4}{a} we see that the
required $\mrh[1]$ values are confined in the $\EeV$ region
whereas from \sFref{fig4}{b} we see that the resulting $\Trh$ lies
at the $\PeV$ region. As a consequence of the latter result, the
$\Gr$ constraint in \Eref{ygrdat} is comfortably satisfied for
$\mgr$ even lower than $1~\TeV$ -- see \Eref{ygrdat}.

Although we may achieve partially acceptable result for a little
larger $\lm$ values than that adopted in \Fref{fig4} we preferred
to fix it to a low enough value same for all the depicted curves
-- e.g., along the solid line we can use $\lm=10^{-7}$. This value
is one or two orders of magnitude lower than the value of the
Yukawa coupling constant in \Eref{wy} related to the up-quark
mass, which is around $10^{-6}$ \cite{fermionM}. Therefore, its
selection signals some tuning in the parameters which can not be
characterized, though, very ugly. Such an adjustment can be
evaded, if we remove $W_\mu$ from $W$ in \Eref{wtot} changing the
$R$ assignments and resolving the $\mu$ problem through a
Peccei-Quinn symmetry -- see e.g. \cref{tfhi} -- or via the
Masiero-Giudice mechanism -- see e.g. \cref{blfhi}. In the first
case, issues related to the axion isocurvature perturbations has
to be arranged whereas the realization of baryogenesis is made
difficult due to the lower resulting $\Trh$ in the latter case --
nonetheless larger $\gmcs$ values than the upper bound in
\Eref{csnano} may be obtained \cite{blfhi} in this case. We here
opted to follow the simplest method at the cost of a mild tuning
related to the realization of nTL. A more detailed investigation
of this topic, however, requires the inclusion of extra Higgs
fields \cite{tfhi} which violate moderately the $\nu_\tau-\tau$
Yukawa unification, predicted by the current simple model -- see
\Eref{wmssm}. This violation allows for a variation of the Dirac
neutrino masses and so of the resulting $\mrh[i]$'s via the
see-saw mechanism.

\begin{figure}[!t]\vspace*{-.12in}
\hspace*{-.19in}
\begin{minipage}{8in}
\epsfig{file=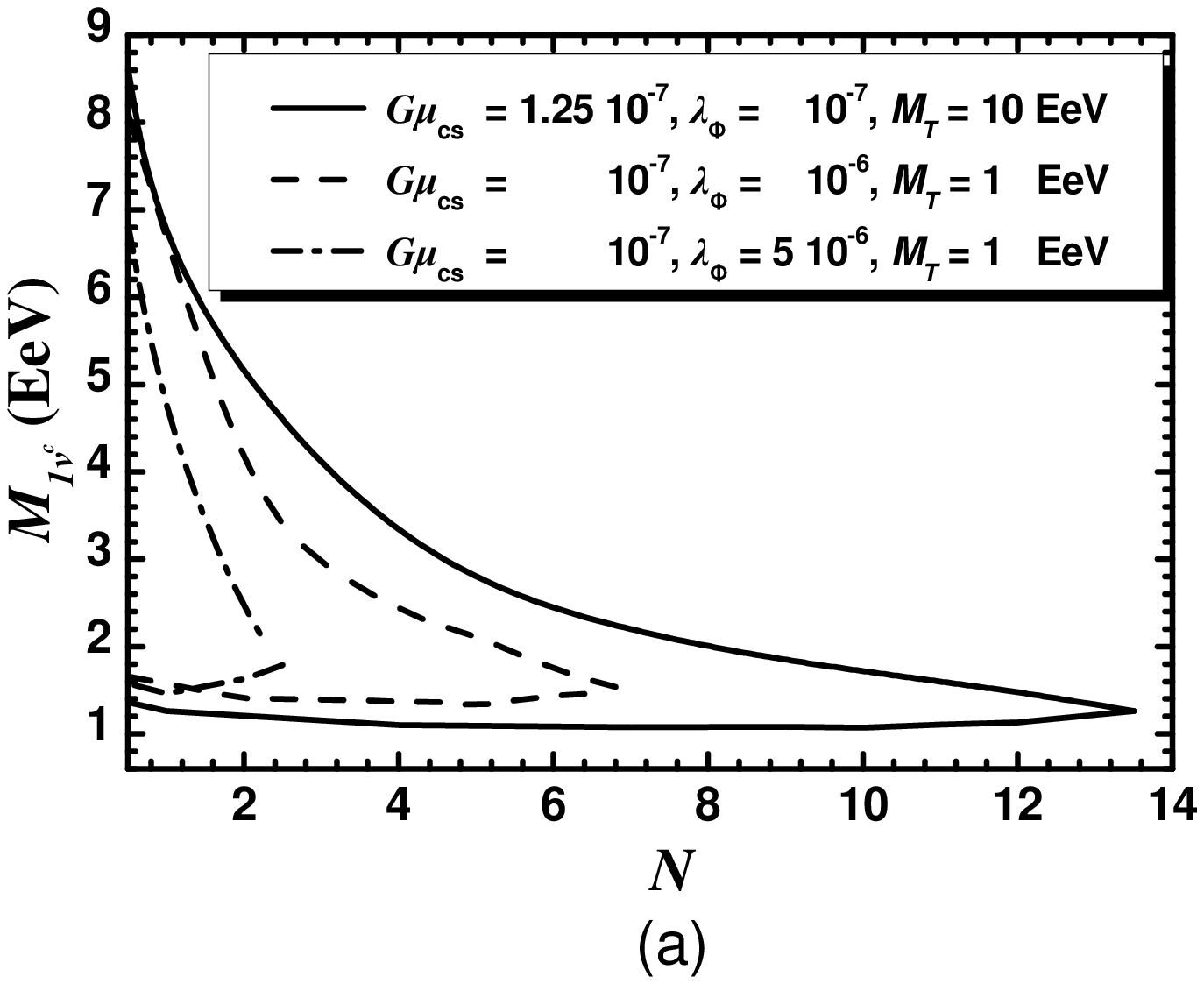,height=3.6in,angle=-90}
\hspace*{-1.2cm}
\epsfig{file=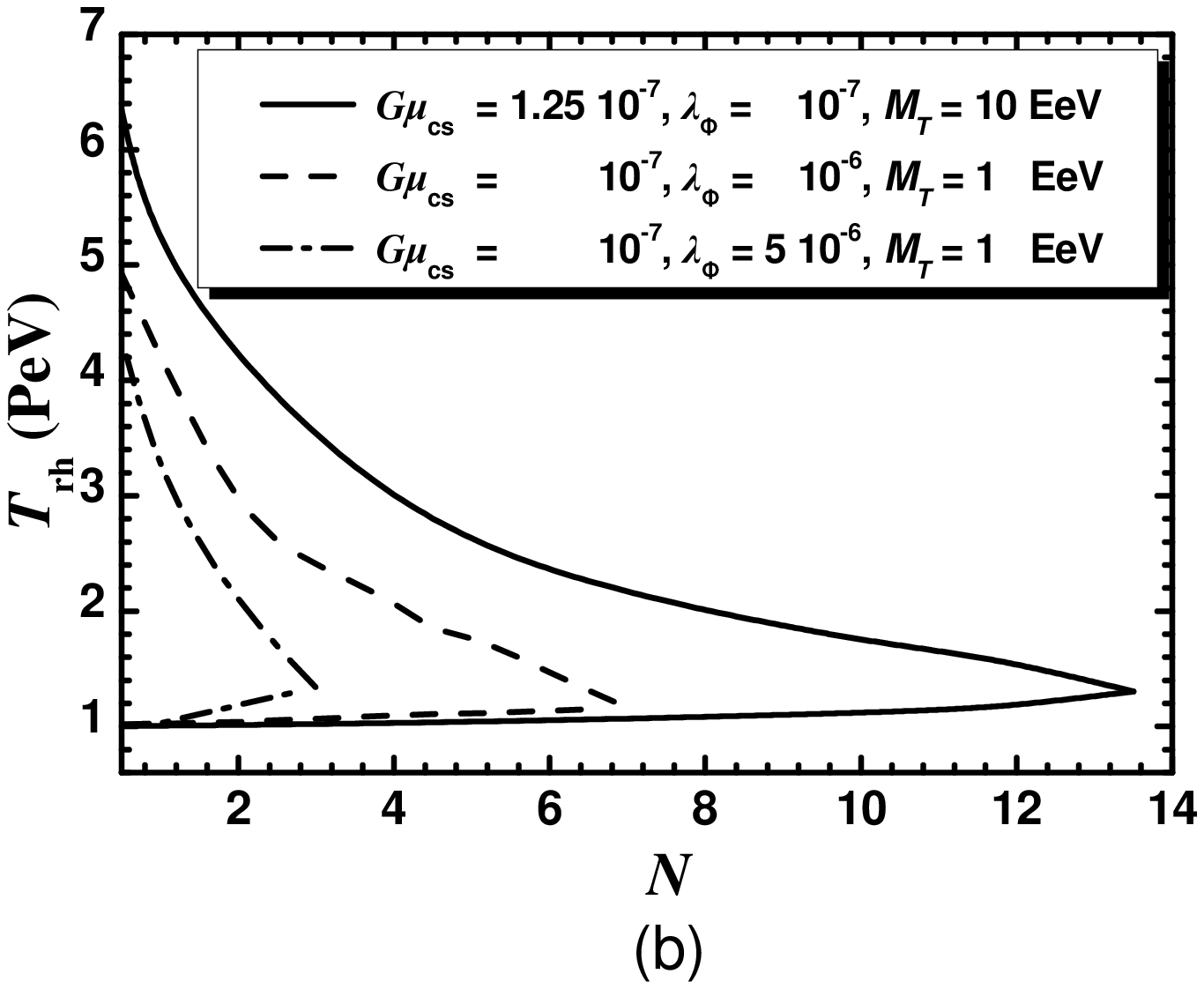,height=3.6in,angle=-90} \hfill
\end{minipage}
\hfill \caption{\sl\small Contours allowed by Eqs.~(\ref{ntot}) --
(\ref{ygrdat}) in the $N-\mrha$ ({\ftn\sffamily a}) and $N-\Trh$
({\ftn\sffamily b})  plane for $\srms\simeq8$, $\lm=10^{-8}$ and
various $\gmcs$, $\mt$ and $\lp$ values indicated in the plots.
}\label{fig4}
\end{figure}





\section{Conclusions}\label{con}

Motivated by the recent PTA results \cite{pta,pta1,pta2,nano},
hinting at a stochastic GW background, we proposed a model which
leads to the formation of metastable CSs. Their decay via MM and
anti-MM pair production may explain the aforementioned data. Our
model is relied on the $\Glr$ gauge group in \Eref{glr} and
employs the super- and \Kaa\ potentials $W$ and $K$ in
\eqs{wtot}{ktot} respectively. It includes two pairs of Higgs
superfields: two $B-L$-neutral \sur\ triplets, $\btd-T$, and two
\sur\ doublets oppositely charged under $B-L$, $\phcb-\phc$. TI
(i.e., T-model inflation) in SUGRA is realized by the radial part
of $T$, is tied to the quartic scalar potential -- see \Eref{Vhi0}
-- and the kinetic mixing in \Eref{VJe3} arising from the adopted
hyperbolic \Kaa\ geometry -- see \Sref{md2}. The predictions of TI
are $\ns\simeq0.963$ with negligible $\as$ and $r$ increasing with
the coefficient $N\lesssim36$ in \Eref{ki}. It remains also
largely immune from one-loop RCs derived in \Sref{hi2b}. As a
byproduct, TI inflates away the early produced MMs and possesses a
natural mechanism for CS formation via a tachyonic instability,
occurring after its termination in the $\phcb-\phc$ system.

Nonetheless, we also specified a post-inflationary completion,
which offers a nice solution to the $\mu$ problem of MSSM and
allows for baryogenesis via nTL.  To properly accommodate these
goals a specific relation is imposed among two $W$ coupling
constants -- see \Eref{lmpcon} -- which implies the relation
$\mu\ll\mgr$. We also worked out the complete particle spectrum of
the model in the SUSY vacuum and determined the reheating system
which is an admixture of the inflaton and $\phcb-\phc$ systems. We
adopted the simplest setting according to which the neutrino
masses are hierarchically ordered and the inflaton decays to the
lightest right-handed neutrino $\wrhn[1]$. We carved out the
parameter region consistent with constraints from \nano, the mass
of the heaviest light neutrino, the correct BAU and the $\Gr$
abundance. The success of nTL scenario clearly favors large
$\gmcs$ and low $N$ and $\lm$ values which confine $\mrh[1]$ and
$\Trh$ in the ranges $(1-9)~\EeV$ and $(1-6.3)~\PeV$,
respectively.

Finally, we would like to point out that, although we have
restricted our discussion on $\Glr$, the proposed mechanism of the
production of metastable CSs has a much wider applicability. It
can be realized within other GUTs, based on other gauge groups
such as the Pati-Salam \cite{ax}, the flipped $SU(5)$ \cite{leont}
or the trinification \cite{tri}. The usage of the adjoint
representation for the first step of SSB seems to be unavoidable
for the generation of the MMs whereas the formation of metastable
CSs may be orchestrated by a similar instability after the end of
TI which dilutes the MMs and is compatible with data. We expect
increasing technical difficulties due to the larger
representations required.


\acknowledgments I would like to thank  G. Leontaris, M.
Malinsk\'y and Q. Shafi for useful discussions.

\appendix{Anticipated Gauge Coupling Unification}\label{app}

\renewenvironment{subequations}{%
\refstepcounter{equation}%
\setcounter{parentequation}{\value{equation}}%
  \setcounter{equation}{0}
  \def\theequation{A.\theparentequation{\sf\small \alph{equation}}}%
  \ignorespaces
}{%
  \setcounter{equation}{\value{parentequation}}%
  \ignorespacesafterend
}

\paragraph{\hspace*{.25cm}} We here discuss the question of gauge coupling unification in our
model. As already stated in \Sref{res2}, the presence of
\Gsm-non-singlets components of $\btd, T, \phcb$ and $\phc$ with
masses below the unification scale $\mun\simeq2\cdot10~\YeV$
raises doubts regarding the validity of the simple unification of
the gauge coupling constants $g_l$ with $l=1,2,3$ within MSSM
\cite{martin}.

Indeed, the interpretation of \nano\ forces the magnitude of $\vx$
and $\vt$ in \Eref{vevs} to be of order $\YeV$ and that of $\mt$
to be of order \EeV. As a consequence, calculating the full mass
spectrum of the model in the SUSY vacuum of \Eref{vevs} -- see
\Sref{phi1} --, one finds that there are \Gsm-non-singlet fields
acquiring mass of order $\EeV$ and others that acquire mass of
order $\YeV$. In particular, the lowest mass eigenstate is $\mtpm$
given by \Eref{m06a} whereas the highest mass scale is identified
with $\mwr$ in \Eref{mgb3}. Therefore, we can introduce an
intermediate mass threshold, $\mint$, and set the (new)
unification scale $\bmu$ according to the prescription
\beq \mint=\mtpm ~~\mbox{and}~~\bmu=\mwr~~\mbox{with}~~\mint<\mun.
\label{thr} \eeq
This situation obviously deviates from the great-desert hypothesis
and so the MSSM unification of $g_l$ is expected to be spoiled. To
restore it, we supplement our model with more superfields with
masses in the interval $\mint-\bmu$ following the analysis of
\cref{magic}, whose the key points are summarized in \Sref{app1}
below -- cf. \cref{ax}. The implications to our setting are then
organized in \Sref{app2}.

\subsection{Intermediate Scale \& Magic Fields}\label{app1}

The \emph{renormalization-group equations} ({\small\sf RGEs}) for
the SM fine structure constants $\ali=g_l^2/4\pi$ admit at one
loop the solution
\beq\label{ali} \alpha_l(Q)^{-1}={\alpha_l(\tZ)}^{-1}-
{\bi}(t-\tZ)/2\pi ~~\mbox{with}~~t=\ln(Q/\GeV),\eeq
where $Q$ is the renormalization-group scale and $\tZ$ corresponds
at the electroweak scale $M_Z$. If we assume that there is
unification logarithmic scale $\tu$ such that $\ali(\tu)=\alu$, we
can determine $\tu$ and $\alu$ solving two of the equations in
\Eref{ali} as follows
\beq\label{alu}
\tu=\tZ+2\pi\frac{\albz^{-1}-\alaz^{-1}}{b_2-b_1}~~\mbox{and}~~\alu^{-1}=\frac{b_1\albz^{-1}-b_2\alaz^{-1}}{b_1-b_2}.
\eeq
If we take into account that the MSSM spectrum gives \cite{martin}
$(b_1,b_2,b_3) = (33/5, 1, -3)$ and the values of $\alpha_l(\tZ)$
we reveal the well-known MSSM predictions $\tu\simeq37.6$ and
$\alu\simeq1/24$. Moreover, If we solve \Eref{ali} w.r.t the
values of $\ali(\tZ)$ we find
\beq\label{rali} r_b :=
\frac{\alcz^{-1}-\albz^{-1}}{\albz^{-1}-\alaz^{-1}}=\frac{b_3-b_2}{b_2-b_1}=\frac57\eeq
in excellent agreement with the experimental values.

If we introduce an intermediate scale $\mint$ in the scheme above,
due to the presence of extra superfields for $t
>\tint=\ln(\mint/\GeV)$, then the running of $\ali$ from $\tZ$ to
$\tint$ remains unchanged but it is modified from $\tint$ until a
(new) unification logarithmic scale $\btu=\ln(\bmu/\GeV)$ as
follows
\beqs\bea\label{ali1} \alpha_l(t)^{-1}&=&{\alpha_l(\tZ)}^{-1}-
{\bi}(t-\tZ)/2\pi ~~\mbox{for}~~\tZ\leq t \leq\tint;
\\ \bali(t)^{-1}&=&{\bali(\tint)}^{-1}-
{\bbi}(t-\tint)/2\pi ~~\mbox{for}~~\tint\leq t \leq\btu,
\label{ali2}\eea\eeqs
with matching condition $\bali(\tint)=\ali(\tint)$. Here, the
coefficients $\bar b_l = b_l + d_l$ include the contribution $d_l$
of the extra superfields with masses above $\mint$ -- for a
comprehensive review in computing the $\di$ coefficients see
\cref{roy}. Upon estimating again $r_b$ in \Eref{rali} we infer
that it remains intact if we impose the condition
\beq \label{rd}
\rd:=\frac{d_3-d_2}{d_2-d_1}=\frac{b_3-b_2}{b_2-b_1}=\frac57.\eeq
If we solve \eqs{ali1}{ali2} w.r.t $\btu$ we find
\beq \label{bmu}\btu=\tu+\brd(\tint-\tu)~~\Rightarrow~~\bmu = \mun
\left({\mint}/{\mun} \right)^{\brd} ~~\mbox{with}~~ \brd=
\frac{d_{3}-d_{2}}{\bar b_3-\bar b_2},\eeq
and the (new) unified value $\balu=\bali(\btu)$ is estimated in
terms of $\alu$ as follows
\beq \label{balu}\balu^{-1}=\alu^{-1}-\frac{(1-\brd) d_l-\brd
b_l}{2\pi}\left(\tu-\tint\right),\eeq
where $\tu$ and $\alu$ are given in \Eref{alu}. All the sets of
superfields which satisfy Eq.~(\ref{rd}) without being complete
GUT multiplets are called ``magic'' superfields \cite{magic}. They
fall into two categories: those with $\brd=0$, which just mimic
the effect of complete GUT multiplets and those with $\brd\neq 0$,
which change the GUT scale according to \Eref{bmu}. In our case,
we focus on $0<\brd< 1$ which leads to anticipated unification of
$\bali$, i.e., $\bmu<\mun$ for $\mint<\mun$ and serves for the
mass scales encountered in the findings of \Sref{res2}.

\subsection{Implications to our Model}\label{app2}

In \Sref{app2a} below we propose a possible completion of the
model in order to be compatible with the $\bali$ unification and
then -- in \Sref{app2} -- we delineate the ramifications induced
to the results of \Sref{res2}.

\subsubsection{Model Modification.}\label{app2a}

As we see below, $5d_1/3$ and $d_2$ may acquire semi-integer
positive values whereas $d_3$ takes only integer positive values.
To achieve $0<\brd<1$, which is imperative for anticipated
unification of $\bali$, we need $d_2>d_3$ since $d_2>0$ and
$d_3>0$ whereas $b_3-b_2<0$. If we fix $d_3=1$ and select
$d_2>d_3$ we can solve \Eref{rd} w.r.t $d_1$. As a result, a
cornucopia of suitable $\di$ combinations arises which lead to the
desired $\brd$ values. Confining ourselves to $d_2\leq9$ we list
in \Tref{tabd} a set of various $d_1$ and $d_2$ for $d_3=1$
together with the resulting $\brd$. The presented $\di$ values are
the initial ones of a class of solutions achieved by adding unity
to each $\di$. Obviously, in that case $\rd$ and $\bmu$ remain
unchanged but $\balu$ increases -- see \eqss{rd}{bmu}{balu}.

\begin{table}[!t]
\begin{center}
\begin{tabular}{|c|ccccccccc|}
\hline
$5d_1$&$11$&$17$&$29$&$41$&$53$&$39$&$77$&$89$&$101$\\
$d_2$&$3/2$&$2$&$3$&$4$&$5$&$6$&$7$&$8$&$9$\\\hline
$\brd$&$1/9$&$1/5$&$1/3$&$3/7$&$1/2$&$5/9$&$3/5$&$7/11$&$2/3$ \\
\hline
\end{tabular}
\end{center}
\caption[]{\sl \small Values of $d_1$ and $d_2$ for $d_3=1$
resulting to various values $0<\brd<1$ consistently with
\Eref{rd}.}\label{tabd}
\end{table}
\renewcommand{\arraystretch}{1.}

From the solutions arranged in Table~\ref{tabd} we can easily
deduce that only those corresponding to $\brd=1/9$ may reconcile
\Eref{thr} with $\mtpm$ and $\mwr$ values found in \Sref{res2}. To
select some specific combination of $\di$ we first find the
contributions from the superfields (besides those of MSSM) which
already exist in the model -- see \Tref{tab1}. These are displayed
in the first three rows of \Tref{tabe} and their contribution is
estimated to be $(3\cdot 6/5,0,0)$ -- note that $Q_Y=0$ for
$\nci$. Therefore, the acceptable solutions with $\brd=1/9$ are
those with $5d_1/3\geq6$. As it turns out, $\gmcs$ less than the
upper bound in \Eref{csnano} can be achieved for
$(d_1,d_2,d_3)=(3\cdot36/5, 13/2, 6)$ or larger $\di$ values. This
is because $g(\btu)\simeq1$ assists us to reduce adequately $\mwr$
without jeopardizing the perturbativity of the model.

Our next task is to propose a particle content which yields the
required $\di$ values. A possible arrangement is shown in the four
last lines of \Tref{tabe}, where we list the extra (calligraphic)
superfields, their representations under $\Glr$, their
decompositions under $\Gsm$, their contributions to $\di$ and
their $R$ charges. To avoid gauge anomalies, we take care so as
the total $B-L$ charge of the new representations to vanish. From
these, $\hh_\al$, $\bcy_\al$, $\cy_\al$ and ${\cal Z}_\al$ may be
motivated by the $SO(10)$ embedding of $\Glr$. Note that two of
the four $\hh_\al$ are already introduced in \cref{tfhi} but with
different $R$ charges. On the other hand, $\cx_\al$ originate from
the $({\bf 1, 2, 1})$ representation of the Pati-Salam gauge group
which is not included in some $SO(10)$ representation up to ${\bf
210}$. However, it appears \cite{koba,kounas} in the context of
heterotic-string constructions as an additional SM exotic state.
The superpotential relevant for these new superfields has the form
\beq  \we=\sum_{\al,\bt = 1}^4  m_{\al\bt} \Tr\lf\hh^\tr_\al\veps
\hh_\bt\veps \rg\ +\ \sum_{\al,\bt = 1}^5  m_{\cx\al\bt} \Tr
(\cx_\al^\tr \veps \cx_\bt)\ +\ \sum_{\al,\bt = 1}^3 m_{\cy\al\bt}
\bcy_\al\cy_\bt \ + m_{\cal Z}\Tr{\cal Z}^2,\label{we} \eeq
where $m_{\al\bt}$, $m_{\cx\al\bt}$, $m_\cx$, $m_\cy$ and
$m_{{\cal Z}\al\bt}$ are mass parameters which may lie between
$\mtpm$ and $\mwr$. Thanks to the representations and the $R$
charges assigned to the extra superfields no renormalizable mixing
term between them and the initial ones -- shown in \Tref{tab1} --
is permitted. As a consequence, these terms do not affect the
stability of the inflationary path in \Eref{inftr} and the
constitution of the inflationary system as analyzed in
\Sref{phi3}.

\renewcommand{\arraystretch}{1.3}

\begin{table}[!t]
\begin{center}
\begin{tabular}{|cccccc|c|}
\hline {\sc Super-}&{\sc Represe-}&{\sc
Decompo-}&\multicolumn{3}{c|}{ {\sc Contributions} }&$R$
\\
\multicolumn{1}{|c}{\sc fields}&{\sc tantions}&{\sc sitions}
&\multicolumn{3}{c|}{\sc to $\di, l=1,2,3$}&{\sc Charges}
\\
\multicolumn{1}{|c}{}&{\sc under $\Glr$} &{\sc under $\Gsm$} &
$5d_1/3$&$d_2$&$d_3$&{}
\\\hline \hline
{$\bar \Phi, \Phi$}&$({\bf 1, 1, 2},\pm1)$&$({\bf 1, 1},\pm1)$ &$2$&$0$&$0$&$0,0$\\
{$T$} &{$({\bf 1, 1, 3},0)$}&$({\bf 1, 1},\pm1,0)$
&$2$&$0$&$0$&$0$\\
{$\btd$} &{$({\bf 1, 1, 3},0)$}&$({\bf 1, 1},\pm1,0)$
&$2$&$0$&$0$&$2$ \\ \hline
{$\hh_\al$} & {$({\bf1, 2, 2},0)$}&$({\bf 1,
2},\pm1/2)$&$4$&$4$&$0$&$1$ \\
{$\cx_\bt$} & {$({\bf1, 2, 1},0)$}&$({\bf 1, 2},0)$
&$0$&$5/2$&$0$&$1$ \\
{$\cy_\gm$} & {$({\bf3, 1, 1},-2/3)$}&$({\bf 3, 1},-1/3)$
&$1$&$0$&$3/2$&$1$ \\
{${\bcy}_\gm$} & {$({\bf \bar 3, 1, 1},2/3)$}&$({\bf \bar 3,
1},1/3)$ &$1$&$0$&$3/2$&$1$\\
${\cal Z}$ & {$({\bf 8, 1, 1},0)$}&$({\bf 8, 1},0)$
&$0$&$0$&$3$&$1$ \\ \hline
\end{tabular}
\end{center}
\caption[]{\sl \small The representations under~~$\Glr$, the
decompositions under~~$\Gsm$, the contributions to the $\di$
coefficients as well as the $R$ charges of the superfields (with
$\al=1,...,4$, $\bt=1,...,5$ and $\gm=1,2,3$) of the model with
$\brd=1/9$ beyond those included in MSSM.}\label{tabe}
\end{table}
\renewcommand{\arraystretch}{1.}

\subsubsection{Numerical Results.}\label{app2b}

To provide a pictorial verification of the gauge coupling
unification achieved in the augmented version of our model
described above, we first show in \Fref{figg} a sample running of
$\ali$ and $\bali$ (solid thick lines) for $N=12$, $\mt=10~\EeV$,
$\ld=4.27 \cdot 10^{-6}$, $\lp=10^{-6}$, $\lm=10^{-7}$ and
$\srms=8$. The selected values yield
\beq
\mtpm=6.97~\EeV,~\mwr\simeq3.83~\YeV,~\balu=1/10.6~~\mbox{and}~~
\gmcs\simeq1.97\cdot10^{-7}.\label{resa1}\eeq
These values satisfy the condition in \Eref{bmu} with $\brd=1/9$
and the identification of \Eref{thr}.  To clarify further the
difference of the anticipated versus the conventional unification
we also depict in \Fref{figg} the RGE evolution of $\ali$ within
MSSM (dashed thin lines). Let us note here that in our RGE running
we do not take into account possible threshold effects -- due to
the dispersion of the masses involved in $W_{\rm E}$ of \Eref{we}
-- and a possible running with the SM $\bi$ values for
$Q<10~\TeV$. Both effects are expected to have negligible impact
to our results. Moreover, possible two loop corrections are mild
(typically $10\%$) especially for $\mint>1~\EeV$ -- cf.
\cref{cards}.

The augmentation of the particle content of our model causes a
number of modifications to the numerical results exposed in
\Sref{res2}. Most notably, taking as input \Eref{thr} with
$\brd=1/9$ we can eliminate one of the free parameters, $\ld$ or
$\mt$, from those shown in \Eref{para} and obtain, for given
$\rms$, a prediction for $\gmcs$. In particular, for any selected
$\mt$, e.g.,  we can adjust $\ld$ so as $\mtpm$ and $\mwr$ in
\eqs{m06a}{mgb3} satisfy \eqs{bmu}{thr}. As a consequence, the
resulting $\gmcs$ is relied on the imposed conditions -- for any
fixed $\mt$ and $\rms$. To investigate if \eqs{thr}{bmu} with
$\brd=1/9$ can be reconciled with the \nano-favored region of
$\gmcs$ in \Eref{csnano} we plot in \Fref{figgmcs} the allowed
(shaded) region in the $N-\gmcs$ plane for $\mt=1~\EeV$,
$\lp=10^{-6}$ and $\lm=10^{-7}$. These values are comparable with
those employed in \sFref{fig2}{a}. The boundaries of the allowed
area are determined by the dashed [dotted] line corresponding to
the upper [lower] bound on $\srms$ in Eq.~(\ref{csnano}). We also
display by solid line the allowed contour for $\srms=8$. The
allowed margin of $\gmcs$ in \Eref{csnano} is also limited by two
thin lines. We remark that $\gmcs$ decreases as $\rms$ and $N$
increase.

If we attempt, in addition, to meet the correct BAU in
\Eref{ybdat} for the inputs mentioned above \Eref{resa1} we obtain
the following solutions
\beq \mrha\simeq 1.34~\EeV~~\mbox{or}~~ \mrha\simeq 2.81~\EeV
~~\mbox{with}~~\Trh\simeq~2.3~\PeV~~\mbox{or}~~
\Trh\simeq~1.2~\PeV, \label{resa2} \eeq
respectively. These values are close to the values used along the
solid lines shown in \Fref{fig4}. On the other hand, $\gmcs$ turns
out to be closer to the upper bound in \Eref{csnano}. Needless to
say, \Eref{ygrdat} is comfortably fulfilled thanks to the low
achieved $\Trh$ values.

\begin{figure}[t]
\hspace*{-.1in}
\begin{minipage}{8in}
\epsfig{file=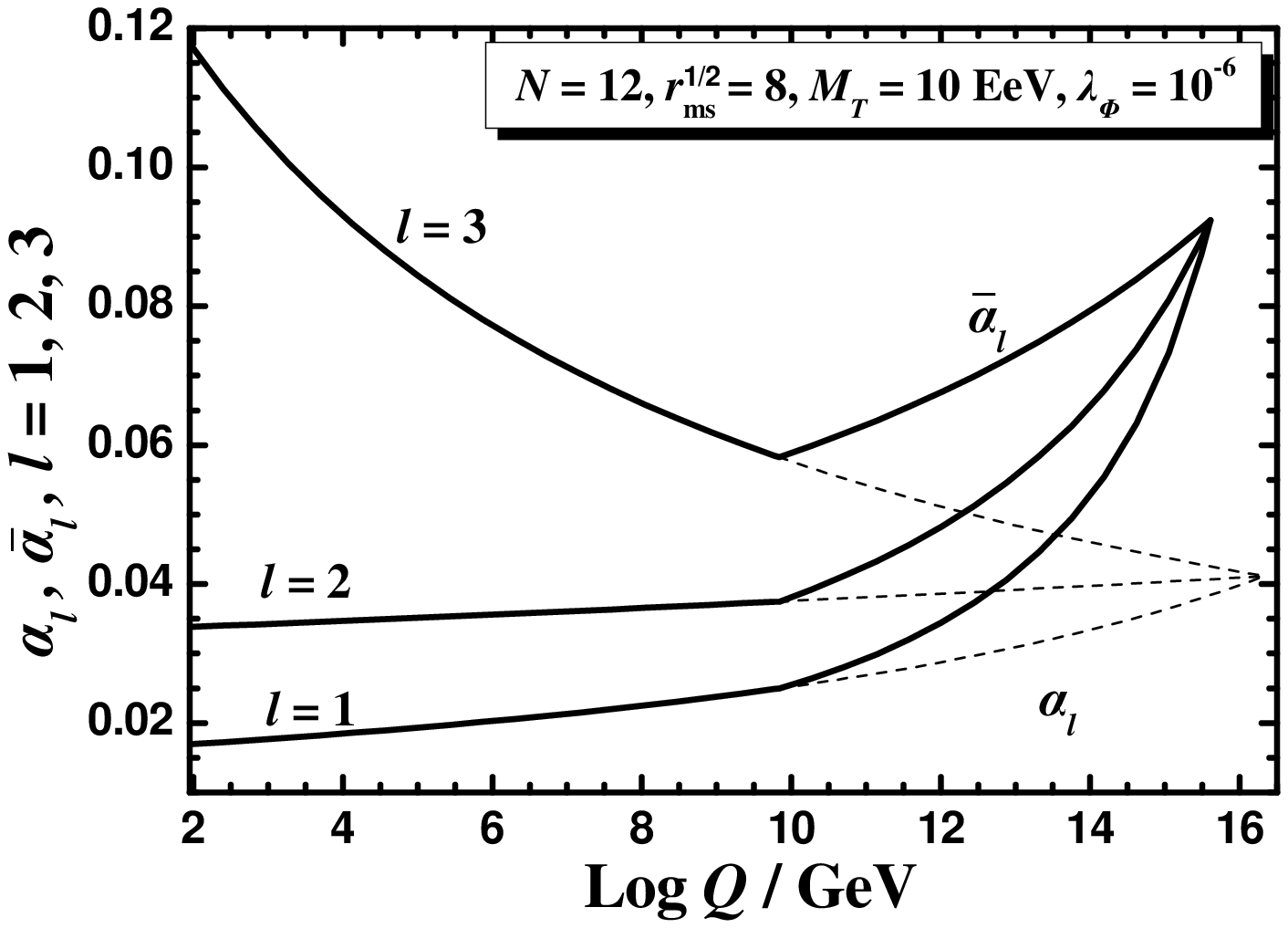,height=3.6in,angle=-90} \hspace*{-1.37
cm}\epsfig{file=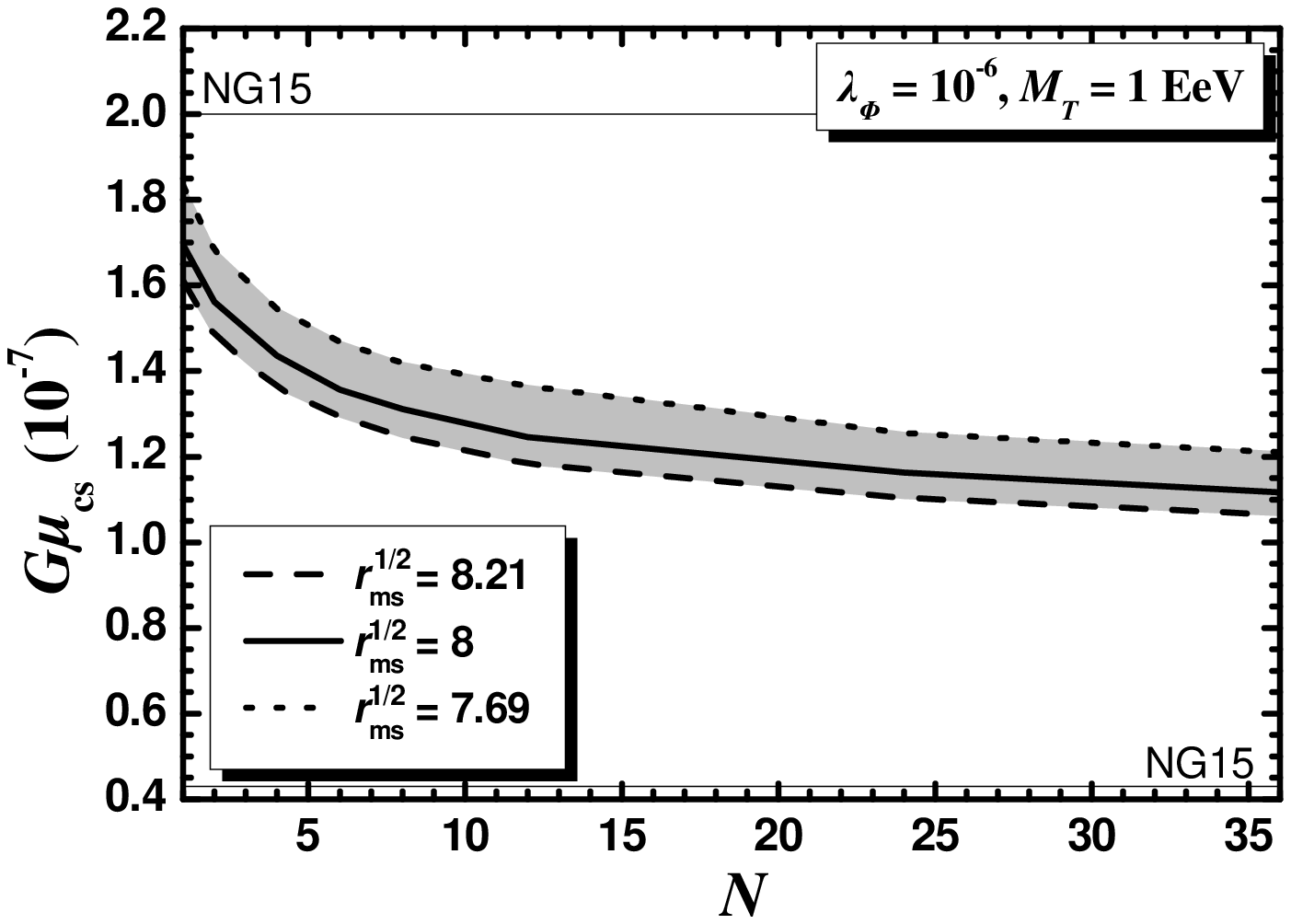,height=3.6in,angle=-90}   \hfill
\end{minipage}\vspace*{-.30in}
\begin{flushleft}
\begin{tabular}[!h]{ll}
\hspace*{-.13in}
\begin{minipage}[t]{7.8cm}\caption[]{\sl\small The RGE evolution of
$\ali$ and $\bali$ within the extended version of our model for
$N=12$, $\mt=10~\EeV$, $\ld=4.15 \cdot 10^{-6}$, $\lp=10^{-6}$,
$\lm=10^{-7}$, $\srms=8$ and $\gmcs\simeq2\cdot10^{-7}$ (solid
thick lines) and the RGE evolution of $\ali$ within MSSM (dashed
thin lines).}\label{figg}\end{minipage}
&\begin{minipage}[t]{7.5cm}\caption[]{\sl\small Allowed (shaded)
region in the $N-\gmcs$ plane for the extended version of our
model $\mt=1~\EeV$, $\lp=10^{-6}$, $\lm=10^{-7}$. The conventions
for the various thick curves are indicated in the legend of the
plot. The bounds from the {\ftn\sffamily NG15}-favored region in
\Eref{csnano} are also depicted.}\label{figgmcs}
\end{minipage}
\end{tabular}
\end{flushleft}\vspace*{-.11in}
\end{figure}


In conclusion, the consideration of the unification hypothesis
enhances the predictability of the model. Despite of this, we
opted to focus on the simplified version of the model  in the main
part of our paper, since our primary scope was the demonstration
of the generation of the metastable CSs after TI. Obviously, this
mechanism may or may not depend on a possible scheme which assures
the gauge coupling unification -- cf.~\cref{ax}.

\renewenvironment{subequations}{%
\refstepcounter{equation}%
\setcounter{parentequation}{\value{equation}}%
  \setcounter{equation}{0}
  \def\theequation{B.\theparentequation{\sf\small \alph{equation}}}%
  \ignorespaces
}{%
  \setcounter{equation}{\value{parentequation}}%
  \ignorespacesafterend
}

\setcounter{equation}{0}

\appendix{Waterfall Phase}\label{appw}

\paragraph{\hspace*{.25cm}} As pointed out in \Sref{hi2b}, a
tachyonic instability occurs in the $\phcb-\phc$ system for $\sg
\simeq \sgc$ with $\sgc$ given in \Eref{sgc}. For the consistent
implementation of our scenario, it is imperative to clarify
whether the SM neutral components of $\phcb$ and $\phc$, $\nhcb$
and $\nhc$ respectively, are moving towards their vacuum in
\Eref{vevs} suddenly or performing a secondary inflationary
period. Indeed, imposing the constraints of \eqs{ntot}{csnano} we
assume a standard cosmological evolution after TI, i.e., an
evolution without a secondary episode of inflation or reheating.
Therefore we have to examine if we obtain a fast or mild
$\nhcb-\nhc$ waterfall regime, adopting the terminology developed
in papers focusing on hybrid inflation \cite{pbha, pbhb, cles,
man}.

To embark on it, we parameterize $\nhcb$ and $\nhc$ as follows
\beq\label{hparw} \nhcb={\phi e^{i\thpb}}\sin\thn/{\sqrt{2}}
~~\mbox{and}~~\nhc={\phi e^{i\thp}}\cos\thn/{\sqrt{2}}
~~~\mbox{where}~~~0\leq\thn\leq\pi/2.\eeq
Then, we select as a waterfall trajectory the D-flat direction
\beq \label{inftr2}
\thp=\thpb=0\>\>\>\mbox{and}\>\>\>\thn={\pi/4}\eeq
with all the other fields in \Eref{inftr} fixed at zero. We can
verify the stability of these fields along the direction above --
cf. \cref{nmh, jhep, nmhk, ighi, univ, unit, sor}. Taking into
account that $\sgc\ll\sgf$ -- see \Fref{fig1b} --, we are allowed
to assume that $\sg$ is confined to its v.e.v in \Eref{vevs}
during the waterfall phase transition. The relevant potential
energy density $V_{\rm w}$ can be derived from \Eref{Vsugra} and
takes the form
\beq V_{\rm w}=\lf 1+\frac{\phi^2}{2N_0\mP^2}\rg^{N_0+1} \lf
V_{\rm
w0}-\frac12m_\phi^2\phi^2+\frac{\phi^4}{32}(2\lp^2+\ld^2)\rg, \eeq
where the various contributions are identified as
\beq V_{\rm
w0}=\mt^2\vt^2+\lt^2(M^2-\vt^2)^2~~\mbox{and}~~m_\phi^2=\frac1{\sqrt{2}}\ld\mt\vt-\lp\lt(\vt^2-M^2).\eeq
The expression above is consistent with \Eref{VF} given that the
prefactor tends to unity for $\phi\ll\mP$. Note that the last term
renders the concave downward $V_{\rm w}(\phi)$ bounded from below,
as it should. Assuming semiclassically an initial displacement at
the onset of the waterfall phase $\phi_{\rm I}\simeq H_{\rm
TIc}/2\pi$ due to quantum diffusion \cite{lindefr, pbha} -- where
$H_{\rm TIc}=\Hhi(\sg = \sgc)$  is estimated from \Eref{hti} --,
we expect that $\phi$ starts rolling down the hilltop $V_{\rm w}$
until to relax at its v.e.v $\vev{\phi}=2\vx$ -- see \Eref{vevs}
-- after performing a series of damped oscillations -- see
\Sref{phi3}. Note that $\phi$ is canonically normalized and the
$\bar\nu_\Phi^c-\nu_\Phi^c$ mixing is negligible since
$\phi\ll\mP$.
%

The slow-roll parameters during the waterfall period can be
determined by the expressions
\beq\epsilon_{\rm w}=-\lf\frac{\dot H_{{\rm w}}}{H_{{\rm
w}0}}\rg^2=\frac12\lf\frac{\phi^\prime}{\mP}\rg^2~~\mbox{and}~~\eta_{\rm
w} = \frac{m_\phi^2}{3H_{{\rm w}0}^2}~~\mbox{where}~~H_{\rm
i}=\frac{\sqrt{V_{\rm i}}}{\sqrt{3}\mP}. \label{srw}\eeq
is the Hubble parameter during the waterfall phase with ${\rm
i}={\rm w}$ or ${\rm w}0$. Also prime denotes derivation w.r.t the
number of e-foldings during the waterfall phase $N_{\rm w}$.
Solving the Klein-Gordon equation for $\phi$ which assumes the
form
\beq \phi''+(3-\epsilon_{\rm w})\phi'-3\eta_{\rm w}\phi=0,\eeq
we obtain the evolution of $\phi$ as a function of $N_{\rm w}$ for
$\phi'(N_{\rm w}=0)=0$ and $\epsilon_{\rm w}\ll1$
\beq \label{phisol} \phi (N_{\rm w}) =  \frac{\phi_{\rm
I}}{2F_{\rm w} + 3} \lf (F_{\rm w} + 3)e^{F_{\rm w} N_{\rm w}} +
F_{\rm w}\ e^{-(F_{\rm w} + 3) N_{\rm w}}\rg ~~\mbox{with}~~F_{\rm
w} = \frac32\lf\sqrt{1 + \frac43\eta_{\rm w}}  - 1\rg. \eeq
Note that the slow-roll approximation \cite{review} is not
convenient for our proposes here since $\eta_{\rm w}\gg1$ as we
verify below. Upon substitution of \Eref{phisol} into the leftmost
expression of \Eref{srw}, we obtain
\beq \label{epw} \epsilon_{\rm w}^{1/2}(N_{\rm
w})=\frac1{\sqrt{2}}\frac{F_{\rm w}(F_{\rm w} + 3)}{2F_{\rm w} +
3} \lf e^{F_{\rm w} N_{\rm w}}-e^{-(F_{\rm w} + 3) N_{\rm w}}\rg
\frac{\phi_{\rm I}}{\mP}.\eeq
The condition of the termination of a possible inflationary stage
allows to determine the total number of e-foldings $\Delta N_{\rm
w}$ elapsed as follows
\beq \label{dNw} \epsilon_{\rm w}(\Delta N_{\rm w})=1~~\Rightarrow
~~\Delta N_{\rm w}\simeq\frac1{F_{\rm w}} \ln\lf\frac{\sqrt{2}(2
F_{\rm w} + 3)}{F_{\rm w}(F_{\rm w} + 3)}\frac{\mP}{\phi_{\rm
I}}\rg,\eeq
where we neglect the decreasing contribution in \Eref{epw}.
Therefore, we may in principle obtain a stage of fast-roll
inflation \cite{lindefr} with $\epsilon_{\rm w}<1$ which generates
a sizable $\Delta N_{\rm w}$ provided that $\eta_{\rm w}\sim1$
which ensures $F_{\rm w}\sim1$.

\begin{figure}[t]
\hspace*{-.12in}
\begin{minipage}{75mm}\renewcommand{\arraystretch}{1.2}
\begin{center} \vspace*{-.2in}{\small
\begin{tabular}{|c||c|c|c|}\hline
{\sc Case}&A&B&C \\ \hline\hline
$\gmcs/10^{-7}$&$0.5$&$1$&$2$\\
$\ld/10^{-7}$&$11.1$&{$7.8$}&{$5.5$}\\ \hline
$m_\phi/\EeV$&$0.98$&$1.25$&{$1.7$}\\
$H_{\rm w0}/\PeV$&{$0.18$}&$3.3$&{$6.2$}\\
$\eta_{\rm w}/10^6$&{$9.9$}&$4.9$&{$2.4$}\\
$F_{\rm w}/10^4$&{$5.4$}&$3.8$&{$2.7$}\\ \hline
$\Delta N_{\rm w}/10^{-3}$&$1.7$&$2.6$&$3.8$ \\ \hline
\end{tabular}}
\end{center}\renewcommand{\arraystretch}{1.0}
\end{minipage} \hfill
\vspace*{-.12in}\begin{minipage}{99mm}
\epsfig{file=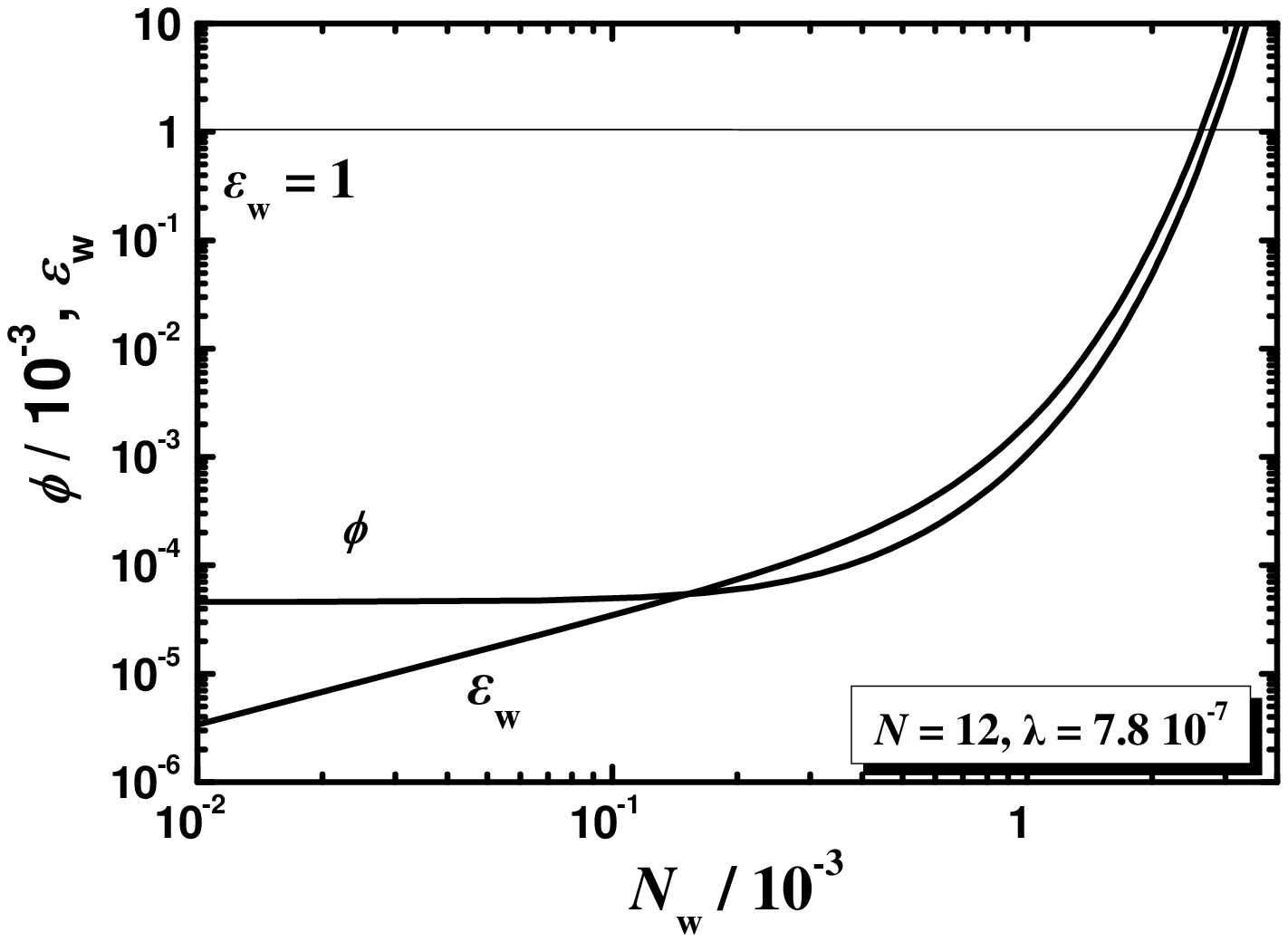,height=3.6in,angle=-90}
\end{minipage}\vspace*{.01in}
\hfill \caption{\sl\small Presented in the Table are the values of
various parameters related to the evolution of $\phi$ during the
waterfall phase for $N=12$, $\mt/\mP=10^{-9}$,
$\lt=2.17\cdot10^{-5}$, $\lp=10^{-6}$, $\lm=10^{-7}$, $\srms=8$
and three selected $\ld$ and $\gmcs$ values. Shown in the plot is
the evolution of $\phi$ and $\epsilon_{\rm w}$ as a function of
$N_{\rm w}$ for the data of Case B. The violation of the inflation
condition occurs when the thin horizontal line is crossed by the
curve $\eph_{\rm w}=\eph_{\rm w}(N_{\rm w})$. }
\label{figw}\end{figure}

To evaluate the duration of a such inflationary period in our
scheme we calculate $\phi$ and $\eph_{\rm w}$ from
\eqs{phisol}{epw} respectively for the representative inputs given
in the Table of \Fref{figw}. Namely, we adopt the parameters
employed in \Fref{fig1b} besides $\ld$ which is varied causing
also a variation to $\gmcs$ within the range of \Eref{csnano}.
Since the parameters related to TI are the same, $\phi_{\rm
I}=4.6\cdot10^{-8}\mP$ is the same for all three cases considered.
We observe that the resulting $\eta_{\rm w}$ and $F_{\rm w} $ are
quite large and so the generated $\Delta N_{\rm w}$, evaluated via
\Eref{dNw}, is well suppressed. For case B  we also depict the
evolution of $\phi$ (rescaled by a factor $10^3$) and $\eph_{\rm
w}$ as a function of $N_{\rm w}$ in the plot of \Fref{figw}. We
see that both $\phi$ and $\eph_{\rm w}$ increase with $N_{\rm w}$
rendering the duration of the waterfall inflation really
negligible since $\eph_{\rm w}$ saturates unity for $\Delta N_{\rm
w}=0.0026$.

Given that the values of the parameters satisfying the
requirements in \Sref{res1} do not change drastically, as shown in
\Fref{fig2}, we can infer that our conclusion regarding the tiny
$\Delta N_{\rm w}$ value is valid within the overall parameter
space of our setting. Therefore, our assumption in
\eqs{ntot}{csnano} respecting a standard cosmological evolution
after TI is self-consistent. As a further consequence, no
primordial black holes \cite{pbha, pbhb} are produced and so the
interpretation of the GW data may be exclusively  done by invoking
the decay of metastable CSs, as detailed in Sec.~\ref{phi2}.

\newpage

\def\ijmp#1#2#3{{\sl Int. Jour. Mod. Phys.}
{\bf #1},~#3~(#2)}
\def\plb#1#2#3{{\sl Phys. Lett. B }{\bf #1}, #3 (#2)}
\def\prl#1#2#3{{\sl Phys. Rev. Lett.}
{\bf #1},~#3~(#2)}
\def\rmp#1#2#3{{Rev. Mod. Phys.}
{\bf #1},~#3~(#2)}
\def\prep#1#2#3{{\sl Phys. Rep. }{\bf #1}, #3 (#2)}
\def\prd#1#2#3{{\sl Phys. Rev. D }{\bf #1}, #3 (#2)}
\def\npb#1#2#3{{\sl Nucl. Phys. }{\bf B#1}, #3 (#2)}
\def\npps#1#2#3{{Nucl. Phys. B (Proc. Sup.)}
{\bf #1},~#3~(#2)}
\def\mpl#1#2#3{{Mod. Phys. Lett.}
{\bf #1},~#3~(#2)}
\def\jetp#1#2#3{{JETP Lett. }{\bf #1}, #3 (#2)}
\def\app#1#2#3{{Acta Phys. Polon.}
{\bf #1},~#3~(#2)}
\def\ptp#1#2#3{{Prog. Theor. Phys.}
{\bf #1},~#3~(#2)}
\def\n#1#2#3{{Nature }{\bf #1},~#3~(#2)}
\def\apj#1#2#3{{Astrophys. J.}
{\bf #1},~#3~(#2)}
\def\mnras#1#2#3{{MNRAS }{\bf #1},~#3~(#2)}
\def\grg#1#2#3{{Gen. Rel. Grav.}
{\bf #1},~#3~(#2)}
\def\s#1#2#3{{Science }{\bf #1},~#3~(#2)}
\def\ibid#1#2#3{{\it ibid. }{\bf #1},~#3~(#2)}
\def\cpc#1#2#3{{Comput. Phys. Commun.}
{\bf #1},~#3~(#2)}
\def\astp#1#2#3{{Astropart. Phys.}
{\bf #1},~#3~(#2)}
\def\epjc#1#2#3{{Eur. Phys. J. C}
{\bf #1},~#3~(#2)}
\def\jhep#1#2#3{{\sl J. High Energy Phys.}
{\bf #1}, #3 (#2)}
\newcommand\njp[3]{{\sl New.\ J.\ Phys.\ }{\bf #1}, #3 (#2)}
\def\prdn#1#2#3#4{{\sl Phys. Rev. D }{\bf #1}, no. #4, #3 (#2)}
\def\jcapn#1#2#3#4{{\sl J. Cosmol. Astropart.
Phys. }{\bf #1}, no. #4, #3 (#2)}
\def\epjcn#1#2#3#4{{\sl Eur. Phys. J. C }{\bf #1}, no. #4, #3 (#2)}

\end{document}